\documentclass[12pt,a4paper]{article}

\usepackage{natbib}
\usepackage[colorlinks,citecolor=blue,urlcolor=blue,filecolor=blue,backref=page]{hyperref}
 \usepackage{doi}
\usepackage{graphicx,psfrag,epsf}
\usepackage{subfigure}
\usepackage{xcolor}
\usepackage{algorithm}
\usepackage{adjustbox}
\usepackage[noend]{algpseudocode}
\usepackage{natbib}
\usepackage{bbm, dsfont}
\usepackage{amsfonts}
\usepackage{siunitx}
\usepackage{amsmath}
\usepackage{bbm}
\usepackage{amsfonts}
\usepackage[flushleft]{threeparttable}
\usepackage{caption}
\newcommand{\be}{\begin{equation}}
	\newcommand{\ee}{\end{equation}}

\usepackage[paperheight=11in,paperwidth=9.5in]{geometry}

\begin{document}
	
	\begin{center}
		\Large \bf   Estimation of Population Size with Heterogeneous Catchability and Behavioural Dependence: Applications to Air and Water Borne Disease Surveillance
	\end{center}
	\begin{center}
		\textbf{Prajamitra Bhuyan$^{\dag}$
			Kiranmoy Chatterjee$^{\ast}$}
		\footnote{Both authors contributed equally to this paper.}\\
		$^{\dag}$Indian Institute of Management, Calcutta\\
		$^{\ast}$Bidhannagar College Kolkata
	\end{center}
	
	\begin{abstract}
		Population size estimation based on the capture-recapture experiment is an interesting problem in various fields including epidemiology, criminology, demography, etc. In many real-life scenarios, there exists inherent heterogeneity among the individuals and dependency between capture and recapture attempts. A novel trivariate Bernoulli model is considered to incorporate these features, and the Bayesian estimation of the model parameters is suggested using data augmentation. Simulation results show robustness under model misspecification and the superiority of the performance of the proposed method over existing competitors. The method is applied to analyse real case studies on epidemiological surveillance. The results provide interesting insight on the heterogeneity and dependence involved in the capture-recapture mechanism. The methodology proposed can assist in effective decision-making and policy formulation.
		\paragraph{}	
	\end{abstract}

	\section{Introduction}\label{Intro}
The knowledge about the true prevalence of a disease in a specified period is an essential requirement for surveillance and effective policy formulation regarding the healthcare system of a state \citep{Bird18}. In general, the available data source fails to cover all the relevant events and that leads to an undercount of the target population suffering from the disease. Therefore, disease ascertainment data are accumulated from multiple sources to increase the coverage and for the estimation of the disease prevalence \citep{Papoz96}. This method is known as multiple system estimation (MSE) which is equivalent to the capture-mark-recapture (CMR) method traditionally applied to estimate the size of wildlife populations \citep{Hook95, WorkingGroup95a}. The MSE is also popularly used for the estimation of demographic counts such as births and deaths \citep{O'Hara}, census undercoverage \citep{Zaslavsky93}, crime incidence \citep{Cruyff17}, number of beneficiaries in a welfare economic study \citep{Bird18}, etc. Lately, the MSE is also applied in clinical settings for screening and preventive studies \citep{Bohning09}.

The MSE involves the matching of the individuals enlisted from different sources and recording the various forms of overlaps among these lists. Traditionally two sources are considered, and it is a common practice to assume the independence between the lists for the estimation of population size which in many cases is affected by the correlation bias when the underlying assumption fails \citep{Chandrasekar49}. In most cases, these sources of information are dependent because an individual’s behaviour changes with the time of subsequent recapture attempts after the initial attempt. Another possible reason for dependence between lists is that capture probability may vary across individuals in each list \citep{Chao01a}. Therefore, more than two data sources are considered to capture more eligible events and to assess the underlying interdependence among the lists. In particular, three samples or lists are commonly considered in epidemiological surveillance \citep{Gallay00, Hest07, Ruche13}. Note that the population of interest is assumed closed during the ascertainment of events from the three sources which generally occurs within a short period. The data obtained from three different sources are summarized in the form of an incomplete $2^3$ contingency table. This data structure, presented in Table S1 of the Supplementary Material, is typically known as the triple record system (TRS). Denote the capture status of an event in the first, second, and third lists by $i$, $j$, and $k$, respectively. The dummy variables $i$, $j$, and $k$ take value 1 for capture and 0 otherwise. The total number of events with a particular capture status, say ($i,j,k$), is denoted by $x_{ijk}$. For example, the count $x_{101}$ associated with the cell ($1,0,1$) represents the number of events that appear in the first and the third lists but are absent in the second list. Note that $N=\sum_{i,j,k}x_{ijk}=n+x_{000}$, where $n=\sum_{i,j,k\neq 0}x_{ijk}$ refers to the total number of individuals that are recorded at least in one list, and $x_{000}$ refers the total count of individuals that are not recorded in any of the three lists. Therefore, the problem of estimating the population size $N=\sum_{i,j,k}x_{ijk}$ is equivalent to that of the unknown cell count $x_{000}$. Interested readers are referred to \cite{Zaslavsky93, Chatterjee20a} for a detailed discussion on TRS and associated estimation methodology. In this article, we consider the problem of estimating disease prevalence motivated by TRS data available from the following case studies on epidemiological surveillance related to fatal illness due to air and water borne bacteria and viruses.

\subsection{Legionnaires' Disease}\label{Legionnaires-data}
Legionnaires' Disease (LD) is an unusual variant of pneumonia that occurs sporadically and in outbreaks caused by the bacterium Legionella which spreads through airborne water droplets \citep{Den02, Lettinga02}. The motivation for considering surveillance on LD is two-fold in light of the current pandemic of COVID-19. Firstly, reports of co-infection with respiratory pathogens are increasing throughout the world \citep{Lai20}. It has also been reported that 50\% of COVID-19 patients who succumbed had secondary bacterial infections like LD \citep{Zhou20}. Secondly, patients with COVID-19 should be screened for LD because the signs and symptoms of both infections are similar. The scientific community very recently found that COVID-19 infections can predispose patients to Legionella co-infections and consequently pose a serious threat to high-risk COVID-19 patients, which can lead to an increase in the severity of LD and mortality \citep{Dey20}.

The Netherlands is one of the worst affected countries by LD in the world. We consider a study on LD in the Netherlands conducted during the period from 2000 to 2001 \citep{Hest08}. The average national annual prevalence rate of LD was 1.4 patients per 100 thousand inhabitants of the Netherlands in 1999. But experts suspect that the conventional LD notification system contains false-positive cases and is often incomplete for true positive cases of LD \citep{Hest08}. In this regard, efficient record-linkage and capture-recapture analysis for assessing the quality and completeness of infectious disease registers are warranted. In this study, information on LD-affected patients is collected from Disease Notifications Register (DNR) from the Health Care Inspectorate, Laboratory results, and Hospital admissions. In the DNR, 373 LD patients are identified through record linkage. On the other hand, a total of 261 patients with a positive test for LD are recorded from the Laboratory survey. In total, 663 patients are enlisted in the hospital records with a relatively large number of 332 patients captured exclusively in this list. The overlap counts associated with these three lists are provided in Section 1 of the Supplementary Material (see Table S2). The presence of dominant interaction between the DNR and the Laboratory register has been indicated in a previous study \citep{Hest08}. Moreover, various factors like geographical region, age, and laboratory diagnostics technique cause heterogeneity in the capture probabilities of the individuals in each list \citep{Nardone03, Den02b}.

\subsection{Hepatitis A Virus}\label{Hepatitis-data}
Viral hepatitis is among the top ten leading causes of mortality worldwide and is the only communicable disease where mortality is increasing \citep{Brown2017}. The hepatitis A virus (HAV) spreads through person-to-person contact and contaminated food or water. In high-income countries, HAV infection usually occurs among drug addicts, and men having sex with men, including patients with HIV and other sexually transmitted infections due to a low prevalence of anti-HAV antibodies \citep{Cuthbert01, Chen19}. The scientific community opines that continuous HAV surveillance and evaluation of long-term vaccine effectiveness among HIV patients are warranted and HAV vaccination for all in childhood will provide more sustainable immunity in the general population. Therefore, it is important to know the actual population size infected with HAV for efficient vaccination policy and disease surveillance.

Approximately 1.4 million infections are reported worldwide each year, of which approximately half occur in Asian countries \citep{Martin2006}. In particular, Taiwan has a history of a high prevalence of HAV infection because of indigenous townships with inadequate water, sanitation, and hygiene infrastructure in the last decade of the twentieth century \citep{Chen19}. We consider a study on HAV outbreak in northern Taiwan from April to July 1995 near a technical college \citep{Chao97}. There are three registered sources of information on the HAV infected college students available: (i) P-list consists of 135 records based on a serum test taken by the Institute of Preventive Medicine, Department of Health of Taiwan, (ii) Q-list includes 122 cases reported by doctors in local hospitals and this list was provided by the National Quarantine Service and (iii) E-list comprise of 126 cases based on questionnaires conducted by epidemiologists. 
\citet{Tsay01} reported the existence of heterogeneity in capture probabilities in all three sources, and low overlap has been observed among the lists. The summarized data are provided in Table S2 in Section 1 of the Supplementary Material.

\subsection{The Challenges in Analysing TRS Data}\label{Challenge}
As mentioned before, the assumption of list independence is often violated in practice mainly due to two reasons: (i) list dependence caused by behavioural response variation over different sources, i.e. the inclusion status of an individual in one list has a direct causal effect on his/her inclusion status in other lists; (ii) heterogeneity in the propensity to be captured in a list over individuals \citep{Chao01a}. Even when no behavioural response variation occurs for individuals over different lists, their ascertainment in the lists becomes positively dependent if the capture probabilities are heterogeneous across individuals in each list \citep{Chandrasekar49, Chao15}. We call this second type of list dependence as heterogeneity-induced list dependence \citep{Darroch93}. In both the case studies, discussed in sections \ref{Legionnaires-data} and \ref{Hepatitis-data}, the associated TRS datasets are potentially affected by the list dependence because of these two aforementioned reasons.

In the literature of MSE for closed population, various models account for the first kind of list dependence i.e. dependence due to the behavioural change in response of each individual at the time of subsequent attempts after the first capture. This is called behavioural list dependence. \citet{Fienberg72b} proposed a log-linear model (LLM) incorporating various possible interactions among the available lists. An overview of the LLMs and their applications for population size estimation are provided in \citet{Cormack89, WorkingGroup95a}. However, the parameters associated with the LLMs are not well interpretable to the practitioners \citep{Coumans17}. In this context, model $M_{tb}$ is frequently used in ecological surveillance to account for the list dependence due to behavioural response variation \citep{Otis78}. 
This model may produce unstable estimates in many instances \citep{Rivest22} and has several other limitations in the context of human population \citep{Chatterjee20b}. 
To model the individual heterogeneity, \citet{Sanathanan72a} adapted the Rasch model and derived associated estimation methodology. Later, \citet{Goodman74} developed latent class models assuming the population is composed of homogeneous strata with independent lists. \citet{Darroch93} provided some variants of the heterogeneity model by extending the log-linear representation of the Rasch model, such as quasi-symmetry and partial quasi-symmetry models \citep{Chao98}. 
However, model selection in the case of LLMs may be difficult because an adequate fit to the observed cells may not necessarily produce an efficient estimate of the unobserved cell \citep{Chao15}. Also, the existence of heterogeneity in data might result in the lack of a reliable estimate \citep{WorkingGroup95a}. Recently, \citet{Vallier16} developed a Bayesian non-parametric method to accommodate individual heterogeneity based on Dirichlet process mixtures assuming independence between the lists.

The most general scenario of the capture-recapture experiment encompasses both the list dependence due to behavioural response variation and individual heterogeneity in the capture probabilities of the individuals. The heterogeneity-induced dependence among lists is generally confounded with the list dependence due to behavioural response variation \citep{Chao01b}. Such a complex scenario may often arise in various fields of applications including epidemiology \citep{Tsay01, Hest08}. It is challenging to incorporate these two factors together in a model and provide an efficient estimate of the population size for TRS data \citep{Chao01a}. In this context, an ecological model $M_{bh}$ has been developed to estimate the abundance of the animal population \citep{Otis78}. However, the reliability of the estimate based on this model depends on several conditions, including the sequential ordering of the capture sampling \citep{Otis78}. These conditions may not be relevant for MSE. As a result, this model may produce an inefficient and/or infeasible estimate (\textit{see} Sections \ref{Simulation}-\ref{Casestudy}). Chao's non-parametric sample coverage approach accounts for both the individual heterogeneity and possible interaction between the lists \citep{Chao98, Chao15}. However, the performance of this estimator is not satisfactory as it may also provide infeasible estimates. See Section \ref{Simulation} for details.

In this article, we propose a novel modeling approach based on the trivariate Bernoulli model that incorporates both the behavioural dependence between the lists and the individual heterogeneity along with the list variation. The parameters of the model are easily interpretable and provide interesting insights into the capture-recapture mechanism. We develop a Bayesian estimation methodology for the estimation of $N$ and associated model parameters. In particular, we demonstrate how Gibbs sampling facilitates Bayesian analysis of capture-recapture experiments in the most general as well as complex scenarios. As a result, formulations that were previously avoided because of analytical intractability and computational time can now be easily considered for practical applications. The description of the proposed model and its special cases are provided in Section \ref{general}. In Section \ref{Inference}, estimation methodologies for the population size $N$ and associated model parameters are discussed. Next, the performance of the proposed method is compared with the existing competitors through an extensive simulation in Section \ref{Simulation}. Sensitivity analyses concerning prior choices and model mis-specifications are also discussed in the same section. Analyses of the real datasets on LD and HAV are presented in Section \ref{Casestudy}. We summarise the key findings and conclude with some discussion on future research in Section \ref{Discussion}.

\section{Modeling Individual Heterogeneity and List Dependence}\label{general}
In this section, we first discuss the Trivariate Bernoulli model (TBM) proposed by \citet{Chatterjee20b} for modeling capture–recapture data under TRS. The TBM is used to incorporate the inherent list dependence, and its parameters possess physical interpretation. We extend this model in a more general setup to account for the `individual heterogeneity' in addition to the list dependence. 

In TRS, some individuals behave independently over the three different capture attempts and behavioural dependence exists for the rest of the population. To model the association among the three lists $L_1$, $L_2$ and $L_3$, we first consider that $\alpha_1$, $\alpha_2$, and $\alpha_3$ proportion of individuals possess pairwise dependence between lists ($L_1$ and $L_2$), ($L_2$ and $L_3$) and ($L_1$ and $L_3$). Further, we consider the second-order dependency among the three lists $L_1$, $L_2$, and $L_3$ for $\alpha_{4}$ proportion of individuals. Therefore,
the remaining $(1-\alpha_0)$ proportion of individuals, where $\alpha_0=\sum_{\omega=1}^{4}\alpha_{\omega}$, behave independently over the three lists. Now we define a triplet ($X_{1h}^{*}, X_{2h}^{*}, X_{3h}^{*}$) which represents the latent capture statuses of the $h$th individual in the first, second and third attempts respectively, for $h=1,2,\ldots, N$. The latent capture status $X_{sh}^{*}$ takes the value 1 or 0, denoting the presence or absence of the $h$th individual in the $s$th list, for $s=1,2,3$. Under this setup,  $X_{2h}^{*}=X_{1h}^{*}$ and $X_{3h}^{*}=X_{2h}^{*}$ for $\alpha_{1}$ and $\alpha_{2}$ proportion of individuals, respectively. Similarly, $X_{3h}^{*}=X_{1h}^{*}$ for $\alpha_{3}$ proportion of individuals, and $X_{3h}^{*}=X_{2h}^{*}=X_{1h}^{*}$ for $\alpha_{4}$ proportion of individuals. Now, denote $Z_h^{(1)}$, $Z_h^{(2)}$, and $Z_h^{(3)}$ as the the inclusion statuses of the $h$th individual in $L_1$, $L_2$, and $L_3$ respectively, for $h=1,2,\ldots,N$. Therefore, we can formally write the model to account for the interdependence among the three lists as:
\begin{eqnarray}
	(Z_h^{(1)},Z_h^{(2)}, Z_h^{(3)}) = \begin{cases} 
		(X_{1h}^{*},X_{1h}^{*},X_{3h}^{*})  & \mbox{ with prob. } \alpha_{1},\\
		(X_{1h}^{*},X_{2h}^{*},X_{2h}^{*})  & \mbox{ with prob. } \alpha_{2},\\
		(X_{1h}^{*},X_{2h}^{*},X_{1h}^{*})  & \mbox{ with prob. } \alpha_{3},\\
		(X_{1h}^{*},X_{1h}^{*},X_{1h}^{*})  & \mbox{ with prob. } \alpha_{4},\\
		(X_{1h}^{*},X_{2h}^{*},X_{3h}^{*}) & \mbox{ with prob. } 1-\alpha_0,
	\end{cases}\label{prob-model}
\end{eqnarray}  
where $X_{1h}^{*}$'s, $X_{2h}^{*}$'s and $X_{3h}^{*}$'s are independently distributed Bernoulli random variables with parameters $\mathcal{P}_{1}$, $\mathcal{P}_{2}$ and $\mathcal{P}_{3}$, respectively, for all $h=1,\ldots,N$. Note that $\mathcal{P}_{s}$ refers to the capture probability of a causally independent individual in the $s$th list. To incorporate heterogeneity in capture probabilities, we now consider $X_{1h}^{*}$'s to follow independent Bernoulli distributions with parameter $\mathcal{P}_{sh}$ and model it as:
\begin{eqnarray}
	logit(\mathcal{P}_{sh})=\log\left(\frac{\mathcal{P}_{sh}}{1-\mathcal{P}_{sh}}\right)&=&b_{sh}, \quad \text{$h=1,\ldots,N$},\label{logit}
\end{eqnarray}
where $b_{sh}$'s are independent and identically distributed realizations of random effect $b_{s}$, for each $s=1,2,3$. We also assume $b_s$'s are independently distributed. We refer to this generic model, given in (\ref{prob-model}) and (\ref{logit}), as the Trivariate Heterogeneous Bernoulli model (THBM).

\subsection{Special Cases}
The generic form of the proposed THBM incorporates the list variation, individual heterogeneity, and behavioural dependence arising from different sources. However, in some cases, the effect of the behavioural response variation is confounded due to heterogeneous catchability and may not provide additional information for the estimation of $N$. In such cases, it is prudent to consider a parsimonious model with some restrictions on the THBM. In particular, one can consider $\alpha_{\omega}=0$, for $\omega=1, \ldots,4$, and this reduced model is applicable for the scenarios where the generalized Rasch model is useful \citep{Darroch93, Chao98} (see Section 2.2 of the Supplementary Material). One can further assume $b_{l}$'s are identically distributed where the catchability is the same over the individuals irrespective of the lists. 
The THBM model reduces to TBM when $b_l$'s are degenerate random variables. This model is useful for TRS when the individuals are equally catchable in each list.  As discussed in \citet{Chatterjee20b}, the TBM can be further reduced to the submodels TBM-1 and TBM-2 if we consider $\alpha_3=0$ and $\alpha_4=0$, respectively. In the absence of list dependence (i.e. $\alpha_{\omega}=0$, for $\omega=1,\ldots,4$) the TBM and the $M_t$ model are equivalent \citep{Otis78}. 

\section{Estimation Methodology}\label{Inference}
A classical approach for estimating $N$ in the context of CMR is based on the likelihood theory, where the vector of observed cell counts $$\boldsymbol{x}=\left\{x_{ijk}:x_{ijk}=\sum_{h=1}^{n}\mathbb{I}\left[Z_h^{(1)}=i, Z_h^{(2)}=j, Z_h^{(3)}=k\right]; i,j,k=0,1;i=j=k\neq0\right\}$$ (as presented in Table S1 of the Supplementary
Material, follow a multinomial distribution with index parameter $N$ and the associated cell probabilities $\boldsymbol{p}=\{p_{ijk}:i,j,k=0,1;i=j=k\neq0\}$ \citep{Sanathanan72b}, where $\mathbb{I}\left[\right]$ denotes indicator function. Therefore, the likelihood function is given by
\begin{eqnarray}\label{Mult_lik}
	L(N,\boldsymbol{p}|\boldsymbol{x})& = & \frac{N!}{\prod_{i,j,k=0,1;i=j=k\neq0}^{}x_{ijk}!(N-n)!}\prod_{i,j,k=0,1}^{}p_{ijk}^{x_{ijk}},\nonumber
\end{eqnarray}
where $x_{000}=N-n$. The TBM is characterized by the parameters $N$, $\alpha_1$, $\alpha_2$ ,$\alpha_3$, $\alpha_4$, $\mathcal{P}_1$, $\mathcal{P}_2$, $\mathcal{P}_3$, and using its relationships with the cell probabilities $p_{ijk}$, the likelihood function is given by
\begin{eqnarray}\label{L_Model_I}
	L(N,\boldsymbol{\alpha},\boldsymbol{\mathcal{P}}|\boldsymbol{x})&\propto & \frac{N!}{(N-n)!} [(1-\alpha_0)\mathcal{P}_1 \mathcal{P}_2 \mathcal{P}_3 + \alpha_1\mathcal{P}_1 \mathcal{P}_3 + \alpha_2 \mathcal{P}_1 \mathcal{P}_2 + \alpha_3 \mathcal{P}_1 \mathcal{P}_2 + \alpha_4 \mathcal{P}_1]^{x_{111}}\nonumber\\
	&&\times
	[(1-\alpha_0) \mathcal{P}_1 \mathcal{P}_2 (1-\mathcal{P}_3) + \alpha_1 \mathcal{P}_1 (1-\mathcal{P}_3)]^{x_{110}}\nonumber\\
	&&\times
	[(1-\alpha_0)(1-\mathcal{P}_1)\mathcal{P}_2 \mathcal{P}_3 + \alpha_2 (1-\mathcal{P}_1) \mathcal{P}_2]^{x_{011}}\nonumber\\
	&&\times
	[(1-\alpha_0)\mathcal{P}_1(1-\mathcal{P}_2)(1-\mathcal{P}_3) + \alpha_2 \mathcal{P}_1 (1-\mathcal{P}_2)]^{x_{100}}\nonumber\\
	&&\times
	[(1-\alpha_0)\mathcal{P}_1(1-\mathcal{P}_2)\mathcal{P}_3 + \alpha_3 \mathcal{P}_1 (1-\mathcal{P}_2)]^{x_{101}}\nonumber\\  
	&&\times
	[(1-\alpha_0)(1-\mathcal{P}_1)\mathcal{P}_2(1-\mathcal{P}_3) + \alpha_3 (1-\mathcal{P}_1) \mathcal{P}_2]^{x_{010}}\nonumber\\
	&&\times
	[(1-\alpha_0)(1-\mathcal{P}_1)(1-\mathcal{P}_2)\mathcal{P}_3 + \alpha_1 (1-\mathcal{P}_1) \mathcal{P}_3]^{x_{001}}\nonumber\\
	&&\times
	[(1-\alpha_0) (1-\mathcal{P}_1) (1-\mathcal{P}_2) (1-\mathcal{P}_3) + \alpha_1 (1-\mathcal{P}_1) (1-\mathcal{P}_3)\nonumber\\ 
	&& +  \alpha_2 (1-\mathcal{P}_1) (1-\mathcal{P}_2) + \alpha_3 (1-\mathcal{P}_1) (1-\mathcal{P}_2)+ \alpha_4 (1-\mathcal{P}_1)]^{N-n},
\end{eqnarray}
where $p_{000}=1-\sum_{i,j,k=0,1;i=j=k\neq0}^{}p_{ijk}$, $\boldsymbol{\alpha}=(\alpha_1, \alpha_2, \alpha_3, \alpha_4)$, and $\boldsymbol{\mathcal{P}}=(\mathcal{P}_1, \mathcal{P}_2, \mathcal{P}_3)$. 
In contrast to TBM, the proposed THBM accounts for the `individual heterogeneity' in terms of the variations in capture probabilities considering $\mathcal{P}_s=\frac{\exp(b_{s})}{1+\exp(b_{s})}$ as a function of random effects $b_s$ for $s=1,2,3$.  Now, integrating the likelihood function, given in (\ref{L_Model_I}), with respect to the distribution of random effects $b_{s}$, we obtain the marginal likelihood function of $\boldsymbol{\theta}$ as
\begin{eqnarray}\label{ultimate_likelihood}
	L(\boldsymbol{\theta}|\boldsymbol{x})& = & \int_{\mathbb{R}^3}^{} L(N,\boldsymbol{\alpha},\boldsymbol{\mathcal{P}}|\boldsymbol{x})\times\left\{\prod_{s=1}^{3} g_{b_s}(b_{s}|\delta_{s})\right\}db_{1}db_{2}db_{3},
\end{eqnarray}
where $\boldsymbol{\theta}=(N,\boldsymbol{\alpha},\boldsymbol{\delta})$, and $g_{b_s}(\cdot|\delta_s)$ is a suitably chosen density function (e.g. Gaussian, logistic, etc.) of $b_s$'s with the unknown real-valued parameter or parameter vector $\delta_s$ for $s=1,2,3$. In the conventional frequentist approach, computational challenges arise in the model fitting due to intractable numerical integration involved in the aforementioned log-likelihood function \citep{Coull99}. Using Gaussian quadrature or Laplace's method, one can consider quasi-likelihood approaches or approximation methods for numerical integration. Alternatively, a Bayesian estimate of the parameters of interest can be obtained using the Metropolis-Hastings algorithm.

\subsection{Identifiability Issues}\label{Ident}
The model parameters associated with THBM are not identifiable with respect to TRS data. In a frequentist setup, the performance of point estimation in the absence of identifiability may yield unsatisfactory results such as non-uniqueness of maximum likelihood estimates \citep{White82, Aldrich02}. To avoid the issues related to identifiability, \citet{Chatterjee20b} considered two submodels TBM-1 and TBM-2 of TBM keeping $\alpha_3$ and $\alpha_4$ fixed as $0$, respectively. Similar restrictions on parameters are also considered in the model $M_{tb}$ and log-linear models as well as in the sample coverage approach proposed by \citet{Chao98}. However, this process of model contraction may lead to a model that involves dubious assumptions or a model that is less realistic in some other way \citep{Gustafson05}. For example, available sources in epidemiological surveillance are not time ordered \citep{Chao01a}, and the capture attempts are supposed to be interdependent between themselves \citep{Tsay01, Ruche13}. Hence, the dependence between the first and third lists may be present, unlike the situation that is modelled by TBM-1. Similarly, TBM-2 is unsuitable for scenarios when the second-order interaction among the lists is present. On the other hand, non-identifiability is not seen as a big obstacle to performing inferential procedures in the Bayesian paradigm \citep{Wechsler13}. Infusing crude prior information into a non-identifiable model is considered an alternative to achieving identifiability through a model contraction. It is important to note that a non-identifiable model may bring information about the parameters of
interest, and identifiable models do not necessarily lead to better decision-making than non-identifiable ones \citep{Gustafson05, Wechsler13}. In the following subsection, we propose a Bayesian estimation methodology of the parameter of interest $N$ and associated parameters of the THBM under different choices of prior specifications on the nuisance parameters. Later, we also demonstrate that the proposed estimate of $N$ is not sensitive to those prior specifications and outperforms existing competitors through an extensive simulation study (see Section \ref{Simulation}).


\subsection{Bayesian Approach}\label{Bayesian} We propose a Bayesian estimation of the model parameters
involved in the THBM given by (\ref{prob-model}) and (\ref{logit}), using the MCMC algorithm based on data augmentation. Unlike the usual frequentist setup, the proposed Bayesian approach provides a natural framework for prediction over unobserved data. Thus, by generating posterior predictive densities, rather than point estimates, we can make probability statements giving greater flexibility in presenting results.
For instance, we can discuss findings concerning specific hypotheses or in terms of credible intervals which can offer a more intuitive understanding for the practitioners.

As mentioned before, the likelihood function $L(\boldsymbol{\theta}|\boldsymbol{x})$, given by (\ref{ultimate_likelihood}), is not mathematically tractable due to the involvement of an integral and the additive structure of the cell probabilities ${p_{ijk}}$ impose further challenges. One can apply the Metropolis-Hastings algorithm to generate samples from the posterior distribution of the model parameters $\boldsymbol{\theta}$, however, such computationally intensive methods are time inefficient and not appealing for the practitioners. To avoid such difficulty, we first attempt to simplify the likelihood function by taking a cue from the data augmentation strategy as suggested by \citet{Tanner87}. Data augmentation refers to a scheme of augmenting the observed data so that it is easier to analyse. From a Bayesian point of view, this strategy may help to obtain an explicitly known form of conditional posterior distributions of the parameters of interest. For this purpose, we partition the cell counts $x_{ijk}$ depending on the various types of list dependence as described in (\ref{prob-model}), and define a vector of latent cell counts as:
$${\boldsymbol{y}=\left\{y_{ijk,u}: \sum_{u=1}^{\nu}y_{ijk,u}=x_{ijk}; i,j,k=0,1; u=1,\ldots,\nu; \nu=5^{\eta}2^{1-\eta}; \eta= \mathcal{I}(i=j=k) \right\}}.$$
where $\mathcal{I}(\cdot)$ is an indicator function. For example, $x_{111}$ can arise from all the five types of dependence structures, whereas $x_{110}$ can be generated only based on the first and last types of dependence structures, as presented in (\ref{prob-model}). We also treat the random effects $\boldsymbol{b}=(b_1,b_2,b_3)$ as unobserved data. Therefore, the likelihood function of $\boldsymbol{\theta}$, based on the complete data $(\boldsymbol{x},\boldsymbol{y},\boldsymbol{b})$, is given by 
\begin{eqnarray}\label{com_likelihood}		\mathcal{L}_c(\boldsymbol{\theta}|\boldsymbol{x},\boldsymbol{y},\boldsymbol{b})& = &  \mathcal{L}_c(N,\boldsymbol{\alpha},\boldsymbol{b}|\boldsymbol{x},\boldsymbol{y})\times\left\{\prod_{s=1}^{3} g_{b_s}(b_{s}|\delta_{s})\right\},
\end{eqnarray}
where
\begin{eqnarray}		\mathcal{L}_c(N,\boldsymbol{\alpha},\boldsymbol{b}|\boldsymbol{x},\boldsymbol{y})&\propto & \frac{N!}{(N-n)!} [(1-\alpha_0)\mathcal{P}_1 \mathcal{P}_2 \mathcal{P}_3]^{y_{111,1}}\times[\alpha_1 \mathcal{P}_1 \mathcal{P}_3]^{y_{111,2}}\times [\alpha_2 \mathcal{P}_1 \mathcal{P}_2]^{y_{111,3}}\times[\alpha_3 \mathcal{P}_1 \mathcal{P}_2]^{y_{111,4}}\nonumber\\
	&&\times[\alpha_4 \mathcal{P}_1]^{x_{111}-\sum_{i=1}^{4}y_{111,i}}\times
	[(1-\alpha_0) \mathcal{P}_1 \mathcal{P}_2 (1-\mathcal{P}_3)]^{y_{110,1}} \times [\alpha_1 \mathcal{P}_1 (1-\mathcal{P}_3)]^{x_{110}-y_{110,1}}\nonumber\\
	&&\times
	[(1-\alpha_0)(1-\mathcal{P}_1)\mathcal{P}_2 \mathcal{P}_3]^{y_{011,1}} \times [\alpha_2 (1-\mathcal{P}_1) \mathcal{P}_2]^{x_{011}-y_{011,1}}\nonumber\\
	&&\times
	[(1-\alpha_0)\mathcal{P}_1(1-\mathcal{P}_2)(1-\mathcal{P}_3)]^{y_{100,1}} \times [\alpha_2 \mathcal{P}_1 (1-\mathcal{P}_2)]^{x_{100}-y_{100,1}}\nonumber\\
	&&\times
	[(1-\alpha_0)\mathcal{P}_1(1-\mathcal{P}_2)\mathcal{P}_3]^{y_{101,1}} \times [\alpha_3 \mathcal{P}_1 (1-\mathcal{P}_2)]^{x_{101}-y_{101,1}}\nonumber\\  
	&&\times
	[(1-\alpha_0)(1-\mathcal{P}_1)\mathcal{P}_2(1-\mathcal{P}_3)]^{y_{010,1}} \times [\alpha_3 (1-\mathcal{P}_1) \mathcal{P}_2]^{x_{010}-y_{010,1}}\nonumber\\
	&&\times
	[(1-\alpha_0)(1-\mathcal{P}_1)(1-\mathcal{P}_2)\mathcal{P}_3]^{y_{001,1}} \times [\alpha_1 (1-\mathcal{P}_1) \mathcal{P}_3]^{x_{001}-y_{001,1}}\nonumber\\
	&&\times
	[(1-\alpha_0) (1-\mathcal{P}_1) (1-\mathcal{P}_2) (1-\mathcal{P}_3)]^{y_{000,1}} \times [\alpha_1 (1-\mathcal{P}_1) (1-\mathcal{P}_3)]^{y_{000,2}}\nonumber\\ 
	&& \times [\alpha_2 (1-\mathcal{P}_1) (1-\mathcal{P}_2)]^{y_{000,3}} \times [\alpha_3 (1-\mathcal{P}_1) (1-\mathcal{P}_2)]^{y_{000,4}} \times [\alpha_4 (1-\mathcal{P}_1)]^{N-n-\sum_{i=1}^{4}y_{000,i}}\nonumber.
\end{eqnarray}
Interestingly, the complete data likelihood, given by (\ref{com_likelihood}), possesses a simple form as a product of the power functions of the parameters associated with the THBM. Hence, the joint posterior density of all the unobserved quantity in the model $\boldsymbol{\theta}$, $\boldsymbol{y}$, and $\boldsymbol{b}$ given the observed data $\boldsymbol{x}$ is provided by
\begin{eqnarray}\label{joint-posterior}	\pi(\boldsymbol{\theta},\boldsymbol{b},\boldsymbol{y}|\boldsymbol{x})\propto \mathcal{L}_c(\boldsymbol{\theta}|\boldsymbol{x},\boldsymbol{y},\boldsymbol{b})\times\pi(\boldsymbol{\theta}), 
\end{eqnarray}
where $\pi(\boldsymbol{\theta})$ denotes the prior for $\boldsymbol{\theta}$. Following \citet{Tanner87}, we employ a simple iterative algorithm to generate samples from the posterior density of $\boldsymbol{\theta}$. To implement the following steps, one must be able to sample from the conditional posterior distributions of $\boldsymbol{\theta}$ given $(\boldsymbol{x},\boldsymbol{y},\boldsymbol{b})$,  $\boldsymbol{y}$ given $(\boldsymbol{\theta},\boldsymbol{x},\boldsymbol{b})$, and  $\boldsymbol{b}$ given $(\boldsymbol{\theta},\boldsymbol{x},\boldsymbol{y})$, denoted by $\pi(\boldsymbol{\theta}|\boldsymbol{x},\boldsymbol{y},\boldsymbol{b})$, $\pi(\boldsymbol{y}|\boldsymbol{\theta},\boldsymbol{x},\boldsymbol{b})$, and $\pi(\boldsymbol{b}|\boldsymbol{\theta},\boldsymbol{x},\boldsymbol{y})$, respectively.

\begin{itemize}
	\item[Step 1.] Set $t=0$ and initialize $(\boldsymbol{y}^{(t)}, \boldsymbol{b}^{(t)})$.
	
	\item[Step 2.] Generate $\boldsymbol{\theta}^{(t+1)}$ from $\pi(\boldsymbol{\theta}|\boldsymbol{x},\boldsymbol{y}^{(t)},\boldsymbol{b}^{(t)})$. 
	
	\item[Step 3.] Generate 
	$\boldsymbol{y}^{(t+1)}$ from $\pi(\boldsymbol{y}|\boldsymbol{\theta}^{(t+1)},\boldsymbol{x},\boldsymbol{b}^{(t)})$, and then $\boldsymbol{b}^{(t+1)}$ from  $\pi(\boldsymbol{b}|\boldsymbol{\theta}^{(t+1)},\boldsymbol{x},\boldsymbol{y}^{(t+1)})$.
	
	\item[Step 4.] Update $( \boldsymbol{y}^{(t)}, \boldsymbol{b}^{(t)})$ with $(\boldsymbol{y}^{(t+1)}, \boldsymbol{b}^{(t+1)})$.
	\item[Step 5.] Repeat Step 2-4 until convergence of $\left\{\boldsymbol{\theta}^{(t)}\right\}_{t\ge 0}$.
\end{itemize}

Note that the conditional distribution $\pi(\boldsymbol{y}|\boldsymbol{\theta},\boldsymbol{x},\boldsymbol{b})$ of the unobserved data $\boldsymbol{y}$, is expressed as a product of multinomial and binomial distributions. See Section 3.1 of the Supplementary Materials for details. Now, one can easily execute the above steps using Gibbs sampling if the full conditional distributions of $\boldsymbol{b}$ and $\boldsymbol{\theta}$ can be obtained in closed forms. In the next subsection, we propose appropriate priors for $\boldsymbol{\theta}$ and suitable distribution of the random effects $\boldsymbol{b}$, so that the conditional posteriors  $\pi(\boldsymbol{\theta}|\boldsymbol{b},\boldsymbol{y},\boldsymbol{x})$ and $\pi(\boldsymbol{b}|\boldsymbol{\theta},\boldsymbol{y},\boldsymbol{x})$ result in standard distributions.

\subsubsection{Prior Specifications}\label{Prior}
We first assume the random effect $b_{s}$ follows generalized logistic type-I distribution \citep{Johnson94} with shape parameter $\delta_{s}$ for $s=1,2,3$. This is among the few distributions which can model both positively and negatively skewed data on the whole real line.  With this specific choice of distribution, the conditional posterior distribution of $\boldsymbol{b}$ given $(\boldsymbol{\theta},\boldsymbol{x},\boldsymbol{y})$ is obtained as
\begin{eqnarray}
	\pi(b_{s}|\boldsymbol{\theta},\boldsymbol{x},\boldsymbol{y}) & \propto & EGB2(n_s+1, m_s+\delta_s), \quad s=1,2,3,\nonumber
\end{eqnarray} 
where $EGB2$ denotes for exponential generalized beta distribution of second kind \citep{Fischer00} with parameters involving $m_1=x_{1\cdot\cdot}$, $m_2=y_{111,1}+y_{111,3}+y_{111,4}+y_{110,1}+x_{011}+x_{010}$, $m_3=y_{111,1}+y_{111,2}+y_{011,1}+y_{101,1}+x_{001}, n_1=N-x_{1\cdot\cdot}$, $n_2=x_{100}+x_{101}+y_{001,1}+y_{000,1}+y_{000,3}+y_{000,4}$,  and $n_3=x_{110}+y_{100,1}+y_{010,1}+y_{000,1}+y_{000,2}$. See Section 3.2 of the Supplementary Materials for detailed derivation. Now, one can easily sample from $\pi(b_{s}|\boldsymbol{\theta},\boldsymbol{x},\boldsymbol{y})$ and compute $\mathcal{P}_s=\left[1+exp(-b_s)\right]^{-1}$ for $s=1,2,3$. 

For our analysis, the prior on $\boldsymbol{\theta}=(N,\boldsymbol{\alpha},\boldsymbol{\delta})$ are assigned in (\ref{joint-posterior}) independently, i.e. $\pi(\boldsymbol{\theta})=\pi(N)\pi(\boldsymbol{\alpha})\pi(\boldsymbol{\delta})$. We propose both informative and noninformative priors for the nuisance parameters $\boldsymbol{\alpha}$ and $\boldsymbol{\delta}$ with noninformative prior on the parameters of interest $N$. These different prior choices enable us to study the robustness of the proposed inferential methodology and provide flexibility to the users. The two different choices of the priors and associated conditional posterior distribution of $\boldsymbol{\theta}$ are provided below.

\subsubsection*{Prior Choice I}
As discussed before, we first consider noninformative priors for the purpose of Bayesian estimation. The Jeffrey's priors for all the parameters involved in THBM are given by $\pi(N)\propto N^{-1}$, $\pi(\alpha)\equiv\pi(\alpha_1,\alpha_2,\alpha_3,\alpha_4,\alpha_5)\propto$ Dirichlet(0.5,0.5,0.5,0.5,0.5), where $\alpha_5=1-\alpha_{0}$ and $\pi(\delta_{s})\propto \delta_{s}^{-1}$ for $s=1,2,3$. Under these choices, the full conditional densities of $\boldsymbol{\alpha}$ and $\delta_{l}$'s are obtained as
\begin{eqnarray}
	\pi(\boldsymbol{\alpha}|\boldsymbol{\theta}_{-\boldsymbol{\alpha}},\boldsymbol{x},\boldsymbol{y}, \boldsymbol{b}) & \propto &  Dirichlet(\boldsymbol{d}),\nonumber\\
	\pi(\delta_{s}|\boldsymbol{\theta}_{-\delta_{s}},\boldsymbol{x},\boldsymbol{y},\boldsymbol{b}) & \propto &  Exponential(w_s),\hspace{0.2in}s=1,2,3,\nonumber
\end{eqnarray} 
where $\boldsymbol{d}=(d_1,d_2,d_3,d_4,d_5)$ with $d_1=y_{111,2}+x_{110}-y_{110,1}+x_{001}-y_{001,1}+y_{000,2}+0.5$, $d_2=y_{111,3}+x_{011}-y_{011,1}+x_{100}-y_{100,1}+y_{000,3}+0.5$, $d_3=y_{111,4}+x_{101}-y_{101,1}+x_{010}-y_{010,1}+y_{000,4}+0.5$, $d_4=x_{111}-\sum_{u=1}^{4}y_{111,u}+N-n-\sum_{u=1}^{4}y_{000,u}+0.5$, $d_5=y_{111,1}+y_{110,1}+y_{011,1}+y_{100,1}+y_{101,1}+y_{010,1}+y_{001,1}+y_{000,1}+0.5$, and $w_s=\log\left[1+\exp(-b_s)\right]$. The full conditional distribution of the parameter of interest $N$ is given by
\begin{eqnarray}
	\pi(N|\boldsymbol{\theta}_{-N},\boldsymbol{x},\boldsymbol{y}_{-y_{000,5}},\boldsymbol{b}) & \propto & \frac{(N-1)!}{(N-n-\sum_{u=1}^{4}y_{000,u})!}[\alpha_4(1-\mathcal{P}_1)]^{N-n-\sum_{u=1}^{4}y_{000,u}},\nonumber\\
	i.e. \hspace{0.2in} \pi(y_{000,5}|\boldsymbol{\theta}_{-N},\boldsymbol{x},\boldsymbol{y}_{-y_{000,5}},\boldsymbol{b}) & \propto & NB\left(n+\sum_{u=1}^{4}y_{000,u},1-\alpha_4(1-\mathcal{P}_1)\right),\label{cond_post_N} 
\end{eqnarray}
where $y_{000,5}=N-n-\sum_{u=1}^{4}y_{000,u}$, and NB denotes for negative binomial distribution. The detailed derivations of the full conditional distributions are provided in Sections 3.3-3.5 of the Supplementary Material. Now, one can easily execute the algorithm provided in Section \ref{Bayesian} to generate a sample from the posterior distribution $\boldsymbol{\theta}$ using Gibbs sampling. Therefore, summary statistics of the posterior distribution can be used to find a point estimate of $N$ based on a suitable choice of loss function. We refer to this estimate as THBM-I. The associated R-programme is provided in Section 5 of the Supplementary Material.

\subsubsection*{Prior Choice II}
Here, we consider informative priors on $\boldsymbol{\alpha}$ and $\boldsymbol{\delta}$, and Jeffrey's prior for the parameter of interest $N$ is considered as before. Based on the subjective choices of the hyperparameters, we propose Dirichlet prior for $\boldsymbol{\alpha}$ with parameter vector $(\beta_1,\beta_2,\beta_3,\beta_4,\beta_5)$, and gamma prior for $\delta_s$ with scale $\lambda_s$ and shape $\gamma_s$, for $s=1,2,3$. Here also, the full conditional distributions of $\boldsymbol{\alpha}$ and $\delta_s$ posses closed form expressions and are given by
\begin{eqnarray}
	\pi(\boldsymbol{\alpha}|\boldsymbol{\theta}_{-\boldsymbol{\alpha}},\boldsymbol{x},\boldsymbol{y},\boldsymbol{b}) & \propto &  Dirichlet(\boldsymbol{d^{\ast}}),\label{cond_post2_alpha}\nonumber\\
	\pi(\delta_{s}|\boldsymbol{\theta}_{-\delta_{s}},\boldsymbol{x},\boldsymbol{y},\boldsymbol{b}) & \propto & Gamma\left(\gamma_s+1,\left(w_s+\frac{1}{\lambda_s}\right)^{-1}\right),\hspace{0.2in}s=1,2,3,\label{cond_post2_delta}\nonumber
\end{eqnarray} 
where $\boldsymbol{d^{*}}=(d^{*}_{1},d^{*}_{2},d^{*}_{3},d^{*}_{4},d^{*}_{5})$ with $d^{\ast}_{u}=d_u-0.5+\beta_u$ for $u=1,2,\ldots,5$. The derivations of the full conditional distributions are provided in Sections 3.3-3.5 of the Supplementary Material. In order to consider diffuse prior on $\delta_{s}$, one can choose the hyperparameters $\gamma_s$ and $\lambda_s$ appropriately such that the variance of the prior distribution is large. The conditional posterior distribution of $N$ remains the same as provided in (\ref{cond_post_N}). Here also, the full conditionals are from a standard family of distributions and one can use a Gibbs sampler for inferential purposes and the resulting estimates are referred to as THBM-II.

\section{Simulation Study}\label{Simulation}
In this section, we evaluate the performances of the proposed estimators under the two prior setups in the presence of list dependence due to behavioural change and individual heterogeneity through an extensive simulation study. We compare the results to the following relevant competitors that account for the list dependence or/and individual heterogeneity:
(I) List dependence only-- log-linear model (LLM) with interaction effects \citep{Fienberg72b} and model $M_{tb}$ \citep{Chao00, Chatterjee20b}; 
(II) Heterogeneity only-- quasi-symmetry model (QSM) \citep{Darroch93}, partial quasi-symmetry model (PQSM) \citep{Darroch93} and Bayesian non-parametric latent class model (NPLCM) \citep{Vallier2016};
(III) Both list dependence and heterogeneity-- $M_{bh}$ model \citep{Rivest01} and non-parametric sample coverage method (SC) \citep{Chao98}; and (IV) List independence and homogeneity: the log-linear model without interaction effects (IM) \citep{Fienberg72b}. The details of all these existing models and associated estimates are briefly presented in Section 2 of the Supplementary Material. 

We consider six different choices of $(\alpha_1,\alpha_2,\alpha_3,\alpha_4,\alpha_5)$ representing varying degrees of list dependence and denote them as P1-P6 the in Table \ref{Population}. We generate TRS data from the proposed model keeping the total population size $N$ fixed at 200 and 500 for each of these six different choices of populations P1-P6 with five different combinations of random effects characterized by $\boldsymbol{\delta}$ given as $(1.6,1.2,0.8)$, $(1.3,1.7,0.9)$, $(1.0,1.4,1.8)$, $(0.8,0.8,0.8)$, $(1.6,1.6,1.6)$. The choices of $N$ are similar to the population sizes of the case studies under consideration. Note that $\boldsymbol{\delta}=(0.8,0.8,0.8)$ and $(1.6,1.6,1.6)$ present the cases where capture probabilities of the causally independent individuals in all the three lists are identically distributed. In contrast, the capture probabilities differ from one list to another for the other three choices of $\boldsymbol{\delta}$.
\begin{table}[h]
	\begin{center}
		\caption{Dependence structure of the simulated populations.}
		\begin{threeparttable}
			\begin{tabular}{|c|c|}
				\hline
				Population & $(\alpha_1,\alpha_2,\alpha_3,\alpha_4,\alpha_5)$\\
				\hline
				P1  &    (0.35, 0.15, 0.25, 0.10, 0.15) \\       
				P2  &    (0.30, 0.30, 0.15, 0.10, 0.15) \\       
				P3  & 0.20, 0.10, 0.20, 0.10, 0.40)  \\      
				P4  &    (0.10, 0.20, 0.30, 0.20, 0.20)  \\       
				P5  &    (0.20, 0.20, 0.20, 0.20, 0.20)  \\       
				P6  &    (0.25, 0.15, 0.35, 0.10, 0.15)  \\ 
				\hline
			\end{tabular}
			\begin{tablenotes}
				\small
				\item Note that $\alpha_5=1-\alpha_0$.
			\end{tablenotes}
		\end{threeparttable}
		\label{Population}
	\end{center}
\end{table}

We consider two different sets of prior as discussed in the Subsection \ref{Prior} for estimation of the proposed model. Jeffrey’s priors (Prior Choice I) are considered in THBM-I, and for THBM-II, we consider $\beta_u=8 \alpha_u$ for $u=1,2,\ldots,5$, and diffuse prior for $\delta_s$ using gamma distribution with mean $\delta_s$ and variance 100, for $s=1,2,3$ (Prior Choice II). We generate a chain with 50,000 samples from the posterior distributions of the parameters associated with THBM using Gibbs sampling and find the posterior median of the population size based on every 10th iterate discarding the first 25,000 iterations as burn-in. The thinning interval of size 10 reduces the dependency among the samples at a satisfactory level, and the convergence of the chains is monitored using Geweke’s diagnostic test \citep{Geweke1992}. This is repeated 1000 times and the average estimates $\hat{N}$ are reported in Tables S3-S8 provided in Section 4 of the Supplementary Material. We also report the $95\%$ highest posterior density (HPD) credible interval along with coverage probability (CP) of $N$ and the Relative Mean Absolute Error, defined as
$RMAE =\frac{1}{1000}\sum_{r=1}^{1000}|\frac{\hat{N}_r-N}{N}|$, where $\hat{N}_r$ denotes the estimate based on data generated at the $r$th replication. Similarly, we compute the RMAE of all the existing competitors. For each of these 1000 replications, $95\%$ confidence intervals (CI) of $N$ for these competitive methods are obtained based on the quantiles from 1000 bootstrap samples. The CP of the population size $N$ is also obtained based on the bootstrap CIs.

The proposed model performs the best in terms of RAME for all the simulated populations P1-P6 for both $N=200$ and $N=500$. Moreover, the CPs of the proposed estimates are higher than all the existing competitors except $M_{tb}$ model which possesses a very wide CI due to high variability. In most of the cases, the CP of QSM and PQSM estimates are comparable for $N=200$ but CP of PQSM estimate is lower than QSM estimate for $N=500$. The CP of LLM estimate is lower than PQSM estimates for both $N=200$ and $N=500$. It is important to note that the CIs associated with the independent model and the SC estimates are very tight and which results in extremely low CPs. In contrast, the CI-width based on $M_{bh}$ are exorbitantly large in most cases and with negative lower limits on many occasions. In particular, the CPs of the SC estimates are less than 6\% for P4 and P5 (see Tables S6-S7 in Section 4 of the Supplementary Material), and the highest is around 55\% attained for only two cases in P3 with $N=200$ (see Table S5 in Section 4 of the Supplementary Material). Similar results for NPLCM and $M_{bh}$ are also observed in a few cases. Also, the IM, SC, and NPLCM consistently underestimate $N$ for all populations P1-P6.  Interestingly, the $M_{tb}$ model underestimates for P4 and P5 and overestimates for the rest of the populations. It is also observed that the RAME of all the estimators except the independent model and SC method decreases as $N$ increases from 200 to 500 for all simulated populations P1-P6.

As discussed in Section 2.3.1 of the Supplementary Material, we have observed that the SC estimate suffers from boundary problems in our simulation study. It is easy to generate data such that the percentage of infeasible estimates is as high as 60\% to 80\%. In some other cases, we have also observed that the QSM, PQSM, and LLM may fail to converge and produce extremely large estimates. For a valid and fair comparison, we have not reported those cases here.

\subsection{Sensitivity Analysis}

\subsubsection{Sensitivity of Prior Specification}\label{Prior_sen}
As discussed before, the proposed estimates of the population size $N$ based on two different sets of priors perform better than the existing competitors. But to understand the sensitivity of the estimate with prior specifications, we now compare the results associated with THBM-I and THBM-II in Tables S3-S8 (see Section 4 of the Supplementary Material). In general, the CPs of the estimates of $N$ with both the prior specifications are very similar, but only in a few cases, CPs of THBM-I are marginally lower than those of THBM-II. Although the performance of the THBM-II is better than THBM-I for all the simulated populations P1-P6 in terms of RMAE, the difference is merely less than $3\%$. The marginal improvement is due to the fact that the prior variance of $\alpha_i$'s considered in THBM-II is considerably smaller compared to the variance of Jeffrey's prior in THBM-I. From these observations, it indicates that the proposed methodology is not sensitive to prior specifications, and these are especially valuable when only limited prior information is available.

\subsubsection{Sensitivity of Random Effects}
\begin{figure}[htp]
	\centering
	\includegraphics[scale=0.8]{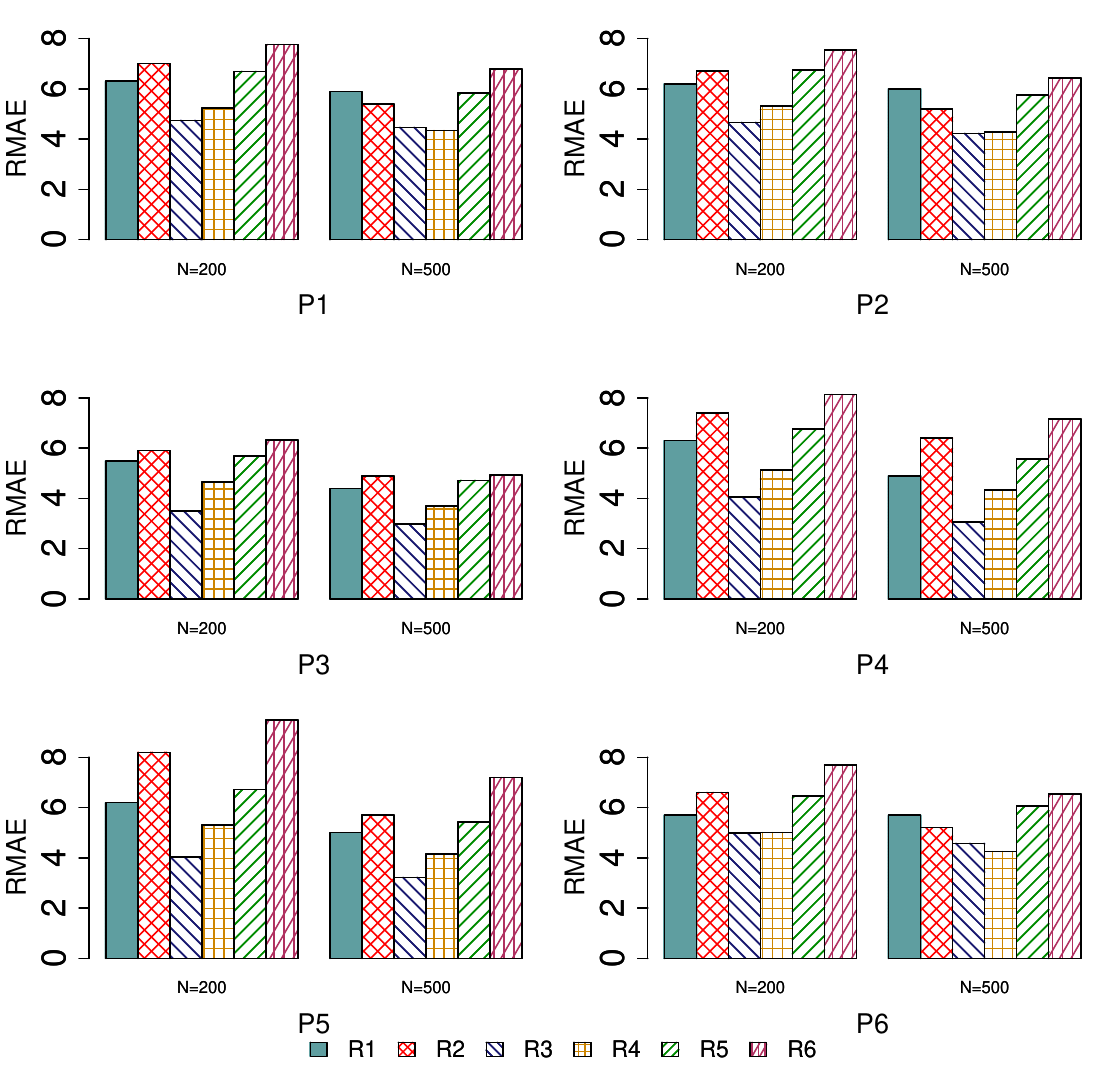}
	\caption{Comparison of RAMEs (in \%) of the proposed estimate under different random effects models for data generation.}
	\label{RAME}
\end{figure}

\begin{figure}[htp]
	\centering
	\includegraphics[scale=0.8]{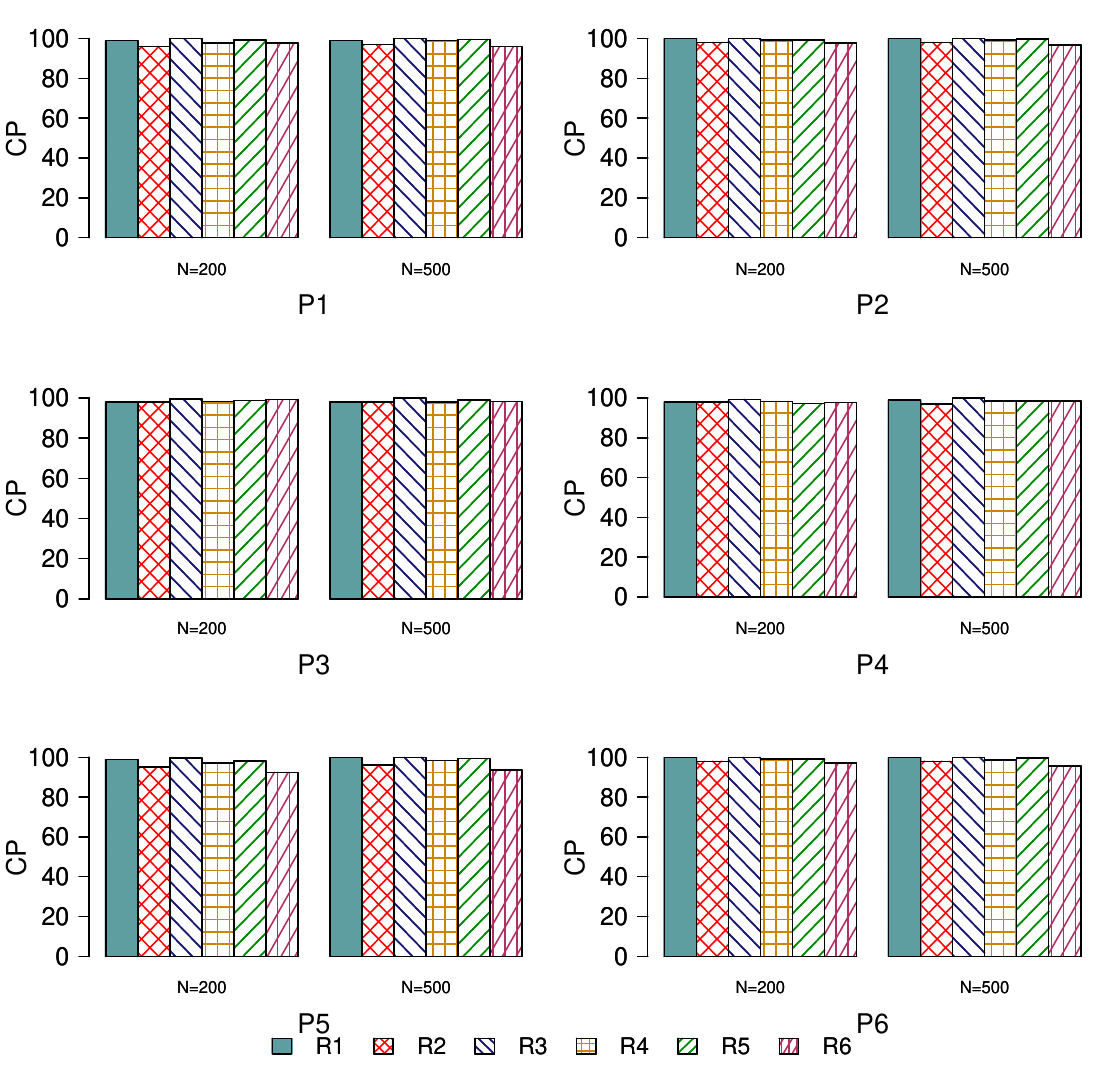}
	\caption{Comparison of CPs (in \%) of the proposed estimate under different random effects models for data generation.}
	\label{CP}
\end{figure}

In the posed estimation methodology, we are obliged to assume that the random effects $b_{s}$'s follow generalized logistic distribution to obtain a closed-form expression of the full conditionals. However, this assumption may not be valid in real applications and it is worth investigating the effect of misspecification of random effects distributions on the performance of the proposed estimate. For this purpose, we generate data for populations P1-P6 considering random effects from generalized logistic type-I ($GL_I$), normal and, gamma distributions with two different parameter choices provided in Table \ref{Ran}. 
The RAMEs and CPs of THBM-I are plotted in Figure \ref{RAME} and Figure \ref{CP}, respectively. Similar patterns are observed for THBM-II (not reported here).
In terms of RAME, we observe minor deterioration due to misspecification when the true random effects are generated from gamma distributions, but in contrast, marginal improvement is observed when random effects are e generated from the normal distributions. In all the six populations P1-P6, the range of differences in RAMEs is only less than $3\%$ (see Figure \ref{RAME}). However, no effect of misspecification is noticed on CPs in any of the six populations (see Figure \ref{CP}). We also consider random effects for discrete distribution, and the results are similar to the aforementioned cases (not reported here).

\begin{table}[h]
	\begin{center}
		\caption{Random effects distributions for analysing sensitivity of misspefication.}
		\begin{threeparttable}
			\begin{tabular}{|c|c|c|c|c|}
				\hline
				Random Effects & $b_1$ & $b_2$ & $b_3$\\
				\hline
				R1  &    $GL_I(1.6)$ & $GL_I(1.6)$ & $GL_I(1.6)$ \\       
				R2  &    $GL_I(1.6)$ & $GL_I(1.2)$ & $GL_I(0.8)$ \\  
				R3  &    $Normal(0.5,1)$ & $Normal(0.5,1)$ & $Normal(0.5,1)$  \\
				R4  &    $Normal(0.5,1)$ & $Normal(0,1)$ & $Normal(-0.5,1)$  \\
				R5  &    $Gamma(2,0.5)$ & $Gamma(2,0.5)$ & $Gamma(2,0.5)$  \\  
				R6  &    $Gamma(2,0.5)$ & $Gamma(1,0.5)$ & $Gamma(0.5,0.5)$  \\
				\hline
			\end{tabular}
			\begin{tablenotes}
				\tiny
				\item $GL_{I}(\eta)$ denotes generalized logistic type-I distribution with parameter $\eta$.
				\item $Normal(\mu, \sigma)$ denotes normal distribution with mean $\mu$ and standard deviation $\sigma$.
				\item $Gamma(a,b)$ denotes gamma distribution with shape parameter $a$ and scale parameter $b$.
			\end{tablenotes}
		\end{threeparttable}
		\label{Ran}
	\end{center}
\end{table}

\subsubsection{Sensitivity of Model Misspecification}
\begin{figure}[htp]
	\centering
	\includegraphics[scale=0.5]{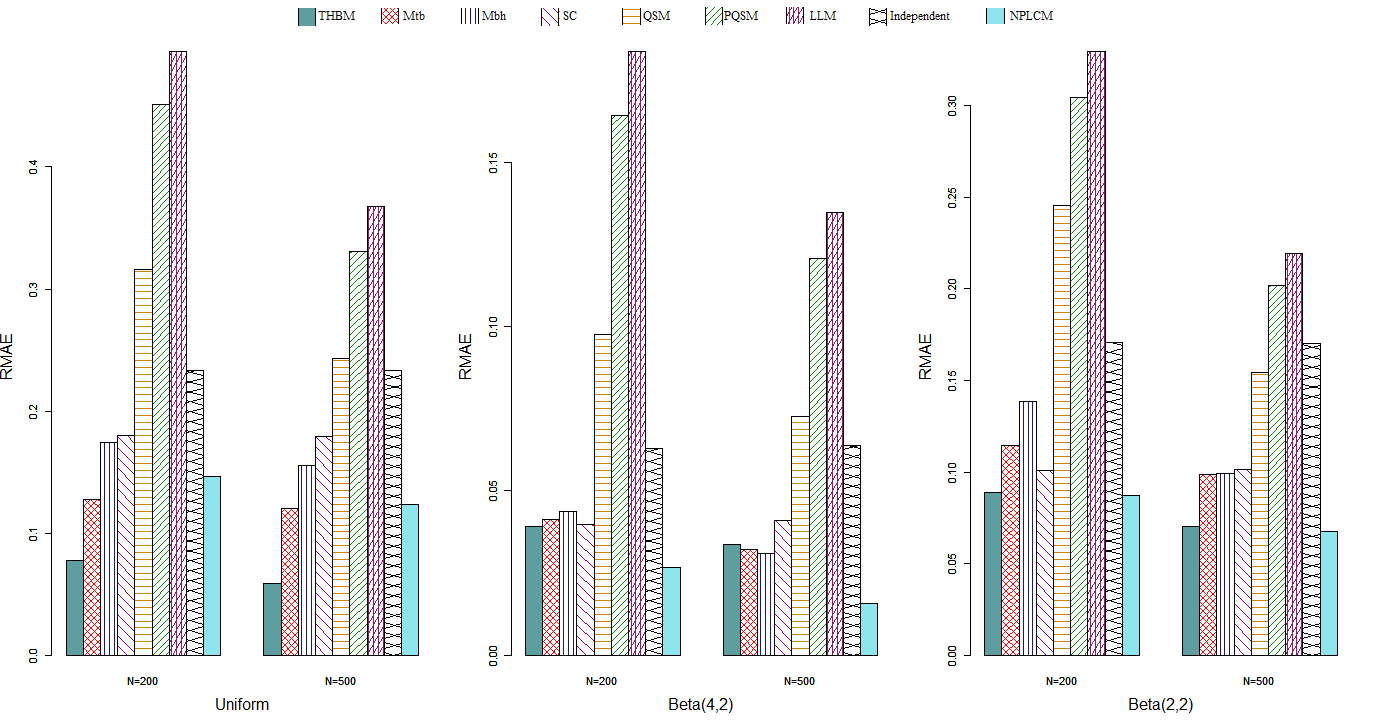}
	\caption{Comparison of the RAMEs (in \%) of the proposed estimate with the existing competitors under model misspecification when data generated from auto-regressive dependence structure with capture probabilities $P_{h}^{(1)}$'s in the list $L_{1}$ follow $(i)$ $Uniform (0,1)$, $(ii)$ $Beta (4,2)$, and $(ii)$ $Beta (2,2)$.}
	\label{RAME_S}
\end{figure}

\begin{figure}[htp]
	\centering
	\includegraphics[scale=0.4]{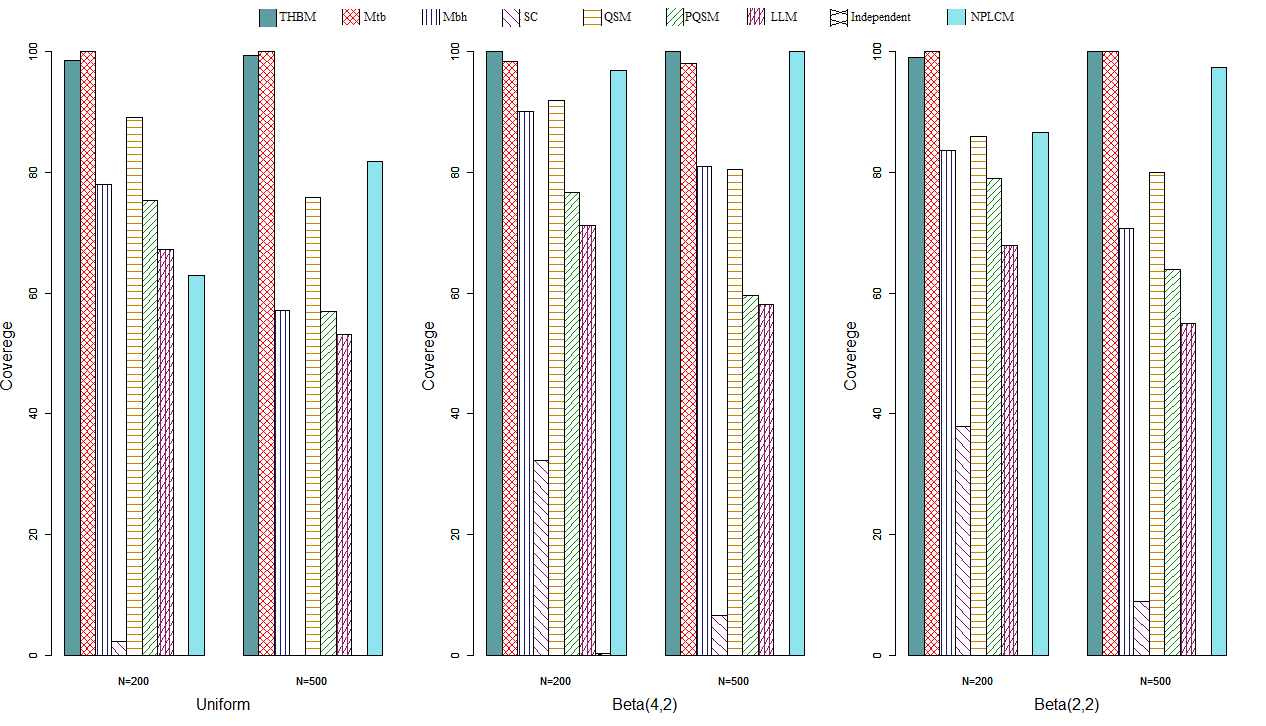}
	\caption{Comparison of the CPs (in \%) of the proposed estimate with the existing competitors under model misspecification when data generated from auto-regressive dependence structure with capture probabilities $P_{h}^{(1)}$'s in the list $L_{1}$ follow $(i)$ $Uniform (0,1)$, $(ii)$ $Beta (4,2)$, and $(ii)$ $Beta (2,2)$.}
	\label{CP_S}
\end{figure}

As discussed in Section \ref{general}, the proposed model represents a realistic mechanism to account for inherent heterogeneity among the individuals and dependencies. Nevertheless, it is important to study the effect of model misspecification when the data-generating mechanism deviates from the assumed structure of the fitted model. For this purpose, we generate the capture statuses of the individuals
$Z_{h}^{(1)}$, $Z_{h}^{(2)}$, $Z_{h}^{(3)}$ from Bernoulli distributions with respective probabilities
$P_{h}^{(1)}$, $P_{h}^{(2)}$, and $P_{h}^{(3)}$, where $P_{h}^{(j)}= \min\{P_{h}^{(j-1)}\mathbb{I}[Z_{h}^{(j-1)}=0]+1.2\mathbb{I}[Z_{h}^{(j-1)}=1], 0.99\}$, for $j=2,3$, and $h=1,\ldots, N$. This mechanism induces a dependence structure among the capture statuses similar to a first-order auto-regressive model. To incorporate heterogeneity in the capture probabilities, we generate $P_{h}^{(1)}$'s from three different choices of distributions: $(i)$ $Uniform (0,1)$, $(ii)$ $Beta (4,2)$, and $(iii)$ $Beta (2,2)$. The RAMEs and CPs of THBM-I and its competitors are plotted in Figure \ref{RAME_S} and Figure \ref{CP_S}, respectively. In most of cases, the proposed estimator outperforms the existing competitors with respect to RAME. The CPs of the proposed estimator and $M_{tb}$ model are close to $100\%$ and much better than all other estimators. In particular, the performance of SC is very poor and its CPs are less than $40\%$ in all the cases.

\section{Case Study}\label{Casestudy}
As discussed before, applications of MSE are commonly found in the domain of epidemiology. In this section, we consider two different case studies on infectious diseases. The results from analyses of the two datasets based on the proposed methodology are discussed below along with the practical implications. We also compare these results with the estimates based on the existing methods.

\subsection{Analysis of Legionnaires’ Disease Surveillance Data}\label{LD}
As discussed in Section \ref{Intro}, we consider a dataset on the LD outbreak in the Netherlands in the year 1999. To apply the proposed methodology to estimate the total number of infected individuals, we consider Jeffrey’s prior for all the parameters associated with THBM-I. For THBM-II, we consider Dirichlet prior for $\alpha$ with parameter $(0.2, 0.1, 0.1, 0.1, 0.5)$ and gamma prior for $\delta_{1}$, $\delta_{2}$, $\delta_{3}$ with mean 0.5, 0.5, 3, respectively, and variance 100. We generate a chain with 1 million samples from the posterior distributions of the associated model parameters using the Gibbs sampling algorithm and discard the first 10\% iterations as burn-in. The convergence of the chain is monitored graphically and using Geweke’s diagnostic test \citep{Geweke1992}.

To reduce the extent of heterogeneity as well as the dependence that is induced by heterogeneity in the population, it is often recommended to employ post-stratification of the population based on appropriate demographic (e.g. age, race or sex) or geographic variables \citep{Wolter86, Islam15}. In this application, we estimate the prevalence of LD in four regions: North (1,671,534 inhabitants), East (4,467,527 inhabitants), West (5,955,299 inhabitants), and South (3,892,715 inhabitants) to facilitate regional surveillance and identification of associated risk factors. The results based on the proposed method are presented in Table \ref{Leg_analysis}. It is quite common to observe that the sum of the strata-wise estimate of the population sizes does not tally with the estimate obtained from the aggregated data if the underlining heterogeneity is not successfully accounted for. In this context of capture-recapture experiments, this phenomenon is popularly known as Simpson's paradox or simply Yule's association paradox \citep{Kadane99}. 
The median, and the 95\% HPD credible interval of $N$ based on the aggregated data are also reported in Table \ref{Leg_analysis}. Interestingly, the estimate of the total LD incidence obtained by summing the four regional estimates approximately matches with the national estimate based on combined data. It is an indication that the proposed method successfully accounted for the heterogeneity present in the population. The estimate of $N$ based on THBM-I is marginally larger than that of THBM-II. Interestingly, the length of the HPD credible interval is marginally shorter for THBM-II due to the use of informative prior for $\alpha$ compared to that of the THBM-I, which in contrast, involves Jeffrey’s prior. This observation is as expected based on the findings from our simulation study discussed in Subsection \ref{Prior_sen}. We also observe that the estimates of $N$ are similar under both Jeffrey’s prior and informative prior for $\delta_{l}$'s while considering Jeffrey’s prior for $\alpha$ (not reported here). We also report the estimates of $N$ based on different existing methods. The CI's are computed based on 10,000 bootstrap samples. The estimates based on the IM, NPLCM, and SC method are considerably lower than other estimates. For all the regions, $M_{bh}$ provides infeasible estimates (smaller than the total observed cell count $n$), and $M_{tb}$ and QSM provide very large estimates compared to other competitors.

\begin{figure}[htp]
	\centering
	\includegraphics[scale=0.76]{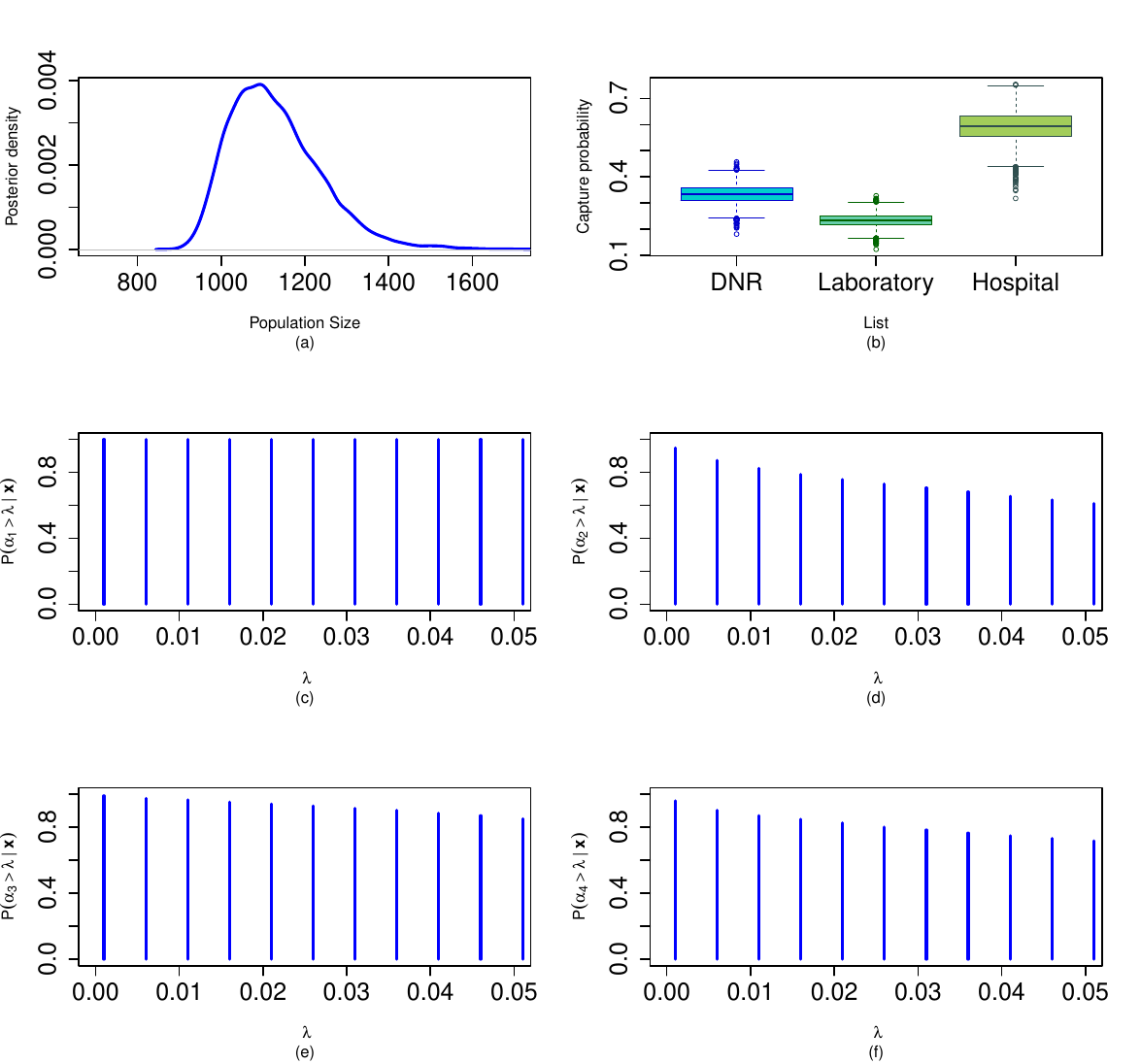}
	\caption{(a) Posterior density of the size of population affected with Legionnaires’ disease; (b) Boxplots of capture probabilities associated with three different lists; (c) Test of significance of $\alpha_1$; (d) Test of significance of $\alpha_2$; (e) Test of significance of $\alpha_3$; (f) Test of significance of $\alpha_4$.}
	\label{Leg_plots}
\end{figure}

To gain insight from our analysis, we present the posterior density of $N$ and boxplot of the capture probabilities associated with DNR, Laboratory, and Hospital records in Figure \ref{Leg_plots}(a), and Figure \ref{Leg_plots}(b), respectively, based on aggregated data using THBM-I. The posterior density of $N$ is slightly skewed toward the right and the catchability of Hospital records is considerably higher than those of DNR and Laboratory records. It is also seen that the heterogeneity in capture probability associated with Hospital records is higher compared to other sources. From Figures \ref{Leg_plots}(c)-\ref{Leg_plots}(f), it is visible that the bulk of the distribution of all the dependence parameters is away from 0 which indicates their significance. The estimates of the dependence parameters suggest that almost 60\% individuals are causally dependent, and 13\% of individuals have perfect positive association among the three lists.

In Table \ref{rate}, we also report the under-reporting rate (UR) and prevalence rate (PR) \footnote{UR$=\frac{(\hat{N}-n)}{\hat{N}}\times100$, \hspace{2mm} PR$= \frac{\hat{N}}{\text{No. of inhabitants}}\times100,000$} for each of these four regions based on our estimate of the prevalence obtained from THBM-I.
The estimated UR in the North and West are similar but marginally lower than that in the South. However, the estimated UR in the East is significantly higher than in other regions. We also observe that the CI associated with the estimate of LD incidence in the East is wide. This encourages further investigation, possibly involving information from other sources, to obtain a more reliable estimate. As mentioned before, the capture probability associated with Laboratory records is very low (see Figure \ref{Leg_plots}(b)), and it is worth considering alternative sources with higher capture probability to improve the coverage. The PR is the highest in the South and least in the North. However, a marginal difference in PR is observed between East and West. Some previous studies suggest that weather and climatic conditions are associated with LD incidence \citep{Karagiannisy09, Brandsema14}. On average, the northern provinces endure lower temperatures compared to the southern provinces. In particular, the east of Brabant and the very north of Limburg are the warmest during summer. Our analyses indicate that warm weather increases LD incidence, which supports the findings by \citet{Karagiannisy09} and \citet{Brandsema14}. Recently, researchers are also interested in the association between LD incidence and precipitation and humidity \citep{Passer20}. In general, there is no dry season in the Netherlands, and precipitation is common throughout the year. This is perhaps one of the major risk factors causing a higher PR in the Netherlands compared to other countries. In particular, the coastal provinces (West) experience the heaviest rain showers after the summer and during the autumn resulting in higher PR compared to the North. Several other factors related to demographics and socioeconomic characteristics may lead to a higher incidence of LD \citep{Farnham14, Passer20}. Following a similar post-stratification strategy, as considered in this article, one can effectively support the general goals of surveillance. It helps to rapidly recognize cases that occur in similar locations and describe incidences and trends. Also, one can identify opportunities for prevention and monitor the effectiveness of interventions implemented as part of an outbreak investigation \citep{CDC}.	

\begin{table}[h]
	\small
	\begin{center}
		\caption{Summary results of the Legionnaires' disease surveillance data analysis.}
		\label{Leg_analysis}
		\begin{threeparttable}
			\resizebox{1\textwidth}{!}{
				\begin{tabular}{|llcccccccccc|}
					\hline
					Region & &  THBM-I & THBM-II & NPLCM & SC & QSM & PQSM & LLM & $M_{tb}$ & $M_{bh}$  & IM\\
					\hline
					North & $\hat{N}$ & 95 & 82 & 79 & 85 & 173 & 98 & 97 & 213 & 20 & 74\\
					& CI\tnote{*} & ( 70, 168 ) & (69 , 137 ) & (69, 105) & (67, 122) &  (81, 873) & (65, 12$\times10^{10}$) & (58, 228) & (80, 475) &  (2, 23) & (65, 76) \\ 
					East & $\hat{N}$ & 318 & 313  & 242 & 243 & 391 & 447 & 720 &  416 & 93 & 198\\
					& CI\tnote{*} & (216, 685) & (214, 537 ) & (189, 368) & (210, 292) &  (253, 796) & (1.7$\times10^{10}$, 70.3$\times10^{10}$) & (1043$\times10^{10}$, 3504$\times10^{10}$) & (201, 790) & (27, 58) & (188, 214) \\ 
					West & $\hat{N}$ & 387 & 363  & 346 & 363 & 672 & 433 & 419 & 714 & 78 & 315 \\
					& CI\tnote{*} & (317, 492) & (307, 455) & (292, 436) & (319, 426) &  437, 1267) & (334, 919) & (303, 1236) & (334, 1284) & (17, 34) & (307, 346) \\ 
					South & $\hat{N}$ & 308 & 283 & 274 &  287 & 540 & 304 & 308 & 674 & 103 & 257\\
					& CI\tnote{*} & (259, 401) & (254, 334) & (235, 346) & (257, 327) &  (352, 1021) & (243, 441) & (256, 671) & (305, 1274) & (31, 56) & (237, 261) \\  
					Total & $\hat{N}$ & 1114 & 1106 & 980 & 992 & 1803 & 1176 & 1253 & 1400 & 351  & 855\\
					& CI\tnote{*} & (948, 1347) & (942, 1290) & (812, 1265) & (922, 1078) &  (1387, 2509) & (923, 1363) & (986, 1614) & (1153, 2104) & (136, 178)  & (829, 878) \\
					\hline
				\end{tabular}
			}
			\begin{tablenotes}
				\tiny
				\item
				THBM: Trivariate Heterogeneous Bernoulli Model; SC: Sample Coverage;
				QSM: Quasi-symmetry Model;
				PQSM: Partial Quasi-symmetry Model;\\
				\hspace*{1mm}LLM: Log-linear Model, $M_{tb}$: Time Behavioural Response Variation Model; $M_{bh}$: Individual Heterogeneity and Behavioural Response Variation Model;\\
			\hspace*{1mm}IM: LLM without interaction effects.
				\item[*] For THBM-I, THBM-II, and NPLCM, CI refers to HPD Credible Interval.
			\end{tablenotes}
		\end{threeparttable}
	\end{center}
\end{table}

\begin{table}[h]
	\small
	\begin{center}
		\caption{Region-wise prevalence rate and underreporting rate of the Legionnaires' disease based on THBM-I.}
		\label{rate}
		\begin{threeparttable}
			\begin{tabular}{|lcccc|}
				\hline
				&  North & South & East & West\\
				\hline
				PR  & 5.7   &  7.9  & 7.1  & 6.5\\
				UR  & 27   & 32  & 42  & 26\\
				\hline
			\end{tabular}
			\begin{tablenotes}
				\tiny
				\item PR: Prevalence rate per 100000 inhabitants; UR: Underreporting rate in \%
			\end{tablenotes}
		\end{threeparttable}
	\end{center}
\end{table}	

\subsection{Analysis of Hepatitis A Virus Surveillance Data}
Here, we consider the surveillance data of the HAV outbreak in northern Taiwan in the year 1995 to estimate the total number of infected individuals. As before, we consider  Jeffrey’s prior for THBM-I. For THBM-II, we consider  Dirichlet prior for $\alpha$ with parameter $(0.1, 0.1, 0.1, 0.2, 0.5)$ and gamma prior for $\delta_{s}$ with mean 0.5 and variance 100 for $s=1,2,3$. For this purpose, we generate 5 million samples using the Gibbs sampling from the posterior distributions of the associated model parameters and discard the first 20\% iterations as burn-in. Here also, no convergence issues are found based on Geweke’s diagnostic test. The estimates of $N$ along with 95\% HPD credible interval based on both THBM-I and THBM-II are reported in Table \ref{HAV_analysis}. Here, the length of the HPD credible interval for $N$ based on THBM-II is considerably smaller than that of THBM-I, in line with our findings in Subsections \ref{Prior_sen} and \ref{LD}. We also provide the results based on the existing methods for comparison. Here also, the estimate based on the IM is considerably lower than other estimates. The proposed estimate is marginally larger than that of NPLCM and $M_{tb}$ but lower than that of $M_{bh}$. The results based on QSM, PQSM, and LLM are much higher than the estimates based on THBM, NPLCM and $M_{tb}$. The estimates based on QSM, PQSM, and LLM are comparable but the CI of both QSM and PQSM is wider than that of LLM.Similar to our findings from simulation study, the CI width of $M_{bh}$ is exorbitantly high. The estimated sample coverage $\hat{C}=51\%$ indicates the estimates provided by the SC method are not reliable. 

\begin{figure}[htp]
	\centering
	\includegraphics[scale=0.78]{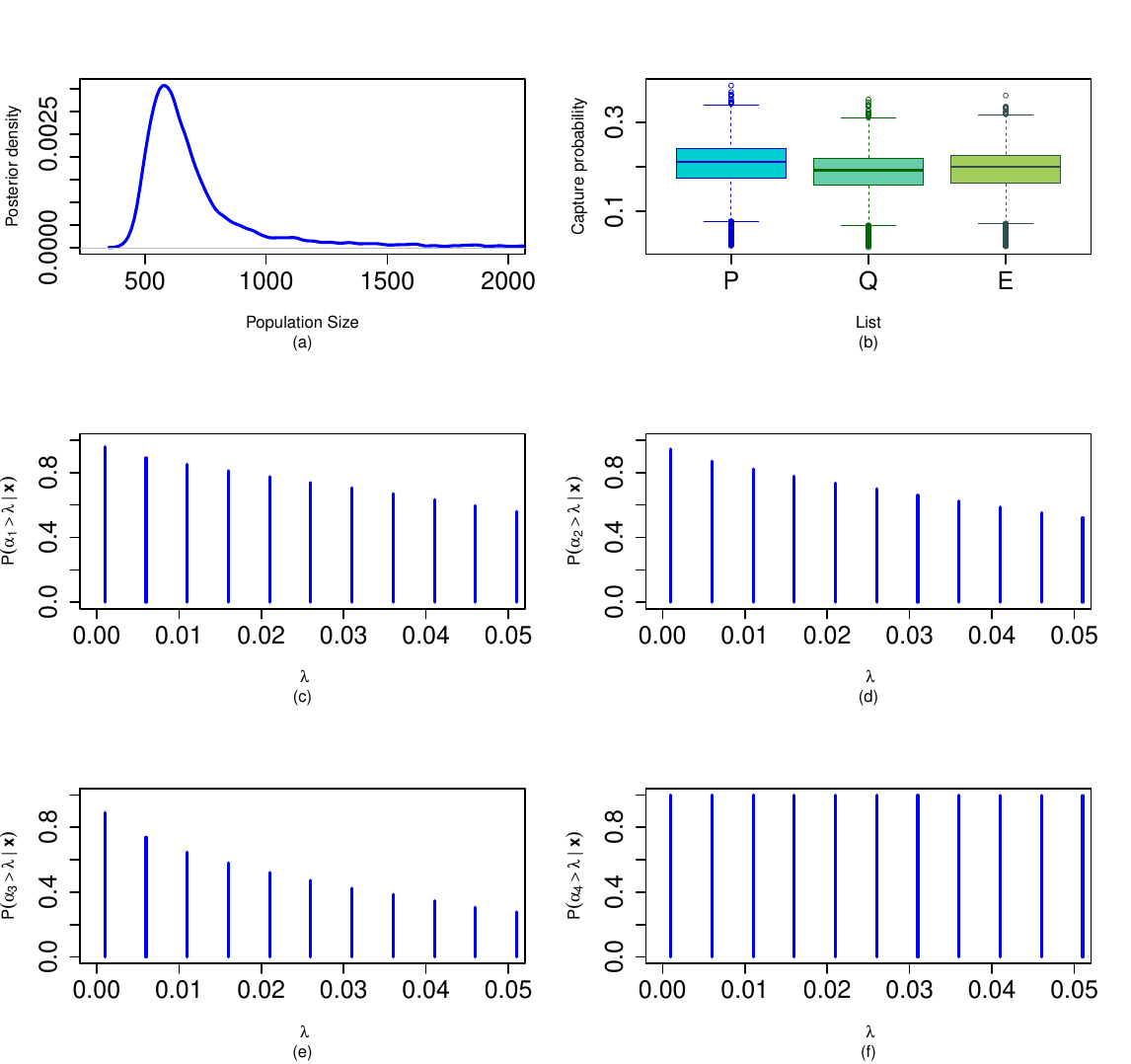}
	\caption{(a) Posterior density of the size of HAV infected population; (b) Box plots of capture probabilities associated with three different lists; (c) Test of significance of $\alpha_1$; (d) Test of significance of $\alpha_2$; (e) Test of significance of $\alpha_3$; (f) Test of significance of $\alpha_4$.}
	\label{HAV}
\end{figure}

Findings based on THBM-I are graphically presented in Figure \ref{HAV}. One can observe that the posterior distribution of $N$ is highly skewed (see \ref{HAV}(a)). The catchability of all three lists is similar but the heterogeneity in capture probability associated with P-list (i.e. records based on a serum test) is marginally higher than that of the other lists (see Figure \ref{HAV}(b)). Cases of HAV are not clinically distinguishable from other types of acute viral hepatitis, which results in low capture probability in all the sources \citep{WHO}. Here also, the bulk of the distribution of all the dependence parameters is significantly away from 0 (see Figure \ref{HAV}(c)-\ref{HAV}(f)). Almost 29\% individuals are causally dependent, and 15\% of individuals have perfect positive associations among the three lists.

In December 1995, the National Quarantine Service of Taiwan conducted a screen serum test for the HAV antibody for all students at the college at which the outbreak of the HAV occurred to determine the number of infected students. Based on their final figure, the estimated UR is $50\%$ \citep{Chao97}, but our estimate of the UR is $57\%$. Nevertheless, the level of underreporting of HAV cases is considerably high. As effective interventions to control the further spread of HAV are available, underreporting leads to serious consequences because it prevents prompt and effective public health action to protect immediate contacts of cases and their communities \citep{Crowcroft01}. After this outbreak, Taiwan started to immunize children in 30 indigenous townships against HAV and further expanded to 19 non-indigenous townships with higher incidence or increased risk of epidemic in 1997-2002. It has been reported that the annual PR of HAV decreased from 2.96 in 1995 to 0.90 in 2003-2008. The PR in vaccinated townships and unvaccinated townships declined 98.3\% and 52.6\%, respectively \citep{Tsou11}. This apparently indicates the long-term efficacy of the HAV vaccine in disease control in the vaccinated population and the out-of-cohort effect in the unvaccinated population. However, these figures do not account for the underreported cases, and any conclusion drawn from these statistics may be misleading. In particular, vaccination efficacy is overestimated when UR is higher in the vaccinated population compared to the unvaccinated population. Therefore, it is essential to estimate the underreported cases and adjust the calculation of PR ( as considered in Subsection \ref{LD}) for an unbiased evaluation of the interventions. For this purpose, the proposed model is a good choice to obtain an efficient estimate of the total number of individuals (or equivalently the underreported cases) infected by a disease such as HAV.

In 2003-2008, PR doubled in unvaccinated townships among those aged more than 30 years \citep{Tsou11}. The lack of demographic or geographical information for this case study hampered our ability to carry out a stratified analysis. At the national level, geographical analysis of outbreaks can support surveillance and identification of associated risk factors. Without detailed risk factor information, it is difficult to develop and implement appropriate policies to contain the spread of this infection \citep{Matin06}.	

\begin{table}[h]
	\small
	\begin{center}
		\caption{Summary results of the HAV surveillance data analysis.}
		\label{HAV_analysis}
		\begin{threeparttable}
			\resizebox{1\textwidth}{!}{
				\begin{tabular}{|lcccccccccc|}
					\hline
					&  THBM-I & THBM-II & NPLCM & SC & QSM & PQSM & LLM & $M_{tb}$ & $M_{bh}$ & IM\\
					\hline
					$\hat{N}$ & 633 & 546 & 516 & 971 & 1313 & 1325 & 1312 &  587 & 668 & 388\\
					CI$^*$ & (400, 1472) & (433, 694) & (337, 784) & (557, 3583) &  (669, 3171) & (607, 2718) & (464, 1961) & (287, 1110) & (339, 1.95$\times10^{11}$) & (381, 510)
					\\
					\hline
				\end{tabular}
			}
			\begin{tablenotes}
				\tiny
				\item
				THBM: Trivariate Heterogeneous Bernoulli Model; NPLCM: Nonparametric Latent Class Model; SC: Sample Coverage;
				QSM: Quasi-symmetry Model;\\
			\hspace*{1mm}	PQSM: Partial Quasi-symmetry Model;
				LLM: Log-linear Model, $M_{tb}$: Time and Behavioural Response Variation Model, \\ 
				\hspace*{1mm} $M_{bh}$: Individual Heterogeneity and Behavioural Response Variation Model, IM: LLM without interaction effects;
				\item[*] For THBM-I, THBM-II, and NPLCM, CI refers to HPD Credible Interval.
			\end{tablenotes}
		\end{threeparttable}
	\end{center}
\end{table}

\section{Discussions}\label{Discussion}
This paper addresses the issue of population size estimation incorporating the possible list dependence arising from behavioural response variation and heterogeneous catchability. For this purpose, a novel model for capture-recapture data is considered and a Bayesian methodology has been developed to estimate the population size based on data augmentation. In particular, the Gibbs sampling method has been developed to make the computation fast and implementation easy for the practitioners. The proposed method seems to have an edge in terms of ease of interpretation and to have a
much wider domain of applicability in the fields of public health, demography, and social sciences. From the simulation study, we can see that the proposed method exhibits a remarkable improvement over the existing models. Moreover, the proposed method is robust with respect to model misspecification. Along with the estimate of the population size, the posterior distribution of the associated dependence parameters and capture probabilities provide specific insights into the capture-recapture mechanism. In particular, the proposed method provides a clear picture of the nature and the extent of behavioural dependence of the individuals in the population under consideration.  In real applications, we exhibit the importance of geospatial analysis of outbreaks in surveillance and identification of associated risk factors. We also emphasized the necessity to adjust the conventional method of calculating PR incorporating the estimated size of the underreported cases for an unbiased evaluation of the interventions. In a frequentist setup, a possible direction for further research could be the development of an estimation methodology based on a suitable pseudo-likelihood approach. Multiple systems estimation strategies have recently been applied to estimate the number of victims of human trafficking and modern slavery. In this context, it is not uncommon to find sparse or even no overlap between some of the lists on which the estimates are based \citep{Chan2020}. It will be an interesting problem to model such data by using the TBM and to develop an associated estimation methodology.

\section*{Acknowledgement}
The authors like to thank Mr. Debjit Majumder for his help in creating the bar diagrams. The work of Dr. Prajamitra Bhuyan is supported in part by the Lloyd’s Register Foundation programme on data-centric engineering at the Alan Turing Institute, UK. The work of Dr. Kiranmoy Chatterjee is funded by (Core Research Grant CRG/2019/003204) the Science and Engineering Research Board (SERB), Department of Science \& Technology, Government of India at Bidhannagar College Kolkata. Both authors contributed equally to this paper.

\section*{Supplementary Information}
Supplementary material is openly available at \doi{10.13140/RG.2.2.14598.34886}.

\bibliographystyle{apalike}
\bibliography{Bibliography_TBM}

\begin{thebibliography}{}

\bibitem[Aldrich, 2002]{Aldrich02}
Aldrich, J. (2002).
\newblock How likelihood and identification went bayesian.
\newblock {\em International Statistical Review}, 70(1):79--98.
\newblock \url{https://doi.org/10.2307/1403727}.

\bibitem[Bird and King, 2018]{Bird18}
Bird, S.~M. and King, R. (2018).
\newblock Multiple systems estimation (or capture-recapture estimation) to
  inform public policy.
\newblock {\em Annu Rev Stat Appl.}, 5:95–118.

\bibitem[Bohning and Heijden, 2009]{Bohning09}
Bohning, D. and Heijden, P. V.~D. (2009).
\newblock Recent developments in life and social science applications of
  capture–recapture methods.
\newblock {\em Advanced Statistical Analysis}, 93:1--3.
\newblock \url{https://doi.org/10.1007/s10182-008-0097-7}.

\bibitem[Brandsema et~al., 2014]{Brandsema14}
Brandsema, P.~S., Euser, S.~M., Karagiannis, I., Den~Boer, J.~W., and Van
  Der~Hoek, W. (2014).
\newblock Summer increase of {L}egionnaires' disease 2010 in the netherlands
  associated with weather conditions and implications for source finding.
\newblock {\em Epidemiology and Infection}, 142(11):2360--2371.

\bibitem[Brown et~al., 2017]{Brown2017}
Brown, C.~R., MacLachlan, J.~H., and Benjamin, C.~C. (2017).
\newblock Addressing the increasing global burden of viral hepatitis.
\newblock {\em Hepatobiliary Surgery and Nutrition}, 6(4):274–276.

\bibitem[{Centers for Disease Control and Prevention}, 2021]{CDC}
{Centers for Disease Control and Prevention} (2021).
\newblock {CDC} surveillance classifications.
\newblock {\em
  \url{https://www.cdc.gov/legionella/health-depts/surv-reporting/surveillance-classifications.html}}.

\bibitem[Chan et~al., 2021]{Chan2020}
Chan, L., Silverman, B.~W., and Vincent, K. (2021).
\newblock Multiple systems estimation for sparse capture data: Inferential
  challenges when there are non overlapping lists.
\newblock {\em Journal of the American Statistical Association},
  116(535):1297--1306.
\newblock \url{https://doi.org/10.1080/01621459.2019.1708748}.

\bibitem[ChandraSekar and Deming, 1949]{Chandrasekar49}
ChandraSekar, C. and Deming, W.~E. (1949).
\newblock On a method of estimating birth and death rates and the extent of
  registration.
\newblock {\em Journal of the American Statistical Association}, 44:101--115.
\newblock \url{https://doi.org/10.2307/2280353}.

\bibitem[Chao, 2001]{Chao01b}
Chao, A. (2001).
\newblock An overview of closed capture–recapture models.
\newblock {\em Journal of Agricultural, Biological, and Environmental
  Statistics}, 6:158–175.

\bibitem[Chao, 2015]{Chao15}
Chao, A. (2015).
\newblock Capture-recapture for human populations.
\newblock {\em Wiley StatsRef: Statistics Reference Online, John Wiley \& Sons,
  Ltd}, page 158–175.

\bibitem[Chao et~al., 2000]{Chao00}
Chao, A., Chu, W., and Chiu, H.~H. (2000).
\newblock Capture-recapture when time and behavioral response affect capture
  probabilities.
\newblock {\em Biometrics}, 56:427--433.

\bibitem[Chao and Tsay, 1998]{Chao98}
Chao, A. and Tsay, P.~K. (1998).
\newblock A sample coverage approach to multiple-system estimation with
  application to census undercount.
\newblock {\em Journal of American Statistical Association}, 93:283--293.

\bibitem[Chao et~al., 2001]{Chao01a}
Chao, A., Tsay, P.~K., Lin, S.~H., Shau, W.~Y., and Chao, D.~Y. (2001).
\newblock The application of capture-recapture models to epidemiological data.
\newblock {\em Statistics in Medicine}, 20:3123--3157.

\bibitem[Chao et~al., 1997]{Chao97}
Chao, D., Shau, W., Lu, C., Chen, K., Chu, C., Shu, H., and Horng, C. (1997).
\newblock A large outbreak of hepatitis {A} in a college school in {T}aiwan:
  associated with contaminated food and water dissemination.
\newblock {\em Epidemiology Bulletin, Department of Health, Executive Yuan,
  Taiwan Government}, page 693–702.

\bibitem[Chatterjee and Bhuyan, 2020a]{Chatterjee20b}
Chatterjee, K. and Bhuyan, P. (2020a).
\newblock On the estimation of population size from a dependent triple-record
  system.
\newblock {\em Journal of Royal Statistical Society, Series A}, 182:1487--1501.

\bibitem[Chatterjee and Bhuyan, 2020b]{Chatterjee20a}
Chatterjee, K. and Bhuyan, P. (2020b).
\newblock On the estimation of population size from a post-stratified
  two-sample capture–recapture data under dependence.
\newblock {\em Journal of Statistical Computation and Simulation}, 819-838.

\bibitem[Chen et~al., 2019]{Chen19}
Chen, W.~C., Chiang, P.~H., Liao, Y.~H., Huang, L.~C., Hsieh, Y.~J., Chiu,
  C.~M., Lo, Y.~C., Yang, C.~H., and Yang, J.~Y. (2019).
\newblock Outbreak of hepatitis {A} virus infection in {T}aiwan, {J}une 2015 to
  {S}eptember 2017.
\newblock {\em Euro Surveill}, Apr 4:24(14).

\bibitem[Cormack, 1989]{Cormack89}
Cormack, R.~M. (1989).
\newblock Log-linear models for capture-recapture.
\newblock {\em Biometrics}, 45:395--413.

\bibitem[Coull and Agresti, 1999]{Coull99}
Coull, B.~A. and Agresti, A. (1999).
\newblock The use of mixed logit models to reflect heterogeneity in
  capture‐recapture studies.
\newblock {\em Biometrics}, 55:294--301.

\bibitem[Coumans et~al., 2017]{Coumans17}
Coumans, A.~M., Cruyff, M., Heijden, P. G.~M., Heijden, V.~D., Wolf, J., and
  Schmeets, H. (2017).
\newblock Estimating homelessness in the {N}etherlands using a
  capture-recapture approach.
\newblock {\em Social Indicators Research}, 130:189--212.

\bibitem[Crowcroft et~al., 2001]{Crowcroft01}
Crowcroft, N., Walsh, B., Davison, K.~L., Gungabissoon, U., and {PHLS Advisory
  Committee on Vaccination and Immunisation} (2001).
\newblock Guidelines for the control of hepatitis a virus infection.
\newblock {\em Communicable Disease and Public Health}, 4(3):213–227.

\bibitem[Cruyff et~al., 2017]{Cruyff17}
Cruyff, M., Van~Dijk, J., and van~der Heijden, P. G.~M. (2017).
\newblock The challenge of counting victims of human trafficking: Not on the
  record: A multiple systems estimation of the numbers of human trafficking
  victims in the {N}etherlands in 2010–2015 by year, age, gender, and type of
  exploitation.
\newblock {\em Chance}, 30:41--49.

\bibitem[Cuthbert, 2001]{Cuthbert01}
Cuthbert, J.~A. (2001).
\newblock Hepatitis {A}: old and new.
\newblock {\em Clin Microbiol Rev.}, 14:38--58.

\bibitem[Darroch et~al., 1993]{Darroch93}
Darroch, J.~N., Fienberg, S.~E., Glonek, G. F.~V., and Junker, B.~W. (1993).
\newblock A three-sample multiple-recapture approach to census population
  estimation with heterogeneous catchability.
\newblock {\em Journal of the American Statistical Association}, 88:1137--1148.

\bibitem[Den et~al., 2002a]{Den02b}
Den, B. J.~W., Friesema, I. H.~M., and Hooi, J.~D. (2002a).
\newblock Reported cases of {L}egionella pneumonia in the {N}etherlands,
  1987-2000 [in dutch].
\newblock {\em Ned Tijdschr Geneeskd.}, 146:315--320.

\bibitem[Den et~al., 2002b]{Den02}
Den, B. J.~W., Yzerman, E. P.~F., Schellekens, J., Lettinga, K.~D., Boshuizen,
  H.~C., Van~Steenbergen, J.~E., Bosman, A., Van~den Hof, S., Van~Vliet, H.~A.,
  Peeters, M.~F., Van~Ketel, R.~J., Speelman, P., Kool, J.~L., , and
  Van~Spaendonck, M. A. E. C.~V. (2002b).
\newblock A large outbreak of {L}egionnaires’ disease at a flower show, the
  {N}etherlands, 1999.
\newblock {\em Emerg Infect Dis.}, 8:37–43.

\bibitem[Dey and Ashbolt, 2020]{Dey20}
Dey, R. and Ashbolt, N.~J. (2020).
\newblock {L}egionella infection during and after the {COVID}-19 pandemic.
\newblock {\em ACS ES$\&$T Water}, acsestwater.0c00151.

\bibitem[Farnham et~al., 2014]{Farnham14}
Farnham, A., Alleyne, L., Cimini, D., and Balter, S. (2014).
\newblock Legionnaires' disease incidence and risk factors, new york, new york,
  usa, 2002-2011.
\newblock {\em Emerging Infectious Diseases}, 20(11):1795--1802.

\bibitem[Fienberg, 1972]{Fienberg72b}
Fienberg, S.~E. (1972).
\newblock The multiple recapture census for closed populations and incomplete
  $2^k$ contingency tables.
\newblock {\em Biometrika}, 59:591--603.

\bibitem[Fischer, 2000]{Fischer00}
Fischer, M.~J. (2000).
\newblock The folded {EGB2} distribution and its application to financial
  return data.
\newblock {\em Discussion Papers 32/2000}, pages Friedrich--Alexander
  University Erlangen--Nuremberg, Chair of Statistics and Econometrics.

\bibitem[Gallay et~al., 2000]{Gallay00}
Gallay, A., Vaillant, V., Bouvet, P., Grimont, P., and Desenclos, J. (2000).
\newblock How many foodborne outbreaks of {S}almonella infection occurred in
  {F}rance in 1995? {A}pplication of the capture–recapture method to three
  surveillance systems.
\newblock {\em Am. J. Epidemiol.}, 152:171–177.

\bibitem[Geweke, 1992]{Geweke1992}
Geweke, J. (1992).
\newblock Evaluating the accuracy of sampling-based approaches to the
  calculation of posterior moments.
\newblock {\em J.M. Bernardo, J.O. Berger, A.P. Dawid and A.F.M. Smith (eds.),
  Bayesian Statistics, Oxford University Press, Oxford}, 4:169--193.

\bibitem[Goodman, 1974]{Goodman74}
Goodman, L.~A. (1974).
\newblock Exploratory latent structure analysis using both identifiable and
  unidentifiable models.
\newblock {\em Biometrika}, 61:215--231.

\bibitem[Gustafson, 2005]{Gustafson05}
Gustafson, P. (2005).
\newblock On model expansion, model contraction, identifiability and prior
  information: Two illustrative scenarios involving mismeasured variables.
\newblock {\em Statistical Science}, 20(1):111--140.

\bibitem[Hook and Regal, 1995]{Hook95}
Hook, E.~B. and Regal, R. (1995).
\newblock Capture–recapture methods in epidemiology: methods and limitations.
\newblock {\em Epidemiol. Rev.}, 17:243–264.

\bibitem[{International Working Group for Disease Monitoring and Forecasting},
  1995]{WorkingGroup95a}
{International Working Group for Disease Monitoring and Forecasting} (1995).
\newblock Capture-recapture and multiple-record systems estimation {I}: History
  and theoretical development.
\newblock {\em American Journal of Epidemiology}, 142:1047--1058.

\bibitem[Islam, 2015]{Islam15}
Islam, H.~W. (2015).
\newblock {\em Estimating the Missing People in the {UK} 1991 Population
  Census}.
\newblock AuthorHouse, UK.

\bibitem[Johnson et~al., 1994]{Johnson94}
Johnson, N.~L., Kotz, S., and Balakrishnan, N. (1994).
\newblock {\em Continuous Univariate Distributions, Band 1}.
\newblock John Wiley \& Sons, New York-Chicester-Brisbane, 2nd edition.

\bibitem[Kadane et~al., 1999]{Kadane99}
Kadane, J.~B., Meyer, M.~M., and Tukey, J.~W. (1999).
\newblock Yule's association paradox and ignored stratum heterogeneity in
  capture-recapture studies.
\newblock {\em Journal of the American Statistical Association}, 94:855--859.

\bibitem[Karagiannis et~al., 2009]{Karagiannisy09}
Karagiannis, I., Brandsema, P., and Van Der~Sande, M. (2009).
\newblock Warm, wet weather associated with increased {L}egionnaires' disease
  incidence in the netherlands.
\newblock {\em Epidemiology and Infection}, 137:181--187.

\bibitem[Lai et~al., 2020]{Lai20}
Lai, C.~C., Wang, C.~Y., and Hsueh, P.~R. (2020).
\newblock Co-infections among patients with {COVID}-19: The need for
  combination therapy with non-anti-{SARS}-{CoV}-2 agents?
\newblock {\em J. Microbiol Immunol Infect}, 53:505–512.

\bibitem[Lettinga et~al., 2002]{Lettinga02}
Lettinga, K.~D., Verbon, A., Weverling, G.-J., Schellekens, J. F.~P., Boer, J.
  W.~D., Yzerman, E. P.~F., Prins, J., Boersma, W.~G., van Ketel, R.~J., Prins,
  J.~M., and Speelman, P. (2002).
\newblock {L}egionnaires’ disease at a dutch flower show: prognostic factors
  and impact of therapy.
\newblock {\em Emerg Infect Dis.}, 8:1448--1454.

\bibitem[Manrique-Vallier, 2016a]{Vallier16}
Manrique-Vallier, D. (2016a).
\newblock Bayesian population size estimation using dirichlet process mixtures.
\newblock {\em Biometrics}, 72:1246--1254.

\bibitem[Manrique-Vallier, 2016b]{Vallier2016}
Manrique-Vallier, D. (2016b).
\newblock Bayesian population size estimation using dirichlet process mixtures.
\newblock {\em Biometrics}, 72:1246--1254.

\bibitem[Martin and Lemon, 2006]{Martin2006}
Martin, A. and Lemon, S.~M. (2006).
\newblock Hepatitis {A} virus: from discovery to vaccines.
\newblock {\em Hepatology}.

\bibitem[Matin et~al., 2006]{Matin06}
Matin, N., Grant, A., Granerod, J., and Crowcroft, N. (2006).
\newblock Hepatitis a surveillance in england – how many cases are not
  reported and does it really matter?
\newblock {\em Epidemiology and Infection}, 134:1299–1302.

\bibitem[Nardone et~al., 2003]{Nardone03}
Nardone, A., Decludt, B., Jarraud, S., Etienne, J., Hubert, B., Infuso, A.,
  Gallay, A., and Desenclos, J.~C. (2003).
\newblock Repeat capture–recapture studies as part of the evaluation of the
  surveillance of {L}egionnaires’ disease in {F}rance.
\newblock {\em Epidemiol Infect.}, 131:647--654.

\bibitem[O'Hara et~al., 2009]{O'Hara}
O'Hara, R.~B., Lampila, S., and Orell, M. (2009).
\newblock Estimation of rates of births, deaths, and immigration from
  mark-recapture data.
\newblock {\em Biometrics}, (65):275--281.

\bibitem[Otis et~al., 1978]{Otis78}
Otis, D.~L., Burnham, K.~P., White, G.~C., and Anderson, D.~R. (1978).
\newblock Statistical inference from capture data on closed animal populations.
\newblock {\em Wildlife Monographs: A Publication of Wildlife Society},
  (62):3--135.

\bibitem[Papoz et~al., 1996]{Papoz96}
Papoz, L., Balkau, B., and Lelleouch, J. (1996).
\newblock Case counting in epidemiology: Limitations of methods based on
  multiple data sources.
\newblock {\em International Journal of Epidemiology}, 25:474--478.

\bibitem[Passer et~al., 2020]{Passer20}
Passer, J.~K., Danila, R.~N., Laine, E.~S., Como-Sabetti, K.~J., Tang, W., and
  Searle, K.~M. (2020).
\newblock The association between sporadic {L}egionnaires’ disease and
  weather and environmental factors, minnesota, 2011–2018.
\newblock {\em Epidemiology and Infection}.

\bibitem[Rivest and Baillargion, 2022]{Rivest22}
Rivest, L.-P. and Baillargion, S. (2022).
\newblock Loglinear models for capture-recapture experiments.
\newblock {\em CRAN}.
\newblock Date of publication: 2022-05-04.

\bibitem[Rivest and Levesque, 2001]{Rivest01}
Rivest, L.-P. and Levesque, T. (2001).
\newblock Improved log-linear model estimators of abundance in
  capture-recapture experiments.
\newblock {\em The Canadian Journal of Statistics}, 29:555--572.

\bibitem[Ruche et~al., 2013]{Ruche13}
Ruche, G.~L., Dejour-Salamanca, D., Bernillon, O.~P., Leparc-Goffart, I.,
  Ledrans, M., Armengaud, A., Debruyne, M., Denoyel, G.-A., Brichler, S.,
  Ninove, L., Desprès, P., and Gastellu-Etchegorry, M. (2013).
\newblock Capture–recapture method for estimating annual incidence of
  imported dengue, {F}rance, 2007–2010.
\newblock {\em Emerging Infectious Diseases}, 19:1740--1748.

\bibitem[Sanathanan, 1972a]{Sanathanan72b}
Sanathanan, L. (1972a).
\newblock Estimating the size of a multonomial population.
\newblock {\em The Annals of Mathematical Statistics}, 43:142--152.

\bibitem[Sanathanan, 1972b]{Sanathanan72a}
Sanathanan, L. (1972b).
\newblock Models and estimation methods in visual scanning experiments.
\newblock {\em Technometrics}, 43:813--829.

\bibitem[Tanner and Wong, 1987]{Tanner87}
Tanner, M.~A. and Wong, W.~H. (1987).
\newblock The calculation of posterior distributions by data augmentation.
\newblock {\em Journal of the American Statistical Association}, 82:528--540.

\bibitem[Tsay and Chao, 2001]{Tsay01}
Tsay, P.~K. and Chao, A. (2001).
\newblock Population size estimation for capture-recapture models with
  applications to epidemiological data.
\newblock {\em Journal of Applied Statistics}, 28:25--36.

\bibitem[Tsou et~al., 2011]{Tsou11}
Tsou, T., Liu, C., Huang, J., Tsai, K., and Chang, H. (2011).
\newblock Change in hepatitis a epidemiology after vaccinating high risk
  children in taiwan, 1995-2008.
\newblock {\em Vaccine}, 29(16):2956--2961.

\bibitem[Van~Hest et~al., 2008]{Hest08}
Van~Hest, N. A.~H., Hoebe, C. J. P.~A., Den~Boer, J.~W., Vermunt, J.~K.,
  Ijzerman, E. P.~F., Boersma, W.~G., and Richardus, J.~H. (2008).
\newblock Incidence and completeness of notification of {L}egionnaires’
  disease in the {N}etherlands: covariate capture–recapture analysis
  acknowledging regional differences.
\newblock {\em Epidemiol Infect}, 136:540--550.

\bibitem[Van~Hest, 2007]{Hest07}
Van~Hest, R. (2007).
\newblock {\em Capture-recapture methods in surveillance of {Tuberculosis} and
  other infectious diseases}.
\newblock Print Partners Ipskamp, Enschede.

\bibitem[Wechsler et~al., 2013]{Wechsler13}
Wechsler, S., Izbicki, R., and Esteves, L.~G. (2013).
\newblock A bayesian look at nonidentifiability: A simple example.
\newblock {\em The American Statistician}, 67(2):1537--2731.

\bibitem[White, 1982]{White82}
White, H. (1982).
\newblock Maximum likelihood estimation of misspecified models.
\newblock {\em Econometrica}, 50:1--25.

\bibitem[Wolter, 1986]{Wolter86}
Wolter, K.~M. (1986).
\newblock Some coverage error models for census data.
\newblock {\em Journal of the American Statistical Association}, 81:338--346.

\bibitem[{World Health Organization}, 2021]{WHO}
{World Health Organization} (2021).
\newblock {WHO} {H}epatitis {A} - {D}iagnosis.
\newblock {\em
  \url{https://www.who.int/news-room/fact-sheets/detail/hepatitis-a}}.

\bibitem[Zaslavsky and Wolfgang, 1993]{Zaslavsky93}
Zaslavsky, A.~M. and Wolfgang, G.~S. (1993).
\newblock Triple-system modeling of census, post-enumeration survey, and
  administrative-list data.
\newblock {\em Journal of Business and Economic Statistics}, 11:279--288.

\bibitem[Zhou et~al., 2020]{Zhou20}
Zhou, F., Yu, T., Du, R., Fan, G., Liu, Y., Liu, Z., Xiang, J., Wang, Y., Song,
  B., Gu, X., Guan, L., Wei, Y., Li, H., Wu, X., Xu, J., Tu, S., Zhang, Y.,
  Chen, H., and Cao, B. (2020).
\newblock Clinical course and risk factors for mortality of adult inpatients
  with {COVID}-19 in {Wuhan, China}: A retrospective cohort study.
\newblock {\em Lancet}, 395:1054--1062.

\end{thebibliography}

\end{document}



\begin{center}
	\large{\textbf{Supplementary Material for ``Estimation of Population Size with Heterogeneous Catchability and Behavioural Dependence: Applications to Air and Water Borne Disease Surveillance'' by Bhuyan and Chatterjee}} 
\end{center}

	\setcounter{page}{1}
	
\section{TRS Data Structure and Real Datasets}\label{Dataset}
\setcounter{table}{0}
\renewcommand{\thetable}{\thesection.\arabic{table}}

\setcounter{figure}{0}
\renewcommand\thefigure{ \thesection. \arabic{figure}} 

\begin{table}[h]
\renewcommand\thetable{S1}
	\begin{center}
		\caption{Data structure associated with a triple record system.}
		\begin{tabular}{lcccccccc}
			\hline
			&\multicolumn{8}{c}{List 3} \\
			\cline{2-9}
			&\multicolumn{3}{c}{In}&&&\multicolumn{3}{c}{Out} \\			
			\cline{2-9}
			&\multicolumn{3}{c}{List 2}&&&\multicolumn{3}{c}{List 2} \\	
			List 1 & In & Out & Total &&& In & Out & Total\\
			\hline \hline
			In & $x_{111}$ & $x_{101}$ & $x_{1\cdot1}$ &&& $x_{110}$ & $x_{100}$ & $x_{1\cdot0}$\\
			Out& $x_{011}$ & $x_{001}$ & $x_{0\cdot1}$ &&& $x_{010}$ & \textbf{$x_{000}$} & $x_{0\cdot0}$\\ 
			\hline
			Total& $x_{\cdot11}$ & $x_{\cdot01}$ & $x_{\cdot\cdot1}$ &&& $x_{\cdot10}$ & $x_{\cdot00}$ & $x_{\cdot\cdot0}$\\
			\hline
		\end{tabular}
		\label{Tab:1}
	\end{center}
\end{table}

\begin{table}[ht]
\renewcommand\thetable{S2}
	\tiny
	\centering
	\caption{Dataset on Legionnaires' disease (LD) and  hepatitis A virus (HAV) surveillance.}
	\resizebox{5.6in}{!}{
		\begin{tabular}{|cccccccccc|}
			\hline
			Disease & Stratum &	$x_{111}$ & $x_{110}$ & $x_{101}$ & $x_{011}$ & $x_{100}$ & $x_{010}$ & $x_{001}$ & Total ($n$)\\
			\hline
			LD & North &	13 & 2 & 6 & 8 & 3 & 2 & 35 & 69\\
			& East &	45 & 3 & 42 & 7 & 13 & 13 & 62 & 185 \\
			& West  &	46 & 7 & 55 & 14 & 23 & 5 & 136 & 286 \\
			& South  &	51 & 19 & 28 & 15 & 13 & 9 & 99 & 234 \\
   			& All &	155 & 31 & 131 & 45 & 56 & 30 & 332 & 780 \\
			\multicolumn{10}{|c|}{}\\
			HAV &  &	28 & 21 & 17 & 18 & 69 & 55 & 63 & 271 \\
			\hline
		\end{tabular}
	}
	\label{case_Data}
\end{table}

\section{Existing Models and Estimates}\label{existing}
\numberwithin{equation}{subsection}
\setcounter{table}{0}
\renewcommand{\thetable}{\thesection.\arabic{table}}

\setcounter{figure}{0}
\renewcommand\thefigure{\thesection.\arabic{figure}} 

As mentioned in Subsection 1.3 of the main article, modeling list dependence and individual heterogeneity is an important problem over the decades for its wide applications. In this section, we present frequently used models for TRS and associated estimates for the population size. The assumptions associated  with these models and their limitations in real applications are also discussed briefly. We consider these estimates to compare the performance of the proposed method through the simulation study discussed in Section 4 of the main article.

\subsection{Models with List Dependence}\label{List-dependence} 
\subsubsection{Log-linear Model}
\citet{Fienberg72b} and \citet{Bishop75} discussed several log-linear models (LLMs) to account for list dependence using interaction effects. The general LLM under TRS is  given by
\begin{eqnarray}
	\log(m_{ijk})=u_0 + u_{1(i)} + u_{2(j)} + u_{3(k)} + u_{12(ij)} + u_{13(ik)} + u_{23(ik)} + u_{123(ijk)},\label{log-linear_gen}
\end{eqnarray} 
where $m_{ijk}=\mathbb{E}[x_{ijk}]$, $u_{s(0)}+u_{s(1)}=0$,  $u_{s s^{\prime} (0j)}+u_{s s^{\prime}(1j)}=0$,
$u_{s s^{\prime} (i0)}+u_{s s^{\prime}(i1)}=0$,
$u_{s s^{\prime}l^{\ast} (0jk)}+u_{s s^{\prime}l^{\ast}(1jk)}=0$,
$u_{s s^{\prime}l^{\ast} (i0k)}+u_{s s^{\prime}l^{\ast}(i1k)}=0$,
$u_{s s^{\prime}l^{\ast} (ij0)}+u_{s s^{\prime}l^{\ast}(ij1)}=0$,
for all $s,s^{\prime},s^{\ast}=1,2,3$, with $s\ne s^{\prime}\ne s^{\ast}$ \citep[p.64]{Bishop75}. The parameters $u_s$, $u_{s s^{\prime}}$ and $u_{123}$ denote the main effects, pairwise interaction effects, and the second order interaction effect for $s,s^{\prime}=1,2,3$. See \citet{Fienberg72a, Fienberg72b} and \citet{WorkingGroup95a} for more details. To ensure the estimability of the model given in (\ref{log-linear_gen}), one needs to consider $u_{123}=0$, i.e. no second-order interaction between lists. Under this assumption, the estimate of $m_{000}$ is given by
\begin{eqnarray}
\hat{m}_{000}=\frac{\hat{m}_{111}\hat{m}_{001}\hat{m}_{100}\hat{m}_{010}}{\hat{m}_{101}\hat{m}_{011}\hat{m}_{110}},\nonumber
	\label{est_no_2nd_order}
\end{eqnarray}
where $\hat{m}_{ijk}$ is the maximum likelihood estimate of $m_{ijk}$ assuming $x_{ijk}$ to be a realization of an independent Poisson random variate for all $(i,j,k)$th cells, except the $(0,0,0)$th cell. Finally, the estimate of $N$ is obtained as $\hat{N}_{LLM}=n+\hat{m}_{000}$ \citep{Fienberg72b}. This is equivalent to the estimator proposed by \citet{Zaslavsky93} with $\alpha_{EPA}=1$. With additional assumption that $u_{12}=u_{13}=u_{23}=0$, LLM in (\ref{log-linear_gen}) reduces to independent model.

\subsubsection{Time \& Behavioural Response Variation Model: $M_{tb}$}\label{Mtb}
It is often observed in TRS that an individual's behavior changes in subsequent recapture attempts after the initial attempt. This change is known as behavioural response variation \citep{Otis78, Chatterjee16c}. When the list dependence, caused by the behavioural response variation, is considered along with the assumption of list variation in capture probabilities, one would have a model known as $M_{tb}$ model \citep{Wolter86}. 

Denote the first-time capture probability of any individual in the $s$-th list by $f_s$ for $s=1,2,3$, and the recapture probability is denoted by $c_s$ for $s=2,3$. Further, $u_s$ and $m_s$ respectively denote the number of first-time captured and recaptured individuals in the $s$-th list. Therefore, based on the the sufficient statistics ($u_1=x_{1\cdot\cdot},u_2=x_{01\cdot},u_3=x_{001},m_2=x_{11\cdot}, m_3=x_{101}+x_{011}+x_{111}$) and the assumption of constant proportionality, i.e., $c_s/f_s=\phi$, for $s=2,3$, likelihood of the model $M_{tb}$ is given by
\begin{eqnarray}
	L(N,\textbf{\textit{f}},\phi)=\frac{N!}{(N-n)}f_1^{u_1}(1-f_1)^{N-u_1}\phi^{m_2+m_3}\prod_{s=2}^{3}f_s^{u_s+m_s}(1-f_s)^{N-M_{s+1}}(1-\phi p_s)^{M_s-m_s},\nonumber
\end{eqnarray}
where $M_s=u_1+u_2+...+u_{s-1}$ denotes the number of individuals captured at least once prior to the $s$-th attempt \citep{Chatterjee20b}. Hence, $M_{s+1}=n$. In this context, \citet{Chao00} discussed various likelihood-based methods for the estimation of $N$. Note that, the $M_{tb}$ model does not utilize full information available from seven known cells in TRS; hence, it loses efficiency in estimating $N$ \citep{Chatterjee20b}. Moreover, the assumption of $c_2/f_2=c_3/f_3=\phi$ is not justifiable in various real-life applications.

\subsection{Models with Individual Heterogeneity}\label{Hetero}
\subsubsection{Rasch Model and its extensions}
In educational statistics and Psychometry, Rasch model \citep{Rasch61} is widely used to account for individual heterogeneity in different sources. The capture probabilities are modeled using logistic regression:

\begin{eqnarray}
	\log\left[\frac{\mathbb{E}[Z_h^{(s)}]}{1-\mathbb{E}[Z_h^{(s)}]}\right]=v_h+e_s, \hspace{0.3in} h=1,2,\cdots,N;\hspace{0.2in}s=1,2,3,\label{Rasch}
\end{eqnarray} 
where $v_h$ is a random effect representing the catchability of $h$th individual and $e_s$ is effect of $s$th list. Even with the absence of behavioural response variation, dependence among the lists are induced is purely due to heterogeneity in capture probabilities \citep{Chao01a}. When we set $v_h=0$ in (\ref{Rasch}), the model reduces to a multiple-recapture model with list independence. We now represent the Rasch model as a mixed-effects generalized linear model that allows for individual heterogeneity and list variation in the following subsections.

\subsubsection{Quasi-symmetry Model}
The marginal cell probabilities for all eight cells in a TRS can be represented as:
\begin{eqnarray}	\log(p_{ijk})=a+ie_1+je_2+ke_3+\gamma(i_+),\label{quasi}
\end{eqnarray} 
where $i_+=i+j+k$ and $\gamma(t)=\log\left[\mathbb{E}\left[\exp (v_h t)|i=j=k=0\right]\right]$. See \citet{Darroch93, Fienberg99} for detailed derivation. Note that the term $\gamma(i_+)$ is invariant with respect to permutations of ($i,j,k$), and hence the model in (\ref{quasi}) is known as quasi-symmetry model (QSM) \citep{Darroch81}. This is equivalent to the following assumption involving two constraints
\begin{eqnarray}
	p_{011}p_{100}=p_{101}p_{010}=p_{110}p_{001}\label{two-constraints}
\end{eqnarray}
which do not involve $p_{000}$. Therefore, an additional assumption is required, such as no second-order interaction exists. Under this assumption, the estimated count of the $(0,0,0)$th cell is given by
\begin{eqnarray}
	\hat{x}_{000}=\frac{\hat{x}_{111}\hat{x}_{100}\hat{x}_{010}\hat{x}_{001}}{\hat{x}_{110}\hat{x}_{101}\hat{x}_{011}},\nonumber\label{no-2nd-order}
\end{eqnarray}
where $\hat{x}_{ijk}$ is the maximum likelihood estimate of $\mathbb{E}[x_{ijk}|n]$ \citep{Darroch93}. 

\subsubsection{Partial Quasi-symmetry Model}
In many situations, the constraint provided in equation (\ref{two-constraints}) associated with quasi-symmetry model is not realistic. However, it is reasonable to assume 
$p_{011}p_{100}=p_{101}p_{010}$
if $x_{011}x_{100}$ and $x_{101}x_{010}$ are close enough \citep{Darroch93}. This condition implies that the patterns of heterogeneity are different for two different group of sources or lists as considered in generalized Rasch model proposed by \citet{Stegelman83}. Under this assumption,
the marginal cell probabilities can be represented as:
\begin{eqnarray}\label{partial-quasi}
	\log(p_{ijk})=a+ie_1+je_2+ke_3+\gamma(i+j,k), \nonumber
\end{eqnarray} 
where $\gamma(t_1,t_2)=\log\left[\mathbb{E}\left[\exp(v_ht_1)\exp(z_h t_2)|i=j=k=0\right]\right]$ with two non-identically distributed random effects $v_h$ and $z_h$ associated with group of lists ($L_1$, $L_2$) and $L_3$, respectively. This model is known as partial quasi-symmetry model (PQSM). See \citet{Darroch93, Fienberg99} for more details.

\subsubsection{Non-parametric Latent Class Model}\label{NPLCM}
\citet{Vallier16} proposed a Bayesian non-parametric method for estimating the population size by accounting for the heterogeneity in capture probabilities considering a latent stratification of the population. In this method, the number of strata is considered unknown and strata identities are treated as missing data. Under these assumptions, the Bayesian non-parametric modeling is outlined below.

Let $\boldsymbol{Z_h}=(Z_{h}^{(1)},Z_{h}^{(2)},Z_{h}^{(3)})\in\{0,1\}^3$ denote the capture status of the $h$-th individual in the three lists of TRS and it follows \textit{iid} $f(\cdot|\theta)$ for $h\in\{1,2,\dots,N\}$. Therefore, the joint distribution of the observed and the structurally unobserved cells in TRS is 
\begin{eqnarray}
p(\boldsymbol{z_1},\boldsymbol{z_2},\dots,\boldsymbol{z_n}|\theta,N)=\binom Nn f(\boldsymbol{0}|\theta)^{N-n}\prod_{h=1}^n f(\boldsymbol{z_h}|\theta)\mathbb{I}(N\geq n), \label{eqn01}
\end{eqnarray}
where $\mathbb{I}(\cdot)$ denotes indicator function, and the capture status $\boldsymbol{z_h}=(z_{h}^{(1)},z_{h}^{(2)},z_{h}^{(3)})$ refers to the realization of $\boldsymbol{Z_h}$. Note that $n=\sum_{h=1}^N \mathbb{I}(\boldsymbol{Z_h}\neq \boldsymbol{0})$. Let us assume that the population admits a partition into $M$ homogeneous strata, such that the list independence holds within each stratum. Let $\boldsymbol{\pi}=(\pi_1,\pi_2,\dots,\pi_M)$ be the vector of stratum probabilities with $\sum_{m=1}^M\pi_m=1$ and $\pi_m>0$ for $m=1,2,\cdots,M$. Then, the probability mass function of the capture status vector of \textit{h}-th individual is
\begin{eqnarray}	p(\boldsymbol{z_h}|\boldsymbol{\lambda},\boldsymbol{\pi})=\sum_{m=1}^M\pi_m \prod_{s=1}^3\lambda_{sm}^{z_h^{(s)}}(1-\lambda_{sm})^{1-z_h^{(s)}},\label{eqn03}
\end{eqnarray}
for $h=1,2,\cdots,n$, where, $\boldsymbol{\lambda}=\{\lambda_{sm}\in(0,1): s=1,2,3; m=1,2,\dots,M\}$. The product-Bernoulli mixture distribution, provided in (\ref{eqn03}), is coined as the Latent Class Model \citep{Goodman74}. To apply the model in practice, we need to solve the model selection problem by means of choosing an appropriate number of latent classes $M$.

An augmented data representation for a mixture model is given by 
\begin{eqnarray}
	Z_s|W=w\sim Bernoulli(\lambda_{sw}) \mbox{ for} \mbox{ $s=1,2,3$}\nonumber\\
	W\sim Discrete(\{1,2,\dots,M\},(\pi_1,\pi_2,\dots,\pi_M)),
\end{eqnarray}
where the latent variable $ W$ explicitly represents the stratum assignment. Replacing $f(\boldsymbol{z_h}|\theta)$ in (\ref{eqn01}) by the probability mass function $p(\boldsymbol{z_h}|\boldsymbol{\lambda},\boldsymbol{\pi})$ given in (\ref{eqn03}), we obtain
\begin{eqnarray}
p(\boldsymbol{z_1},\dots,\boldsymbol{z_n}|\boldsymbol{\lambda},\boldsymbol{\pi},N)=\binom Nn \left[\sum_{m=1}^M\pi_m \prod_{s=1}^3(1-\lambda_{sm})\right]^{N-n}\times \prod_{h=1}^n\sum_{m=1}^M\pi_m \prod_{s=1}^3\lambda_{sm}^{z_{h}^{(s)}}(1-\lambda_{sm})^{1-z_{h}^{(s)}}.\nonumber
\end{eqnarray}
The above equation can be written in a marginalized version of the augmented data representation as:
\begin{eqnarray}
p(\boldsymbol{z_1},\dots,\boldsymbol{z_n},\boldsymbol{w},\boldsymbol{w^0}|\boldsymbol{\lambda},\boldsymbol{\pi},N)=\binom Nn \left[\prod_{h=1}^{n_0}\pi_{w_h^0} \prod_{s=1}^3(1-\lambda_{sw_h^0})\right]\times\prod_{h=1}^n\pi_{w_h} \prod_{s=1}^3\lambda_{sw_h}^{z_{h}^{(s)}}(1-\lambda_{sw_h})^{1-z_{h}^{(s)}},\nonumber
\end{eqnarray}
where $\boldsymbol{w}=(w_1,w_2,\dots,w_n)$, $\boldsymbol{w^0}=(w_1^0,w_2^0,\dots,w_{n_0}^0)$, $n_0=N-n$, and both $w_i$, $w_i^0$ take values on the set $\{1,2,\dots,M^*\}$ for each $i=1,2,\dots,n$. Here, $M^*$ refers to a large enough upper bound for the number
of latent classes in order to practically implement this method with finite-dimensional approximation. 
Considering diffuse and Jeffrey's prior on parameter space, Gibbs sampling algorithm is used to obtain samples from the posterior distribution of $N$. Interested readers are referred to \citet{Vallier16} for details of the prior specifications and sampling algorithm.

\subsection{Models accounting Individual Heterogeneity \& List Dependence}
\subsubsection{Sample Coverage Approach}\label{SCM}
Sample Coverage is a non-parametric approach. The idea of the sample coverage of a given sample is the probability-weighted fraction of the population captured in that sample. The main principle relies on the fact that it is difficult to estimate the number of undercounts directly, but the sample coverage can be well estimated in any non-independent situation with heterogeneous capture probability \citep{Chao92, Chao98}. In the context of TRS, the sample coverage is defined based on three hypothetical populations I, II, and III with capture probabilities $\left\{\mathbb{E}\left[Z_h^{(1)}\mid Z_h^{(2)},Z_h^{(3)}\right]; h=1,\ldots, N \right\}$, $\left\{\mathbb{E}\left[Z_h^{(2)} \mid Z_h^{(1)},Z_h^{(3)}\right]; h=1,\ldots, N \right\}$, and \\
$\left\{\mathbb{E}\left[Z_h^{(3)} \mid Z_h^{(1)},Z_h^{(2)}\right]; h=1,\ldots, N \right\}$ corresponding to lists $L_1$, $L_2$, and $L_3$, respectively. In particular, the sample coverage of first two lists $L_1\cup L_2$ with respect to the population III is given by
\begin{eqnarray}
	C_{III}(L_1\cup L_2)=\frac{\sum_{h=1}^{N}\mathbb{E}\left[Z_h^{(3)} \mid Z_h^{(1)},Z_h^{(2)}\right]\mathbb{I}[Z_h^{(1)}+Z_h^{(2)}>0]}{\sum_{h=1}^{N}\mathbb{E}\left[Z_h^{(3)} \mid Z_h^{(1)},Z_h^{(2)}\right]}. \nonumber
\end{eqnarray}
Similarly, one can obtain the sample coverages of other two possibles combinations of samples, $C_{II}(L_1\cup L_3)$ and $C_{I}(L_2\cup L_3)$ with respect to the populations II and I, respectively. Now, the sample coverage of the three lists is defined as $C=\frac{1}{3}\left\{C_{III}(L_1\cup L_2)+C_{II}(L_1\cup L_3)+C_{I}(L_2\cup L_3)\right\}$, and estimated by $\hat{C}=1-\frac{1}{3}\left[\frac{x_{100}}{n_1}+\frac{x_{010}}{n_2}+\frac{x_{001}}{n_3}\right]$. Finally, the estimator of $N$ on based sample coverage approach is given by
\begin{eqnarray}
	\hat{N}_{sc}=\frac{x_{\cdot 11}+x_{1\cdot 1}+x_{11\cdot}}{3\hat{C}}\left\{1-\frac{1}{3\hat{C}}\left[
	\frac{(x_{1\cdot 0}+x_{\cdot 10})x_{11\cdot}}{n_1n_2}+
	\frac{(x_{10\cdot}+x_{\cdot 01})x_{1\cdot 1}}{n_1n_3}+
	\frac{(x_{0\cdot 1}+x_{01\cdot})x_{\cdot 11}}{n_2n_3}\right]\right\}^{-1}.\nonumber
\end{eqnarray}
We refer this estimate as SC. From empirical studies, it is found that the performance of $\hat{N}_{sc}$ is satisfactory if the sample coverage is over 55\%. The performance also depends on the remainder term involved in the derivation of $\hat{N}_{sc}$, and the bias increases with its magnitude \citep{Chao98}. Moreover, this method may produce infeasible estimates, i.e. $\hat{N}_{sc}< n$, unlike other methods. 

\subsubsection{Behavioural Response Variation \& Heterogeneous Catchability Model: $M_{bh}$}\label{Mbh}
Model $M_{bh}$ is widely used for population size estimation, accounting for both the heterogeneous catchability and list dependence due to behavioural response variation. An individual's behavior changes after the first capture, with the capture probability changing from $p$ to $c$, where $p$ denotes the first-time capture probability of individuals and $c$ denotes the recapture probability \citep{Otis78}. Here, $c$ is a nuisance parameter. Therefore, the sufficient statistic for ($N,p$) is $\{u_s: s=1,2,3\}$, where $u_s$ denotes the number of individuals that are first-time included in the $s$-th list. One can express $u_s$ with respect to TRS data as: $u_1=x_{1\cdot\cdot}=(x_{111}+x_{110}+x_{101}+x_{100}), u_2=x_{01\cdot}=(x_{011}+x_{010})$ and $u_3=x_{001}$ (\textit{see} Table \ref{Tab:1}). The log-linear model involving only behavioural response variation is given by
\begin{eqnarray}
	\log(\mu_{s})=log(Np)+(s-1)log(1-p)=\gamma+(s-1)\beta\label{u_s},
\end{eqnarray}
where $\mu_{s}=E(u_{s})$, $\gamma=log(Np)$ and $\beta=log(1-p)$ for $s=1,2,3$ \citep{Rivest01}. If the capture probabilities are heterogeneous, the individuals captured early are not representative of the individuals not captured up to $t_0$th ($t_0=0,1,2$) capture attempts. This can be included in the model by assuming that $\mu_{1},\cdots,\mu_{t_0}$ do not follow the model (\ref{u_s}) for $t_0\geq0$. Therefore, $N=\sum_{s=1}^{t_0} E(u_s)+N_1$, where $N_1$ is the size of the population that is not captured up to $t_0$ capture attempts. Note that these $ N_1$ individuals follow the model (\ref{u_s}). Hence, 
\begin{eqnarray}
	\log(\mu_{t_0+s})=log(N_1p)+(s-1)log(1-p)=\gamma+(s-1)\beta,
\end{eqnarray}
where $\gamma=log(N_1p)$, $\beta=log(1-p)$ for $s=1,2,(3-t_0)$. Therefore, $N=\sum_{s=1}^{t_0} E(u_s)+\frac{e^{\gamma}}{1-e^{\beta}}$. This modeling of heterogeneity with a log-linear for model $M_{bh}$ is an alternative approach to the generalized removal model discussed in \citet{Otis78}. Based on the above model, estimates of the population size and associated model parameters are obtained by maximizing a Poisson log-likelihood. Interested readers are referred to \citet{Rivest01} for more details. The `Rcapture' package available in R software is used to perform the simulation experiments and data analyses based on model $M_{bh}$ \citep{Baillargeon07}. Similar to the sample coverage method, model $M_{bh}$ may also produce infeasible estimates.

\section{Derivations of Full Conditionals}\label{Derev}
The conditional posterior density of all the model parameters and latent variables are obtained from the joint posterior density of the unobserved quantities $\boldsymbol{\theta}$, $\boldsymbol{y}$, and $\boldsymbol{b}$ provided in (5) and (6) of the main article.
\numberwithin{equation}{subsection}
\setcounter{table}{0}
\renewcommand{\thetable}{\thesection.\arabic{table}}

\setcounter{figure}{0}
\renewcommand\thefigure{\thesection.\arabic{figure}} 

\subsection{Derivations of Full Conditionals of the Latent Cell Counts}\label{Derev_y}
The conditional posterior density of the latent cell counts $\boldsymbol{y}$, given $(N, \boldsymbol{\alpha}, \boldsymbol{b})$, is obtained as
$$
\pi(\boldsymbol{y}|N,\boldsymbol{\alpha},\boldsymbol{b},\boldsymbol{x})\propto \prod_{i,j,k=0,1}\pi(y_{ijk}|\boldsymbol{\theta},\boldsymbol{b},\boldsymbol{x}),
$$
where
\begin{eqnarray}
	\pi(\boldsymbol{y_{111}}|\boldsymbol{\theta},\boldsymbol{x}) & \propto & Multinomial(x_{111};Q_{111,1}^*,Q_{111,2}^*,Q_{111,3}^*,Q_{111,4}^*,Q_{111,5}^*),\nonumber\\
	\pi(y_{110,1}|\boldsymbol{\theta},\boldsymbol{x}) & \propto & Binomial(x_{110};(1-\alpha_0) \mathcal{P}_1 \mathcal{P}_2 (1-\mathcal{P}_3)),\nonumber\\
	\pi(y_{011,1}|\boldsymbol{\theta},\boldsymbol{x}) & \propto & Binomial(x_{011};(1-\alpha_0)(1-\mathcal{P}_1)\mathcal{P}_2 \mathcal{P}_3),\nonumber\\
	\pi(y_{100,1}|\boldsymbol{\theta},\boldsymbol{x}) & \propto & Binomial(x_{100};(1-\alpha_0)\mathcal{P}_1(1-\mathcal{P}_2)(1-\mathcal{P}_3)),\nonumber\\
	\pi(y_{101,1}|\boldsymbol{\theta},\boldsymbol{x}) & \propto & Binomial(x_{101};(1-\alpha_0)\mathcal{P}_1(1-\mathcal{P}_2)\mathcal{P}_3),\nonumber\\
	\pi(y_{010,1}|\boldsymbol{\theta},\boldsymbol{x}) & \propto & Binomial(x_{010};(1-\alpha_0)(1-\mathcal{P}_1)\mathcal{P}_2(1-\mathcal{P}_3)),\nonumber\\
	\pi(y_{001,1}|\boldsymbol{\theta},\boldsymbol{x}) & \propto & Binomial\left(x_{001};(1-\alpha_0)(1-\mathcal{P}_1)(1-\mathcal{P}_2)\mathcal{P}_3\right),\nonumber\\
	\pi(\boldsymbol{y_{000}}|\boldsymbol{\theta},\boldsymbol{x}) & \propto & Multinomial(N-n;Q_{000,1}^*,Q_{000,2}^*,Q_{000,3}^*,Q_{000,4}^*,Q_{000,5}^*)\nonumber,\label{sampling_y}
\end{eqnarray}   
with
$\boldsymbol{y_{111}}=\left(y_{111,1},y_{111,2},y_{111,3},y_{111,4},y_{111,5}\right)$, $\boldsymbol{y_{000}}=\left(y_{000,1},y_{000,2},y_{000,3},y_{000,4},y_{000,5}\right)$,
$Q_{111,1}^*=(1-\alpha_0)\mathcal{P}_1 \mathcal{P}_2 \mathcal{P}_3/Q_{111}$, $Q_{111,2}^*=\alpha_1 \mathcal{P}_1 \mathcal{P}_3/Q_{111} $, $Q_{111,3}^*=\alpha_2 \mathcal{P}_1 \mathcal{P}_2/Q_{111}$, $Q_{111,4}^*=\alpha_3 \mathcal{P}_1 \mathcal{P}_2/Q_{111}$, $Q_{111,5}^*=\alpha_4\mathcal{P}_1/Q_{111},$ $Q_{111}=(1-\alpha_0)\mathcal{P}_1 \mathcal{P}_2 \mathcal{P}_3+\alpha_1 \mathcal{P}_1 \mathcal{P}_3+\alpha_2 \mathcal{P}_1 \mathcal{P}_2+\alpha_3 \mathcal{P}_1 \mathcal{P}_2+\alpha_4\mathcal{P}_1$;\\
$Q_{000,1}=(1-\alpha_0) (1-\mathcal{P}_1) (1-\mathcal{P}_2) (1-\mathcal{P}_3)/Q_{000}$, $Q_{000,2}=\alpha_1 (1-\mathcal{P}_1) (1-\mathcal{P}_3)/Q_{000}$, $Q_{000,3}=\alpha_2 (1-\mathcal{P}_1) (1-\mathcal{P}_2)/Q_{000}$, $Q_{000,4}=\alpha_3 (1-\mathcal{P}_1) (1-\mathcal{P}_2)/Q_{000}, Q_{000,5}=\alpha_4(1-\mathcal{P}_1)/Q_{000},$ $Q_{000}=
{(1-\alpha_0) (1-\mathcal{P}_1) (1-\mathcal{P}_2) (1-\mathcal{P}_3)}+\alpha_1 (1-\mathcal{P}_1) (1-\mathcal{P}_3)+\alpha_2 (1-\mathcal{P}_1) (1-\mathcal{P}_2)+{\alpha_3 (1-\mathcal{P}_1) (1-\mathcal{P}_2)}+\alpha_4(1-\mathcal{P}_1)$. 


\subsection{Derivations of Full Conditionals of the Random Effects}\label{Derev_b}
Note that the random effect $b_s$ follows the generalized logistic distribution of type-I with shape parameter $\delta_s$, and its density function is given by
\begin{equation}
	g_{b_s}(z|\delta_s)=\frac{\delta_se^{-z}}{[1+e^{-z}]^{\delta_s+1}}, -\infty< z < \infty , \delta_{s}>0\nonumber
\end{equation}
for $s=1,2,3$. Now, the conditional posterior density of $b_1$ is given by
\begin{eqnarray}
	\pi(b_1|\boldsymbol{\theta},\boldsymbol{x},\boldsymbol{y}) & \propto & \mathcal{P}_1^{x_{1\cdot\cdot}}(1-\mathcal{P}_1)^{N-x_{1\cdot\cdot}}\times g_{b_1}(z|\delta_1)\nonumber\\
	& = & \left[\frac{1}{1+e^{-b_1}}\right]^{x_{1\cdot\cdot}}\left[\frac{e^{-b_1}}{1+e^{-b_1}}\right]^{N-x_{1\cdot\cdot}}\times \frac{\delta_1e^{-b_1}}{[1+e^{-b_1}]^{\delta_1+1}}\nonumber\\
	& \propto & \frac{e^{-b_1[N-x_{1\cdot\cdot}+1]}}{\left[1+e^{-b_1}\right]^{N+\delta_1+1}}\nonumber\\
	& \equiv & EGB2(n_1+1,m_1+\delta_1),\nonumber
\end{eqnarray}
where $EGB2$ denotes exponential generalized beta distribution of second kind with parameters involving
$m_1=x_{1\cdot\cdot}$, $n_1=(N-x_{1\cdot\cdot})$. Similarly, the conditional posterior densities of $b_2$ and $b_3$ can be obtained as 
\begin{eqnarray}
	\pi(b_2|\boldsymbol{\theta},\boldsymbol{x},\boldsymbol{y}) & \propto & EGB2(n_2+1,m_2+\delta_2),\nonumber\\
	\pi(b_3|\boldsymbol{\theta},\boldsymbol{x},\boldsymbol{y}) & \propto & EGB2(n_3+1,m_3+\delta_3),\nonumber
\end{eqnarray}
respectively, where $m_2=y_{111,1}+y_{111,3}+y_{111,4}+y_{110,1}+x_{011}+x_{010}$, $m_3=y_{111,1}+y_{111,2}+y_{011,1}+y_{101,1}+x_{001}, n_1=N-x_{1++}$, $n_2=x_{100}+x_{101}+y_{001,1}+y_{000,1}+y_{000,3}+y_{000,4}$,  and $n_3=x_{110}+y_{100,1}+y_{010,1}+y_{000,1}+y_{000,2}$.

\subsection{Derivations of Full Conditionals of the Dependence Parameter}\label{Derev_alpha}
The conditional posterior of $\boldsymbol{\alpha}$ is obtained as
$$\pi(\boldsymbol{\alpha}|\boldsymbol{\theta}_{-\boldsymbol{\alpha}},\boldsymbol{x},\boldsymbol{y},\boldsymbol{b})\propto\alpha_1^{k_1}\alpha_2^{k_2}\alpha_3^{k_3}\alpha_4^{k_4}(1-\alpha_0)^{k_5}\times\pi(\boldsymbol{\alpha}),$$
where $k_1=y_{111,2}+x_{110}-y_{110,1}+x_{001}-y_{001,1}+y_{000,2}, k_2=y_{111,3}+x_{011}-y_{011,1}+x_{100}-y_{100,1}+y_{000,3}, k_3=y_{111,4}+x_{101}-y_{101,1}+x_{010}-y_{010,1}+y_{000,4}, k_4=x_{111}-\sum_{i=1}^{4}y_{111,i}+N-n-\sum_{i=1}^{4}y_{000,i}, k_5=y_{111,1}+y_{110,1}+y_{011,1}+y_{100,1}+y_{101,1}+y_{010,1}+y_{001,1}+y_{000,1}$, and $\alpha_0=\sum_{\omega=1}^{4}\alpha_{\omega}$. Under the prior choice I, $\pi(\boldsymbol{\alpha})\equiv\pi(\alpha_1,\alpha_2,\alpha_3,\alpha_4,1-\alpha_0)= Dirichlet(0.5,0.5,0.5,0.5,0.5)$ and the resulting posterior distribution of $\boldsymbol{\alpha}$ is given by
$$\pi(\boldsymbol{\alpha}|\boldsymbol{\theta}_{-\boldsymbol{\alpha}},\boldsymbol{x},\boldsymbol{y},\boldsymbol{b})\equiv\pi(\alpha_1,\alpha_2,\alpha_3,\alpha_4,1-\alpha_0|\boldsymbol{\theta}_{-\boldsymbol{\alpha}},\boldsymbol{x},\boldsymbol{y},\boldsymbol{b})\sim Dirichlet(d_1,d_2,d_3,d_4,d_5),$$
where $d_u=k_u+0.5$, for $u=1,\ldots,5$. 

Under prior choice II, $\pi(\boldsymbol{\alpha})\equiv\pi(\alpha_1,\alpha_2,\alpha_3,\alpha_4,1-\alpha_0)=\equiv Dirichlet(\beta_1,\beta_2,\beta_3,\beta_4,\beta_5)$, and the posterior distribution of $\boldsymbol{\alpha}$ is given by 
$$\pi(\boldsymbol{\alpha}|\boldsymbol{\theta}_{-\boldsymbol{\alpha}},\boldsymbol{x},\boldsymbol{y},\boldsymbol{b})\equiv\pi(\alpha_1,\alpha_2,\alpha_3,\alpha_4,1-\alpha_0|\boldsymbol{\theta}_{-\boldsymbol{\alpha}},\boldsymbol{x},\boldsymbol{y},\boldsymbol{b})\sim Dirichlet(d_1^*,d_2^*,d_3^*,d_4^*,d_5^*),$$
where $d_u^*=k_u+\beta_u$, for $u=1,\ldots,5$.

\subsection{Derivations of Full Conditionals of the Shape Parameter Associated with the Random Effects Distribution}\label{Derev_delta}
Under the prior choice I, we consider Jeffrey's prior for $\delta_s$, i.e. $\pi(\delta_s)\propto\delta_s^{-1}$ for $s=1,2,3$, and the conditional posterior distribution of $\delta_s$ is given by
\begin{eqnarray}
	\pi(\delta_{s}|\boldsymbol{\theta}_{-\delta_{s}},\boldsymbol{x},\boldsymbol{y},\boldsymbol{b}) &\propto & g_{b_s}(b_s|\delta_s)\times\pi(\delta_s)\nonumber\\
	&\propto & \left[1+e^{-b_s}\right]^{-\delta_s}\nonumber\\
	& = & e^{-\omega_s\delta_s}\nonumber\\
	&\propto & Exponential(\omega_s),\nonumber
\end{eqnarray}
where $\omega_s=\log\left[1+e^{-b_s}\right]$ for $s=1,2,3$.

Under the prior choice II, $\pi(\delta_s)\propto Gamma\left(\gamma_s,\lambda_s\right)$ for $s=1,2,3$. Then, the conditional posterior distribution of $\delta_s$ is given by
\begin{eqnarray}
	\pi(\delta_{s}|\boldsymbol{\theta}_{-\delta_{s}},\boldsymbol{x},\boldsymbol{y},\boldsymbol{b}) &\propto & g_{b_s}(b_s|\delta_s)\times\pi(\delta_s)\nonumber\\
	&\propto & \left[1+e^{-b_s}\right]^{-\delta_s} \delta_s^{\gamma_s}e^{-\frac{\delta_s}{\lambda_s}} \nonumber\\
	& = & e^{-\delta_s\left[\omega_s+\lambda_s^{-1}\right]} \delta_s^{\gamma_s}\nonumber\\
	&\propto & Gamma\left(\gamma_s+1,[\omega_s+\lambda_s^{-1}]^{-1}\right)\nonumber
\end{eqnarray}
for $l=1,2,3$.

\subsection{Derivations of Full Conditional of the Population Size}\label{Derev_N}
Note that Jeffrey's prior for $N$ ( i.e. $\pi(N)\propto N^{-1}$) is considered under both the prior choices I and II. The conditional posterior distribution of $N$ given $(\boldsymbol{\theta}_{-N},\boldsymbol{x},\boldsymbol{y_{-y_{000,5}}},\boldsymbol{b})$ is obtained as
\begin{eqnarray}
	\pi(N|\boldsymbol{\theta}_{-N},\boldsymbol{x},\boldsymbol{y}_{-y_{000,5}},\boldsymbol{b}) & \propto & \frac{(N-1)!}{(N-n-\sum_{u=1}^{4}y_{000,u})!}[\alpha_4(1-\mathcal{P}_1)]^{N-n-\sum_{u=1}^{4}y_{000,u}}.\nonumber
\end{eqnarray}
To generate samples from the conditional posterior distribution of $N$, one can alternatively simulate from the conditional distribution of $y_{000,5}=N-n-\sum_{u=1}^{4}y_{000,u}$, given $(\boldsymbol{\theta}_{-N},\boldsymbol{x},\boldsymbol{y}_{-y_{000,5}},\boldsymbol{b})$, provided as 
\begin{eqnarray}
	\pi(y_{000,5}|\boldsymbol{\theta}_{-N},\boldsymbol{x},\boldsymbol{y}_{-y_{000,5}},\boldsymbol{b}) & \propto & \frac{(y_{000,5}+ n+\sum_{u=1}^{4}y_{000,u}-1)!}{y_{000,5}!}[\alpha_4(1-\mathcal{P}_1)]^{y_{000,5}}\nonumber\\
	\vspace{0.2in}  & \propto & NB \left(n+\sum_{u=1}^{4}y_{000,u},1-\alpha_4(1-\mathcal{P}_1)\right),\nonumber
\end{eqnarray}
and then compute $N=y_{000,5}+n+\sum_{u=1}^{4}y_{000,u}$.


\section{Results from Simulation Study}\label{Simtab}
\setcounter{table}{0}
\renewcommand{\thetable}{\thesection.\arabic{table}}

\setcounter{figure}{0}
\renewcommand\thefigure{\thesection.\arabic{figure}} 

The performance of the proposed estimators along with the seven existing competitors based on TRS data simulated from six choices of populations P1-P6 (See Table 1) with five different combinations of random-effects $\boldsymbol{\delta}$ as mentioned in Section 4 of the main article, is provided in the following tables. The abbreviations used in the tables are described below. Note that the CI corresponding to THBM and NPCLM refers to the HPD credible interval.\\

\noindent
THBM: Trivariate Heterogeneous Bernoulli Model; SC: Sample Coverage;
QSM: Quasi-symmetry Model;
PQSM: Partial Quasi-symmetry Model; LLM: Log-linear Model, $M_{tb}$: Time and Behavioural Response Variation Model; Independent: LLM without interaction effects.
$M_{bh}$: Behavioural Response Variation and Heterogeneous Catchability Model; NPLCM: Non-parametric Latent Class Model.

\renewcommand{\baselinestretch}{1}

\begin{table}[h]
\renewcommand\thetable{S3}
	\tiny
	\caption{Comparison of various estimators of the population size $N$ under P1.}
	\label{Tab:2}
	\begin{center}
		\begin{tabular}{|lrccccc|}
			\hline
			\multicolumn{7}{|c|}{}\\
			Model & & $\delta=(1.6,1.2,0.8)$ & $\delta=(1.3, 1.7, 0.9)$ & $\delta=(1,1.4,1.8) $ & $\delta=(0.8,0.8,0.8)$ & $\delta=(1.6,1.6,1.6)$\\
			\hline
			\multicolumn{7}{|c|}{}\\
			\multicolumn{7}{|c|}{}\\
			\multicolumn{7}{|c|}{$N=200$}\\
			\multicolumn{7}{|c|}{}\\
			THBM-I  & $\hat{N}$(RMAE) & 194 (0.070) & 199 (0.070) & 209 (0.086) & 187 (0.106) & 203 (0.063)\\
			& CI  & $(164, 318)$& (165, 329) & $(166, 359)$ & $(145, 354)$ & $(172, 321)$\\
			& Coverage  & 96 & 99 & 99 & 96 & 99\\
			\multicolumn{7}{|c|}{}\\
			THBM-II  & $\hat{N}$(RMAE) & 196 (0.050) & 196 (0.050) & 196 (0.051) &  191 (0.079) & 197 (0.041)\\
			& CI  & $(168, 271)$& $(168, 263)$ & $(168, 262)$ & $(149, 315)$ & $(174, 250)$\\
			& Coverage  & 99 & 99 & 99 & 99 & 99\\
			\multicolumn{7}{|c|}{}\\
			SC  & $\hat{N}$(RMAE) & 175 (0.125) & 176 (0.120) & 180 (0.098) & 163 (0.186) & 183 (0.086)\\
			&  CI  & $(164, 187)$ &$(164, 189)$ & $(166, 198)$  & $(146, 185)$ & $(172, 194)$ \\
			& Coverage  & 10 & 14 & 40 & 16 & 24\\
			\multicolumn{7}{|c|}{}\\
			QSM & $\hat{N}$(RMAE) & 191 (0.096) & 209 (0.122) & 288 (0.148) & 194 (0.155) & 238 (0.208)\\
			&  CI  & $(166, 263)$ & $(170, 330)$ & $(192, 631)$ & $(149, 342)$ & $(184, 415)$\\
			& Coverage  & 87 & 94 & 72 & 93 & 87\\
			\multicolumn{7}{|c|}{}\\
			PQSM & $\hat{N}$(RMAE) & 198 (0.105)& 216 (0.139) & 266 (0.343)  & 201 (0.173) & 233 (0.186)\\
			& CI  & $(167, 297)$& $(171, 365)$ & $(184, 514)$  & $(151, 366)$ & $(182, 404)$\\
			& Coverage  & 92 & 96 & 83 & 92 & 91\\
			\multicolumn{7}{|c|}{}\\
			LLM  & $\hat{N}$(RMAE) & 204 (0.120) & 212 (0.132) & 254 (0.292) & 202 (0.189) & 233 (0.188)\\
			&  CI  & $(169, 369)$ & $(171, 405)$ & $(182, 518)$ & $(152, 414)$ & $(181, 439)$\\
			& Coverage  & 84 & 84 & 78 & 82 & 81\\
			\multicolumn{7}{|c|}{}\\
			$M_{tb}$  & $\hat{N}$(RMAE) & 240 (0.215) & 237 (0.213) & 262 (0.313) & 256 (0.291) & 249 (0.255)\\
			&  CI  & $(167, 471)$ & $(167, 491)$ & $(171, 563)$ & $(147, 544)$ & $(177, 494)$\\
			& Coverage  & 100 & 100 & 100 & 100 & 100\\
			\multicolumn{7}{|c|}{}\\
			Independent & $\hat{N}$(RMAE) & 166 (0.172) & 3
			165 (0.175) & 165 (0.173)  & 146 (0.269) & 173 (0.137)\\
			&  CI  & $(160, 170)$& $(159, 170)$ & $(159, 172)$  & $(138, 154)$ & $(167, 177)$\\
			& Coverage & 0 & 0 & 0 & 0 & 0\\
   			\multicolumn{7}{|c|}{}\\
			NPLCM & $\hat{N}$(RMAE) & 178 (0.127) & 181 (0.116) & 190 (0.083)  & 161 (0.217) & 192 (0.067)\\
			&  CI  & $(161, 203)$& $(161, 211)$ & $(161, 231)$  & $(139, 194)$ & $(169, 224)$\\
			& Coverage & 56 & 69 & 94 & 35 & 93\\
      			\multicolumn{7}{|c|}{}\\
			$M_{bh}$ & $\hat{N}$(RMAE) & 203 (0.191) & 194 (0.147) & 307 (0.573)  & 230 (0.322) & 254 (0.323)\\
			&  CI  & $(-5, 1.5 \times 10^{13})$& $(13, 6.6 \times 10^{12})$ & $(-404, 1.9 \times 10^{13})$  & $(-228, 1.7 \times 10^{13})$ & $(-163, 2 \times 10^{13})$\\
			& Coverage & 83 & 84 & 75 & 77 & 83\\
			\multicolumn{7}{|c|}{}\\
			\multicolumn{7}{|c|}{}\\
			\multicolumn{7}{|c|}{$N=500$}\\
			\multicolumn{7}{|c|}{}\\
			THBM-I  & $\hat{N}$(RMAE) & 487 (0.054) & 499 (0.054) & 522 (0.072) & 472 (0.084) & 519 (0.059)\\
			& CI  & $(423, 729)$& $(425, 772)$ & $(431, 815)$ & $(381, 869)$ & $(446, 783)$\\
			& Coverage  & 97 & 98 & 100 & 96 & 99\\
			\multicolumn{7}{|c|}{}\\
			THBM-II  & $\hat{N}$(RMAE) & 489 (0.042) & 490 (0.042) & 493 (0.040) & 481 (0.066) & 494 (0.031)\\
			& CI  & $(429, 642)$& (330, 633) & $(434, 636)$ & $(391, 754)$ & $(448, 604)$\\
			& Coverage  & 99 & 99 & 99 & 98 & 100\\
			\multicolumn{7}{|c|}{}\\
			SC  & $\hat{N}$(RMAE) & 437 (0.126) & 440 (0.120) & 450 (0.100) & 405 (0.191) & 457 (0.086)\\
			&  CI  & $(419, 455)$ &$(421, 460)$ & $(427, 476)$  & $(378, 436)$ & $(440, 474)$ \\
			& Coverage  & 0 & 0 & 7 & 0 & 2\\
			\multicolumn{7}{|c|}{}\\
			QSM & $\hat{N}$(RMAE) & 468 (0.075) & 507 (0.068) & 498 (0.077) & 471 (0.104) & 568 (0.140)\\
			&  CI  & $(427, 547)$ & $(445, 633)$ & $(427, 640)$ & $(397, 625)$ & $(483, 745)$\\
			& Coverage  & 77 & 95 & 94 & 89 & 77\\
			\multicolumn{7}{|c|}{}\\
			PQSM & $\hat{N}$(RMAE) & 485 (0.066) & 528 (0.090) & 515 (0.088)  & 473 (0.107) & 567 (0.138)\\
			& CI  & $(437, 612)$& $(456, 716)$ & $(435, 704)$  & $(401, 663)$ & $(484, 774)$\\
			& Coverage  & 78 & 84 & 86 & 77 & 72\\
			\multicolumn{7}{|c|}{}\\
			LLM  & $\hat{N}$(RMAE) & 495 (0.068) & 514 (0.076) & 500 (0.082) & 474 (0.108) & 566 (0.138)\\
			&  CI  & $(445, 598)$ & $(448, 695)$ & $(430, 712)$ & $(404, 718)$ & $(487, 839)$\\
			& Coverage  & 72 & 82 & 75 & 71 & 66\\
			\multicolumn{7}{|c|}{}\\
			$M_{tb}$  & $\hat{N}$(RMAE) & 619 (0.239) & 611 (0.224) & 614 (0.230) & 626 (0.251) & 622 (0.244)\\
			&  CI  & $(418, 913)$ & $(420, 919)$ & $(398, 943)$ & $(368, 989)$ & $(441, 913)$\\
			& Coverage  & 100 & 100 & 100 & 100 & 100\\
			\multicolumn{7}{|c|}{}\\
			Independent  & $\hat{N}$(RMAE) & 413 (0.174)& 413 (0.175) & 393 (0.215)  & 366 (0.268) & 431 (0.137)\\
			&  CI  & $(404, 420)$& $(404, 420)$ & $(383, 402)$  & $(354, 378)$ & $(424, 438)$\\
			& Coverage & 0 & 0 & 0 & 0 & 0\\
      			\multicolumn{7}{|c|}{}\\
			NPLCM & $\hat{N}$(RMAE) & 454 (0.110) & 462 (0.097) & 483 (0.063)  & 415 (0.192) & 487 (0.049)\\
			&  CI  & $(408, 518)$& $(408, 537)$ & $(409, 584)$  & $(354, 501)$ & $(431, 567)$\\
			& Coverage & 70 & 87 & 98 & 47 & 99\\
      			\multicolumn{7}{|c|}{}\\
			$M_{bh}$ & $\hat{N}$(RMAE) & 501 (0.154) & 468 (0.102) & 951 (0.908)  & 673 (0.438) & 645 (0.314)\\
			&  CI  & $(50, 1.1 \times 10^{12})$& $(283, 2.6 \times 10^{10})$ & $(-1831, 3.6 \times 10^{12})$  & $(-1100, 1.4 \times 10^{12})$ & $(-514, 8.7 \times 10^{10})$\\
			& Coverage & 85 & 83 & 85 & 88 & 89\\
			\multicolumn{7}{|c|}{}\\
			\hline
		\end{tabular}
	\end{center}
\end{table}

\begin{table}[h]
\renewcommand\thetable{S4}
	\tiny
	\caption{Comparison of various estimators of the population size $N$ under P2.}
	\label{Tab:3}
	\begin{center}
		\begin{tabular}{|lrccccc|}
			\hline
			\multicolumn{7}{|c|}{}\\
			Model & & $\delta=(1.6,1.2,0.8)$ & $\delta=(1.3, 1.7, 0.9)$ & $\delta=(1,1.4,1.8) $ & $\delta=(0.8,0.8,0.8)$ & $\delta=(1.6,1.6,1.6)$\\
			\hline
			\multicolumn{7}{|c|}{}\\
			\multicolumn{7}{|c|}{}\\
			\multicolumn{7}{|c|}{$N=200$}\\
			\multicolumn{7}{|c|}{}\\
			THBM-I  & $\hat{N}$(RMAE) & 195 (0.067) & 199 (0.070) & 206 (0.080) & 187 (0.102) & 205 (0.062)\\
			& CI  & $(165, 327)$& $(165, 329)$ & $(165, 353)$ & $(145, 349)$ & $(173, 321)$\\
			& Coverage  & 98 & 99 & 99 & 97 & 100\\
			\multicolumn{7}{|c|}{}\\
			THBM-II  & $\hat{N}$(RMAE) & 195 (0.049) & 195 (0.049) & 196 (0.050) &  193 (0.081) & 197 (0.039)\\
			& CI  & $(168, 266)$& $(168, 262)$ & $(168, 266)$ & $(150, 318)$ & $(175, 250)$\\
			& Coverage  & 99 & 99 & 99 & 99 & 99\\
			\multicolumn{7}{|c|}{}\\
			SC  & $\hat{N}$(RMAE) & 175 (0.124) & 175 (0.126) & 179 (0.105) & 162 (0.190) & 183 (0.087)\\
			&  CI  & $(164, 188)$ &$(164, 188)$ & $(165, 195)$  & $(145, 183)$ & $(172, 194)$ \\
			& Coverage  & 10 & 11 & 32 & 15 & 24\\
			\multicolumn{7}{|c|}{}\\
			QSM & $\hat{N}$(RMAE) & 203 (0.102) & 203 (0.102) & 248 (0.263) & 195 (0.162) & 235 (0.195)\\
			&  CI  & $(169, 306)$ & $(169, 306)$ & $(180, 475)$ & $(150, 344)$ & $(183, 406)$\\
			& Coverage  & 95 & 95 & 86 & 91 & 89\\
			\multicolumn{7}{|c|}{}\\
			PQSM & $\hat{N}$(RMAE) & 208 (0.119)& 207 (0.116) & 239 (0.227)  & 195 (0.157) & 235 (0.197)\\
			& CI  & $(169, 332)$& $(169, 325)$ & $(177, 441)$  & $(149, 345)$ & $(182, 404)$\\
			& Coverage  & 94 & 94 & 90 & 93 & 90\\
			\multicolumn{7}{|c|}{}\\
			LLM  & $\hat{N}$(RMAE) & 203 (0.113) & 216 (0.150) & 255 (0.300) & 197 (0.170) & 234 (0.200)\\
			&  CI  & $(169, 364)$ & $(172, 405)$ & $(180, 526)$ & $(150, 406)$ & $(183, 453)$\\
			& Coverage  & 84 & 83 & 78 & 82 & 81\\
			\multicolumn{7}{|c|}{}\\
			$M_{tb}$  & $\hat{N}$(RMAE) & 224 (0.176) & 216 (0.161) & 242 (0.233) & 232 (0.211) & 230 (0.181)\\
			&  CI  & $(166, 452)$ & $(166, 460)$ & $(168, 540)$ & $(145, 524)$ & $(175, 470)$\\
			& Coverage  & 100 & 100 & 100 & 100 & 100\\
			\multicolumn{7}{|c|}{}\\
			Independent  & $\hat{N}$(RMAE) & 166 (0.171)& 166 (0.172) & 165 (0.177)  & 146 (0.270) & 173 (0.137)\\
			&  CI  & $(163, 174)$& $(160, 170)$ & $(158, 170)$  & $(139, 154)$ & $(167, 177)$\\
			& Coverage & 0 & 0 & 0 & 0 & 0\\
      			\multicolumn{7}{|c|}{}\\
			NPLCM & $\hat{N}$(RMAE) & 180 (0.120) & 181 (0.116) & 187 (0.095)  & 161 (0.218) & 191 (0.069)\\
			&  CI  & $(162, 207)$& $(162, 209)$ & $(161, 224)$  & $(139, 193)$ & $(170, 223)$\\
			& Coverage & 60 & 66 & 87 & 33 & 91\\
				\multicolumn{7}{|c|}{}\\
			$M_{bh}$ & $\hat{N}$(RMAE) & 184 (0.151) & 177 (0.135) & 240 (0.287)  & 195 (0.261) & 217 (0.186)\\
			&  CI  & $(62, 2.5 \times 10^{13})$ & $(105, 2.3 \times 10^{12})$ & $(-201, 5.3 \times 10^{13})$  & $(-96, 5.7 \times 10^{13})$ & $(-26, 1.8 \times 10^{14})$\\
			& Coverage & 78 & 72 & 90 & 81 & 91\\
			\multicolumn{7}{|c|}{}\\
			\multicolumn{7}{|c|}{}\\
			\multicolumn{7}{|c|}{}\\
			\multicolumn{7}{|c|}{$N=500$}\\
			\multicolumn{7}{|c|}{}\\
			THBM-I  & $\hat{N}$(RMAE) & 492 (0.052) & 506 (0.061) & 524 (0.072) & 473 (0.090) & 519 (0.060)\\
			& CI  & $(425, 770)$& $(427, 804)$ & $(431, 822)$ & $(381, 845)$ & $(445, 786)$\\
			& Coverage  & 98 & 98 & 100 & 95 & 100\\
			\multicolumn{7}{|c|}{}\\
			THBM-II  & $\hat{N}$(RMAE) & 490 (0.039) & 493 (0.039) & 493 (0.039) & 482 (0.063) & 494 (0.031)\\
			& CI  & $(431, 645)$ & $(432, 639)$ & $(433, 637)$ & $(391, 745)$ & $(447, 602)$\\
			& Coverage  & 99 & 100 & 99 & 99 & 99\\
			\multicolumn{7}{|c|}{}\\
			SC  & $\hat{N}$(RMAE) & 437 (0.125) & 439 (0.123) & 447 (0.106) & 405 (0.190) & 456 (0.089)\\
			&  CI  & $(419, 456)$ &$(421, 458)$ & $(425, 458)$  & $(378, 436)$ & $(439, 473)$ \\
			& Coverage  & 0 & 0 & 2 & 1 & 2\\
			\multicolumn{7}{|c|}{}\\
			QSM & $\hat{N}$(RMAE) & 496 (0.063) & 495 (0.063) & 475 (0.079) & 472 (0.103) & 572 (0.148)\\
			&  CI  & $(441, 606)$ & $(439, 605)$ & $(418, 589)$ & $(397, 626)$ & $(484, 754)$\\
			& Coverage  & 93 & 94 & 88 & 89 & 75\\
			\multicolumn{7}{|c|}{}\\
			PQSM & $\hat{N}$(RMAE) & 509 (0.071) & 502 (0.066) & 479 (0.078)  & 473 (0.103) & 571 (0.147)\\
			& CI  & $(448, 661)$& $(444, 639)$ & $(422, 622)$  & $(401, 659)$ & $(490, 794)$\\
			& Coverage  & 82 & 85 & 79 & 78 & 68\\
			\multicolumn{7}{|c|}{}\\
			LLM  & $\hat{N}$(RMAE) & 496 (0.066) & 518 (0.077) & 603 (0.214) & 475 (0.106) & 569 (0.145)\\
			&  CI  & $(443, 666)$ & $(447, 674)$ & $(482, 881)$ & $(404, 733)$ & $(495, 885)$\\
			& Coverage  & 73 & 96 & 72 & 71 & 64\\
			\multicolumn{7}{|c|}{}\\
			$M_{tb}$  & $\hat{N}$(RMAE) & 608 (0.219) & 565 (0.151) & 572 (0.165) & 620 (0.241) & 602 (0.206)\\
			&  CI  & $(416, 900)$ & $(416, 866)$ & $(394, 898)$ & $(363, 979)$ & $(437, 893)$\\
			& Coverage  & 100 & 100 & 100 & 100 & 100\\
			\multicolumn{7}{|c|}{}\\
			Independent  & $\hat{N}$(RMAE) & 415 (0.171) & 413 (0.173) & 393 (0.213)  & 366 (0.268) & 431 (0.137)\\
			&  CI  & $(406, 422)$& $(405, 421)$ & $(383, 402)$  & $(354, 377)$ & $(424, 438)$\\
			& Coverage & 0 & 0 & 0 & 0 & 0\\
      			\multicolumn{7}{|c|}{}\\
			NPLCM & $\hat{N}$(RMAE) & 457 (0.104) & 461 (0.098) & 478 (0.072)  & 414 (0.194) & 186 (0.050)\\
			&  CI  & $(409, 525)$& $(410, 534)$ & $(408, 576)$  & $(354, 499)$ & $(430, 566)$\\
			& Coverage & 79 & 87 & 99 & 45 & 99\\
			\multicolumn{7}{|c|}{}\\
			$M_{bh}$ & $\hat{N}$(RMAE) & 454 (0.128) & 438 (0.126) & 576 (0.208)  & 459 (0.187) & 504 (0.090)\\
			&  CI  & $(327, 5.6 \times 10^{8})$ & $(401, 557)$ & $(-264, 1.9 \times 10^{12})$  & $(-1, 1.1 \times 10^{11})$ & $(192, 2.5 \times 10^{9})$\\
			& Coverage & 89 & 49 & 96 & 85 & 94\\
			\multicolumn{7}{|c|}{}\\
			\hline
		\end{tabular}
	\end{center}
\end{table}

\begin{table}[h]
\renewcommand\thetable{S5}
	\tiny
	\caption{Comparison of various estimators of the population size $N$ under P3.}
	\label{Tab:4}
	\begin{center}
		\begin{tabular}{|lrccccc|}
			\hline
			\multicolumn{7}{|c|}{}\\
			Model & & $\delta=(1.6,1.2,0.8)$ & $\delta=(1.3, 1.7, 0.9)$ & $\delta=(1,1.4,1.8) $ & $\delta=(0.8,0.8,0.8)$ & $\delta=(1.6,1.6,1.6)$\\
			\hline
			\multicolumn{7}{|c|}{}\\
			\multicolumn{7}{|c|}{}\\
			\multicolumn{7}{|c|}{$N=200$}\\
			\multicolumn{7}{|c|}{}\\
			THBM-I  & $\hat{N}$(RMAE) & 201 (0.059) & 204 (0.065) & 209 (0.086) & 200 (0.092) & 204 (0.055)\\
			& CI  & $(170, 331)$& $(170, 335)$ & $(166, 359)$ & $(154, 390)$ & $(177, 281)$\\
			& Coverage  & 98 & 99 & 99 & 98 & 98\\
			\multicolumn{7}{|c|}{}\\
			THBM-II  & $\hat{N}$(RMAE) & 196 (0.050) & 197 (0.049) & 196 (0.050) &  193 (0.075) & 198 (0.039)\\
			& CI  & $(172, 244)$& $(173, 241)$ & $(168, 266)$ & $(158, 269)$ & $(178, 234)$\\
			& Coverage  & 97 & 98 & 99 & 96 & 97\\
			\multicolumn{7}{|c|}{}\\
			SC  & $\hat{N}$(RMAE) & 175 (0.124) & 184 (0.083) & 186 (0.075) & 176 (0.122) & 188 (0.063)\\
			&  CI  & $(164, 188)$ & $(170, 199)$ & $(171, 203)$  & $(155, 205)$ & $(176, 200)$ \\
			& Coverage  & 10 & 43 & 55 & 55 & 48\\
			\multicolumn{7}{|c|}{}\\
			QSM & $\hat{N}$(RMAE) & 199 (0.084) & 205 (0.088) & 288 (0.448) & 207 (0.154) & 225 (0.162)\\
			&  CI  & $(172, 272)$ & $(174, 293)$ & $(192, 631)$ & $(159, 356)$ & $(177, 390)$\\
			& Coverage  & 94 & 96 & 72 & 93 & 84\\
			\multicolumn{7}{|c|}{}\\
			PQSM & $\hat{N}$(RMAE) & 200 (0.084)& 207 (0.102) & 266 (0.343)  & 207 (0.149) & 212 (0.095)\\
			& CI  & $(172, 282)$& $(173,306)$ & $(184, 514)$  & $(159, 355)$ & $(180, 304)$\\
			& Coverage  & 94 & 95 & 83 & 95 & 95\\
			\multicolumn{7}{|c|}{}\\
			LLM  & $\hat{N}$(RMAE) & 204 (0.095) & 204 (0.099) & 254 (0.292) & 208 (0.155) & 211 (0.096)\\
			&  CI  & $(174, 343)$ & $(174, 341)$ & $(182, 518)$ & $(161, 409)$ & $(181, 339)$\\
			& Coverage  & 85 & 82 & 78 & 86 & 84\\
			\multicolumn{7}{|c|}{}\\
			$M_{tb}$  & $\hat{N}$(RMAE) & 218 (0.167) & 214 (0.160) & 262 (0.313) & 227 (0.209) & 226 (0.167)\\
			&  CI  & $(171, 466)$ & $(171, 474)$ & $(154, 340)$ & $(154, 564)$ & $(179, 481)$\\
			& Coverage  & 100 & 100 & 100 & 100 & 100\\
			\multicolumn{7}{|c|}{}\\
			Independent  & $\hat{N}$(RMAE) & 173 (0.135)& 173 (0.137) & 165 (0.173)  & 157 (0.213) & 178 (0.0.109)\\
			&  CI  & $(166, 179)$& $(166, 179)$ & $(159, 172)$  & $(148, 168)$ & $(172, 183)$\\
			& Coverage & 1 & 1 & 0 & 2 & 1\\
      		\multicolumn{7}{|c|}{}\\
			NPLCM & $\hat{N}$(RMAE) & 184 (0.097) & 186 (0.089) & 188 (0.081)  & 170 (0.169) & 192 (0.061)\\
			&  CI  & $(167, 209)$& $(167, 212)$ & $(167, 218)$  & $(147, 203)$ & $(174, 217)$\\
			& Coverage & 66 & 72 & 84 & 48 & 86\\
			\multicolumn{7}{|c|}{}\\
			$M_{bh}$ & $\hat{N}$(RMAE) & 180 (0.118) & 177 (0.123) & 222 (0.186)  & 196 (0.213) & 203 (0.112)\\
			&  CI  & $(115, 3.5 \times 10^{12})$ & $(146, 5.2 \times 10^{10})$ & $(-65, 2.5 \times 10^{12})$  & $(-87, 5.3 \times 10^{12})$ & $(47, 1.4 \times 10^{12})$\\
			& Coverage & 72 & 57 & 94 & 85 & 89\\
			\multicolumn{7}{|c|}{}\\
			\multicolumn{7}{|c|}{}\\
			\multicolumn{7}{|c|}{}\\
			\multicolumn{7}{|c|}{$N=500$}\\
			\multicolumn{7}{|c|}{}\\
			THBM-I  & $\hat{N}$(RMAE) & 504 (0.049) & 512 (0.052) & 522 (0.072) & 505 (0.073) & 510 (0.044)\\
			& CI  & $(538, 779)$& $(440, 784)$ & $(431, 815)$ & $(405, 901)$ & $(456, 626)$\\
			& Coverage  & 98 & 99 & 100 & 98 & 98\\
			\multicolumn{7}{|c|}{}\\
			THBM-II  & $\hat{N}$(RMAE) & 491 (0.042) & 494 (0.042) & 495 (0.039) & 485 (0.063) & 498 (0.031)\\
			& CI  & $(443, 574)$& $(445, 571)$ & $(444, 582)$ & $(412, 616)$ & $(457, 561)$\\
			& Coverage  & 95 & 96 & 100 & 96 & 98\\
			\multicolumn{7}{|c|}{}\\
			SC  & $\hat{N}$(RMAE) & 458 (0.083) & 459 (0.082) & 447 (0.106) & 439 (0.122) & 469 (0.063)\\
			&  CI  & $(437, 481)$ & $(438, 482)$ & $(424, 471)$  & $(405, 480)$ & $(451, 487)$ \\
			& Coverage  & 10 & 11 & 4 & 20 & 16\\
			\multicolumn{7}{|c|}{}\\
			QSM & $\hat{N}$(RMAE) & 488 (0.052) & 504 (0.055) & 505 (0.067) & 499 (0.091) & 517 (0.057)\\
			&  CI  & $(444, 568)$ & $(453, 602)$ & $(442, 627)$ & $(422, 655)$ & $(468, 612)$\\
			& Coverage  & 92 & 95 & 95 & 93 & 98\\
			\multicolumn{7}{|c|}{}\\
			PQSM & $\hat{N}$(RMAE) & 493 (0.051) & 511 (0.060) & 511 (0.072)  & 500 (0.091) & 517 (0.057)\\
			& CI  & $(448, 594)$& $(458, 637)$ & $(449, 670)$  & $(427, 691)$ & $(469, 623)$\\
			& Coverage  & 82 & 85 & 83 & 82 & 84\\
			\multicolumn{7}{|c|}{}\\
			LLM  & $\hat{N}$(RMAE) & 501 (0.054) & 505 (0.057) & 502 (0.068) & 502 (0.093) & 516 (0.056)\\
			&  CI  & $(454, 644)$ & $(458, 652)$ & $(449, 690)$ & $(433, 744)$ & $(470, 681)$\\
			& Coverage  & 75 & 75 & 74 & 69 & 74\\
			\multicolumn{7}{|c|}{}\\
			$M_{tb}$  & $\hat{N}$(RMAE) & 575 (0.164) & 555 (0.148) & 578 (0.182) & 619 (0.239) & 567 (0.149)\\
			&  CI  & $(426, 885)$ & $(429, 871)$ & $(410, 934)$ & $(378, 1021)$ & $(447, 874)$\\
			& Coverage  & 100 & 100 & 100 & 100 & 100\\
			\multicolumn{7}{|c|}{}\\
			Independent  & $\hat{N}$(RMAE) & 431 (0.137) & 432 (0.136) & 416 (0.168)  & 393 (0.213) & 446 (0.108)\\
			&  CI  & $(421, 441)$ & $(421, 441)$ & $(385, 401)$  & $(378, 408)$ & $(437, 554)$\\
			& Coverage & 0 & 0 & 0 & 0 & 0\\
      			\multicolumn{7}{|c|}{}\\
			NPLCM & $\hat{N}$(RMAE) & 470 (0.076) & 473 (0.072) & 480 (0.061)  & 441 (0.138) & 486 (0.045)\\
			&  CI  & $(425, 533)$& $(425, 538)$ & $(424, 554)$  & $(379, 523)$ & $(442, 547)$\\
			& Coverage & 86 & 91 & 97 & 69 & 97\\
			\multicolumn{7}{|c|}{}\\
			$M_{bh}$ & $\hat{N}$(RMAE) & 444 (0.113) & 438 (0.124) & 506 (0.077)  & 466 (0.156) & 490 (0.069)\\
			&  CI  & $(410, 554)$ & $(420, 479)$ & $(345, 1.3 \times 10^{9})$  & $(197, 3 \times 10^{9})$ & $(388, 8.1 \times 10^{9})$\\
			& Coverage & 49 & 19 & 96 & 83 & 88\\
			\multicolumn{7}{|c|}{}\\
			\hline
		\end{tabular}
	\end{center}
\end{table}

\begin{table}[h]
\renewcommand\thetable{S6}
	\tiny
	\caption{Comparison of various estimators of the population size $N$ under P4.}
	\label{Tab:5}
	\begin{center}
		\begin{tabular}{|lrccccc|}
			\hline
			\multicolumn{7}{|c|}{}\\
			Model & & $\delta=(1.6,1.2,0.8)$ & $\delta=(1.3, 1.7, 0.9)$ & $\delta=(1,1.4,1.8) $ & $\delta=(0.8,0.8,0.8)$ & $\delta=(1.6,1.6,1.6)$\\
			\hline
			\multicolumn{7}{|c|}{}\\
			\multicolumn{7}{|c|}{}\\
			\multicolumn{7}{|c|}{$N=200$}\\
			\multicolumn{7}{|c|}{}\\
			THBM-I  & $\hat{N}$(RMAE) & 192 (0.074) & 195 (0.076) & 192 (0.080) & 183 (0.122) & 194 (0.063)\\
			& CI  & $(163, 320)$& (163, 326) & $(158, 332)$ & $(141, 352)$ & $(167, 297)$\\
			& Coverage  & 98 & 98 & 97 & 94 & 98\\
			\multicolumn{7}{|c|}{}\\
			THBM-II  & $\hat{N}$(RMAE) & 197 (0.062) & 197 (0.059) & 197 (0.066) &  190 (0.098) & 198 (0.050)\\
			& CI  & $(165, 287)$& $(165, 275)$ & $(161, 286)$ & $(144, 336)$ & $(170, 268)$\\
			& Coverage  & 99 & 99 & 99 & 98 & 99\\
			\multicolumn{7}{|c|}{}\\
			SC  & $\hat{N}$(RMAE) & 172 (0.139) & 171 (0.144) & 166 (0.172) & 156 (0.221) & 176 (0.118)\\
			&  CI  & $(162, 183)$ &$(161, 182)$ & $(153, 180)$  & $(140, 175)$ & $(167, 186)$ \\
			& Coverage  & 4 & 3 & 3 & 6 & 5\\
			\multicolumn{7}{|c|}{}\\
			QSM & $\hat{N}$(RMAE) & 211 (0.125) & 222 (0.164) & 227 (0.186) & 200 (0.172) & 225 (0.162)\\
			&  CI  & $(169, 348)$ & $(172, 391)$ & $(169, 430)$ & $(148, 387)$ & $(177, 390)$\\
			& Coverage  & 96 & 94 & 96 & 94 & 94\\
			\multicolumn{7}{|c|}{}\\
			PQSM & $\hat{N}$(RMAE) & 200 (0.115)& 212 (0.138) & 254 (0.305)  & 205 (0.190) & 226 (0.175)\\
			& CI  & (164, 318)& $(166, 363)$ & $(173, 523)$  & $(147, 713)$ & $(175, 408)$\\
			& Coverage  & 92 & 95 & 91 & 95 & 94\\
			\multicolumn{7}{|c|}{}\\
			LLM  & $\hat{N}$(RMAE) & 204 (0.126) & 208 (0.130) & 244 (0.267) & 205 (0.195) & 226 (0.176)\\
			&  CI  & $(165, 373)$ & $(166, 390)$ & $(169, 501)$ & $(148, 447)$ & $(175, 439)$\\
			& Coverage  & 83 & 82 & 83 & 84 & 84\\
			\multicolumn{7}{|c|}{}\\
			$M_{tb}$  & $\hat{N}$(RMAE) & 165 (0.180) & 179 (0.146) & 166 (0.186) & 168 (0.217) & 170 (0.155)\\
			&  CI  & $(159, 282)$ & $(160, 378)$ & $(154, 340)$ & $(136, 384)$ & $(165, 289)$\\
			& Coverage  & 83 & 100 & 92 & 94 & 88\\
			\multicolumn{7}{|c|}{}\\
			Independent  & $\hat{N}$(RMAE) & 163 (0.186)& 161 (0.195) & 157 (0.213)  & 141 (0.294) & 168 (0.161)\\
			&  CI  & $(158, 167)$& $(156, 165)$ & $(152, 162)$  & $(134, 148)$ & $(163, 171)$\\
			& Coverage & 0 & 0 & 0 & 0 & 0\\
			NPLCM & $\hat{N}$(RMAE) & 179  (0.126) & 179  (0.125) & 178 (0.135)  & 158 (0.231) & 185  (0.096)\\
			&  CI  & $(160,207)$& $(159,208)$ & $   (154, 214)$  & $(135, 193)$ & $(165,214)$\\
			& Coverage & 61 & 63 & 73 & 32 & 79\\
			\multicolumn{7}{|c|}{}\\
			$M_{bh}$ & $\hat{N}$(RMAE) & 161 (0.193) & 160 (0.199) & 158 (0.208)  & 140 (0.302) & 168 (0.158)\\
			&  CI  & $(157, 175)$ & $(157, 168)$ & $(149, 187)$  & $(128, 5.1 \times 10^{8})$ & $(160, 5.7 \times 10^{7})$\\
			& Coverage & 9 & 3 & 17 & 13 & 18\\
			\multicolumn{7}{|c|}{}\\
			\multicolumn{7}{|c|}{}\\
			\multicolumn{7}{|c|}{}\\
			\multicolumn{7}{|c|}{$N=500$}\\
			\multicolumn{7}{|c|}{}\\
			THBM-I  & $\hat{N}$(RMAE) & 485 (0.064) & 490 (0.064) & 483 (0.064) & 468 (0.101) & 491 (0.049)\\
			& CI  & $(418, 783)$& $(417, 790)$ & $(408, 750)$ & $(371,866)$ & $(430, 708)$\\
			& Coverage  & 97 & 98 & 99 & 95 & 99\\
			\multicolumn{7}{|c|}{}\\
			THBM-II  & $\hat{N}$(RMAE) & 493 (0.054) & 495 (0.051) & 497 (0.057) & 479 (0.084) & 499 (0.043)\\
			& CI  & $(424, 673)$& (424, 649) & $(416, 665)$ & $(379, 791)$ & $(436, 626)$\\
			& Coverage  & 99 & 99 & 99 & 98 & 99\\
			\multicolumn{7}{|c|}{}\\
			SC  & $\hat{N}$(RMAE) & 430 (0.352) & 429 (0.143) & 414 (0.171) & 390 (0.221) & 441 (0.119)\\
			&  CI  & $(413, 446)$ &$(412, 445)$ & $(395, 435)$  & $(365, 417)$ & $(426, 456)$ \\
			& Coverage  & 0 & 0 & 0 & 0 & 0\\
			\multicolumn{7}{|c|}{}\\
			QSM & $\hat{N}$(RMAE) & 514 (0.079) & 541 (0.108) & 551 (0.129) & 491 (0.112) & 548 (0.112)\\
			&  CI  & $(444, 659)$ & $(456, 722)$ & $(452, 762)$ & $(399, 690)$ & $(466, 723)$\\
			& Coverage  & 93 & 92 & 91 & 93 & 88\\
			\multicolumn{7}{|c|}{}\\
			PQSM & $\hat{N}$(RMAE) & 486 (0.074) & 513 (0.082) & 589 (0.192)  & 494 (0.116) & 545 (0.110)\\
			& CI  & $(430, 620)$& $(441, 686)$ & $(473, 978)$  & $(404, 777)$ & $(466, 770)$\\
			& Coverage  & 79 & 85 & 72 & 83 & 79\\
			\multicolumn{7}{|c|}{}\\
			LLM  & $\hat{N}$(RMAE) & 494 (0.075) & 504 (0.078) & 571 (0.158) & 494 (0.116) & 543 (0.109)\\
			&  CI  & $(438, 687)$ & $(440, 701)$ & $(464, 914)$ & $(397, 723)$ & $(460, 729)$\\
			& Coverage  & 71 & 73 & 77 & 94 & 91\\
			\multicolumn{7}{|c|}{}\\
			$M_{tb}$  & $\hat{N}$(RMAE) & 422 (0.169) & 457 (0.128) & 472 (0.121) & 490 (0.138) & 445 (0.130)\\
			&  CI  & $(399, 539)$ & $(399, 618)$ & $(387, 722)$ & $(340, 805)$ & $(414, 590)$\\
			& Coverage  & 52 & 59 & 85 & 92 & 68\\
			\multicolumn{7}{|c|}{}\\
			Independent  & $\hat{N}$(RMAE) & 407 (0.186)& 406 (0.188) & 393 (0.214)  & 353 (0.295) & 419 (0.162)\\
			&  CI  & $(399, 414)$& $(398, 413)$ & $(385, 401)$  & $(342, 363)$ & $(412, 424)$\\
			& Coverage & 0 & 0 & 0 & 0 & 0\\
			NPLCM & $\hat{N}$(RMAE) & 452  (0.1138) & 452  (0.114) & 453  (0.118)  & 407  (0.210) & 469 (0.082)\\
			&  CI  & $(403,522)$& $(403,522)$ & $   (391,544)$  & $(344,497)$ & $(419,541)$\\
			& Coverage & 75 & 75 & 87 & 42 & 91\\
			\multicolumn{7}{|c|}{}\\
			$M_{bh}$ & $\hat{N}$(RMAE) & 403 (0.194) & 400 (0.199) & 395 (0.209)  & 348 (0.303) & 420 (0.161)\\
			&  CI  & $(396, 415)$ & $(395, 408)$ & $(384, 416)$  & $(337, 371)$ & $(411, 435)$\\
			& Coverage & 0 & 0 & 1 & 0 & 10\\
			\multicolumn{7}{|c|}{}\\
			\hline
		\end{tabular}
	\end{center}
\end{table}

\begin{table}[h]
\renewcommand\thetable{S7}
	\tiny
	\caption{Comparison of various estimators of the population size $N$ under P5.}
	\label{Tab:6}
	\begin{center}
		\begin{tabular}{|lrccccc|}
			\hline
			\multicolumn{7}{|c|}{}\\
			Model & & $\delta=(1.6,1.2,0.8)$ & $\delta=(1.3, 1.7, 0.9)$ & $\delta=(1,1.4,1.8) $ & $\delta=(0.8,0.8,0.8)$ & $\delta=(1.6,1.6,1.6)$\\
			\hline
			\multicolumn{7}{|c|}{}\\
			\multicolumn{7}{|c|}{}\\
			\multicolumn{7}{|c|}{$N=200$}\\
			\multicolumn{7}{|c|}{}\\
			THBM-I  & $\hat{N}$(RMAE) & 189 (0.080) & 193 (0.083) & 195 (0.079) & 179 (0.132) & 197 (0.062)\\
			& CI  & $(161, 291)$& (160, 305) & $(158, 321)$ & $(140, 325)$ & $(168, 299)$\\
			& Coverage  & 95 & 96 & 99 & 92 & 99\\
			\multicolumn{7}{|c|}{}\\
			THBM-II  & $\hat{N}$(RMAE) & 196 (0.065) & 196 (0.064) & 197 (0.063) &  191 (0.094) & 198 (0.049)\\
			& CI  & $(163, 293)$& $(163, 284)$ & $(161, 286)$ & $(144, 349)$ & $(170, 267)$\\
			& Coverage  & 99 & 99 & 100 & 99 & 99\\
			\multicolumn{7}{|c|}{}\\
			SC  & $\hat{N}$(RMAE) & 171 (0.145) & 170 (0.149) & 164 (0.181) & 155 (0.225) & 176 (0.120)\\
			&  CI  & $(161, 182)$ &$(160, 181)$ & $(151, 178)$  & $(140, 174)$ & $(166, 186)$ \\
			& Coverage  & 2 & 2 & 4 & 6 & 5\\
			\multicolumn{7}{|c|}{}\\
			QSM & $\hat{N}$(RMAE) & 203 (0.114) & 209 (0.130) & 240 (0.240) & 201 (0.175) & 226 (0.171)\\
			&  CI  & $(167, 323)$ & $(167, 346)$ & $(173, 479)$ & $(148, 389)$ & $(177, 398)$\\
			& Coverage  & 94 & 93 & 92 & 95 & 93\\
			\multicolumn{7}{|c|}{}\\
			PQSM & $\hat{N}$(RMAE) & 203 (0.115)& 210 (0.136) & 242 (0.255)  & 207 (0.194) & 226 (0.169)\\
			& CI  & $(167,385)$& $(167, 353)$ & $(172, 470)$  & $(149, 398)$ & $(176, 391)$\\
			& Coverage  & 80 & 94 & 91 & 94 & 94\\
			\multicolumn{7}{|c|}{}\\
			LLM  & $\hat{N}$(RMAE) & 205 (0.129) & 211 (0.139) & 241 (0.256) & 208 (0.201) & 226 (0.173)\\
			&  CI  & $(166, 377)$ & $(168, 401)$ & $(171, 489)$ & $(150, 436)$ & $(178, 446)$\\
			& Coverage  & 83 & 84 & 84 & 85 & 82\\
			\multicolumn{7}{|c|}{}\\
			$M_{tb}$  & $\hat{N}$(RMAE) & 183 (0.120) & 179 (0.130) & 196 (0.101) & 192 (0.136) & 188 (0.101)\\
			&  CI  & $(161, 373)$ & $(160, 378)$ & $(159, 457)$ & $(140, 465)$ & $(168, 383)$\\
			& Coverage  & 99 & 100 & 100 & 100 & 100\\
			\multicolumn{7}{|c|}{}\\
			Independent  & $\hat{N}$(RMAE) & 162 (0.189)& 161 (0.195) & 158 (0.211)  & 141 (0.294) & 168 (0.162)\\
			&  CI  & $(157, 166)$& $(156, 165)$ & $(152, 163)$  & $(134, 148)$ & $(163, 171)$\\
			& Coverage & 0 & 0 & 0 & 0 & 0\\
			NPLCM & $\hat{N}$(RMAE) & 176  (0.138) & 176  (0.136) & 178  (0.133)  & 157  (0.234) & 184  (0.0980)\\
			&  CI  & $(159,199)$& $(159,201)$ & $   (155,208)$  & $(136,186)$ & $(166,210)$\\
			& Coverage & 43 & 48 & 65 & 23 & 71\\
						\multicolumn{7}{|c|}{}\\
			$M_{bh}$ & $\hat{N}$(RMAE) & 170 (0.167) & 166 (0.173) & 181 (0.159)  & 166 (0.261) & 184 (0.138)\\
			&  CI  & $(116, 2.3 \times 10^{12})$ & $(140, 5.1 \times 10^{11})$ & $(43, 1.1 \times 10^{12} )$  & $(15, 5.1 \times 10^{13})$ & $(94, 1.9 \times 10^{13})$\\
			& Coverage & 57 & 41 & 80 & 70 & 72\\
			\multicolumn{7}{|c|}{}\\
			\multicolumn{7}{|c|}{}\\
			\multicolumn{7}{|c|}{$N=500$}\\
			\multicolumn{7}{|c|}{}\\
			THBM-I  & $\hat{N}$(RMAE) & 485 (0.057) & 494 (0.059) & 504 (0.061) & 460 (0.097) & 505 (0.050)\\
			& CI  & $(417, 735)$& $(415, 727)$ & $(411, 754)$ & $(370, 787)$ & $(431, 695)$\\
			& Coverage  & 96 & 99 & 100 & 93 & 100\\
			\multicolumn{7}{|c|}{}\\
			THBM-II  & $\hat{N}$(RMAE) & 491 (0.056) & 496 (0.053) & 496 (0.051) & 483 (0.086) & 497 (0.043)\\
			& CI  & $(421, 683)$& (421, 680) & $(417, 664)$ & $(379, 818)$ & $(435, 629)$\\
			& Coverage  & 98 & 99 & 99 & 99 & 99\\
			\multicolumn{7}{|c|}{}\\
			SC  & $\hat{N}$(RMAE) & 426 (0.148) & 425 (0.151) & 409 (0.182) & 388 (0.225) & 440 (0.120)\\
			&  CI  & $(410, 442)$ &$(408, 442)$ & $(389, 429)$  & $(363, 415)$ & $(425, 455)$ \\
			& Coverage  & 0 & 0 & 0 & 0 & 0\\
			\multicolumn{7}{|c|}{}\\
			QSM & $\hat{N}$(RMAE) & 496 (0.071) & 507 (0.079) & 568 (0.153) & 489 (0.113) & 546 (0.108)\\
			&  CI  & $(436, 620)$ & $(439, 650)$ & $(461, 799)$ & $(398, 686)$ & $(465, 718)$\\
			& Coverage  & 93 & 94 & 86 & 92 & 79\\
			\multicolumn{7}{|c|}{}\\
			PQSM & $\hat{N}$(RMAE) & 496 (0.072) & 509 (0.081) & 572 (0.158)  & 489 (0.114) & 546 (0.108)\\
			& CI  & $(437, 645)$& $(442, 682)$ & $(470, 878)$  & $(404, 733)$ & $(470, 757)$\\
			& Coverage  & 83 & 79 & 76 & 80 & 80\\
			\multicolumn{7}{|c|}{}\\
			LLM  & $\hat{N}$(RMAE) & 497 (0.073) & 509 (0.081) & 572 (0.160) & 490 (0.115) & 545 (0.108)\\
			&  CI  & $(441, 688)$ & $(447, 743)$ & $(475, 964)$ & $(411, 803)$ & $(476, 817)$\\
			& Coverage  & 73 & 72 & 71 & 70 & 71\\
			\multicolumn{7}{|c|}{}\\
			$M_{tb}$  & $\hat{N}$(RMAE) & 453 (0.116) & 458 (0.117) & 493 (0.078) & 521 (0.093) & 474 (0.080)\\
			&  CI  & $(404, 693)$ & $(402, 710)$ & $(399, 812)$ & $(347, 872)$ & $(421, 725)$\\
			& Coverage  & 99 & 99 & 100 & 100 & 100\\
			\multicolumn{7}{|c|}{}\\
			Independent  & $\hat{N}$(RMAE) & 406 (0.189)& 403 (0.195) & 395 (0.209)  & 352 (0.295) & 419 (0.162)\\
			&  CI  & $(398, 412)$& $(395, 409)$ & $(387, 403)$  & $(341, 362)$ & $(412, 425)$\\
			& Coverage & 0 & 0 & 0 & 0 & 0\\
			NPLCM & $\hat{N}$(RMAE) & 445  (0.126) & 446  (0.125) & 451  (0.118)  & 402  (0.214) & 465  (0.087)\\
			&  CI  & $(405,501)$& $(403,505)$ & $   (398,526)$  & $(350,474)$ & $(421,526)$\\
			& Coverage & 48 & 55 & 79 & 21 & 85\\
			\multicolumn{7}{|c|}{}\\
			$M_{bh}$ & $\hat{N}$(RMAE) & 414 (0.172) & 410 (0.181) & 444 (0.131)  & 387 (0.235) & 448 (0.111)\\
			&  CI  & $(392, 1.5 \times 10^{8})$ & $(396, 443)$ & $(322, 6.1 \times 10^{7} )$  & $(274, 4.1 \times 10^{7})$ & $(386, 1.5 \times 10^{9})$\\
			& Coverage & 17 & 6 & 65 & 45 & 55\\
			\multicolumn{7}{|c|}{}\\
			\hline
		\end{tabular}
	\end{center}
\end{table}

\begin{table}[h]
\renewcommand\thetable{S8}
	\tiny
	\caption{Comparison of various estimators of the population size $N$ under P6.}
	\label{Tab:7}
	\begin{center}
		\begin{tabular}{|lrccccc|}
			\hline
			\multicolumn{7}{|c|}{}\\
			Model & & $\delta=(1.6,1.2,0.8)$ & $\delta=(1.3, 1.7, 0.9)$ & $\delta=(1,1.4,1.8) $ & $\delta=(0.8,0.8,0.8)$ & $\delta=(1.6,1.6,1.6)$\\
			\hline
			\multicolumn{7}{|c|}{}\\
			\multicolumn{7}{|c|}{}\\
			\multicolumn{7}{|c|}{$N=200$}\\
			\multicolumn{7}{|c|}{}\\
			THBM-I  & $\hat{N}$(RMAE) & 194 (0.066) & 200 (0.067) & 207 (0.079) & 191 (0.099) & 203 (0.057)\\
			& CI  & $(165, 317)$& $(166, 327)$ & $(165, 363)$ & $(146, 397)$ & $(172, 347)$\\
			& Coverage  & 98 & 99 & 99 & 99 & 100\\
			\multicolumn{7}{|c|}{}\\
			THBM-II  & $\hat{N}$(RMAE) & 195 (0.050) & 196 (0.047) & 196 (0.053) &  192 (0.081) & 196 (0.040)\\
			& CI  & $(168, 270)$& $(169, 259)$ & $(167, 265)$ & $(150, 317)$ & $(174, 248)$\\
			& Coverage  & 100 & 99 & 99 & 99 & 99\\
			\multicolumn{7}{|c|}{}\\
			SC  & $\hat{N}$(RMAE) & 176 (0.122) & 177 (0.116) & 180 (0.103) & 162 (0.190) & 183 (0.086)\\
			&  CI  & $(164, 187)$ &$(165, 190)$ & $(165, 197)$  & $(145, 184)$ & $(172, 194)$ \\
			& Coverage  & 10 & 15 & 35 & 16 & 25\\
			\multicolumn{7}{|c|}{}\\
			QSM & $\hat{N}$(RMAE) & 196 (0.095) & 221 (0.152) & 275 (0.384) & 199 (0.168) & 236 (0.203)\\
			&  CI  & $(167, 279)$ & $(175, 371)$ & $(188, 578)$ & $(151, 361)$ & $(183, 411)$\\
			& Coverage  & 90 & 94 & 77 & 91 & 87\\
			\multicolumn{7}{|c|}{}\\
			PQSM & $\hat{N}$(RMAE) & 195 (0.098) & 222 (0.161) & 277 (0.396)  & 195 (0.161) & 238 (0.207)\\
			& CI  & $(168, 319)$& $(178, 475)$ & $(187, 542)$  & $(149, 346)$ & $(183, 410)$\\
			& Coverage  & 80 & 81 & 79 & 91 & 88\\
			\multicolumn{7}{|c|}{}\\
			LLM  & $\hat{N}$(RMAE) & 204 (0.112) & 214 (0.144) & 253 (0.294) & 197 (0.175) & 237 (0.206)\\
			&  CI  & $(169, 364)$ & $(172, 399)$ & $(176, 504)$ & $(150, 405)$ & $(183, 453)$\\
			& Coverage  & 84 & 84 & 83 & 82 & 80\\
			\multicolumn{7}{|c|}{}\\
			$M_{tb}$  & $\hat{N}$(RMAE) & 185 (0.109) & 186 (0.104) & 201 (0.091) & 199 (0.116) & 193 (0.083)\\
			&  CI  & $(165, 388)$ & $(165, 407)$ & $(166, 484)$ & $(144, 496)$ & $(173, 410)$\\
			& Coverage  & 100 & 100 & 100 & 100 & 100\\
			\multicolumn{7}{|c|}{}\\
			Independent  & $\hat{N}$(RMAE) & 166 (0.168)& 167 (0.167) & 165 (0.176)  & 147 (0.267) & 172 (0.139)\\
			&  CI  & $(161, 171)$& $(161, 171)$ & $(158, 171)$  & $(139, 155)$ & $(167, 176)$\\
			& Coverage & 0 & 0 & 0 & 0 & 0\\
			NPLCM & $\hat{N}$(RMAE) & 180  (0.119) & 183  (0.107) & 189  (0.088)  & 162  (0.214) & 192  (0.065)\\
			&  CI  & $(162,206)$& $(162,213)$ & $   (160,230)$  & $(139,195)$ & $(170,225)$\\
			& Coverage & 60 & 75 & 93 & 34 & 94\\
			\multicolumn{7}{|c|}{}\\
			$M_{bh}$ & $\hat{N}$(RMAE) & 172 (0.149) & 171 (0.149) & 196 (0.156)  & 172 (0.252) & 193 (0.128)\\
			&  CI  & $(119, 5.9 \times 10^{11})$ & $(144, 2.2 \times 10^{10})$ & $(24, 3.4 \times 10^{12} )$  & $(13, 1.5 \times 10^{12})$ & $(76, 2.2 \times 10^{12})$\\
			& Coverage & 55 & 48 & 87 & 72 & 81\\
			\multicolumn{7}{|c|}{}\\
			\multicolumn{7}{|c|}{}\\
			\multicolumn{7}{|c|}{}\\
			\multicolumn{7}{|c|}{$N=500$}\\
			\multicolumn{7}{|c|}{}\\
			THBM-I  & $\hat{N}$(RMAE) & 489 (0.052) & 506 (0.059) & 524 (0.073) & 479 (0.086) & 518 (0.057)\\
			& CI  & $(424, 773)$& $(428, 807)$ & $(431, 834)$ & $(383, 882)$ & $(446, 767)$\\
			& Coverage  & 98 & 99 & 100 & 98 & 100\\
			\multicolumn{7}{|c|}{}\\
			THBM-II  & $\hat{N}$(RMAE) & 489 (0.041) & 490 (0.040) & 492 (0.039) & 479 (0.067) & 496 (0.031)\\
			& CI  & $(431, 654)$& (432, 624) & $(433, 640)$ & $(390, 729)$ & $(448, 615)$\\
			& Coverage  & 99 & 99 & 100 & 98 & 99\\
			\multicolumn{7}{|c|}{}\\
			SC  & $\hat{N}$(RMAE) & 437 (0.125) & 443 (0.115) & 450 (0.101) & 405 (0.190) & 457 (0.086)\\
			&  CI  & $(420, 455)$ &$(424, 462)$ & $(426, 475)$  & $(378, 436)$ & $(440, 474)$ \\
			& Coverage  & 0 & 1 & 7 & 1 & 3\\
			\multicolumn{7}{|c|}{}\\
			QSM & $\hat{N}$(RMAE) & 480 (0.065) & 533 (0.090) & 653 (0.307) & 473 (0.100) & 571 (0.149)\\
			&  CI  & $(433, 570)$ & $(458, 685)$ & $(509, 964)$ & $(398, 628)$ & $(484, 752)$\\
			& Coverage  & 87 & 93 & 50 & 90 & 75\\
			\multicolumn{7}{|c|}{}\\
			PQSM & $\hat{N}$(RMAE) & 479 (0.067) & 534 (0.094) & 657 (0.314)  & 473 (0.101) & 571 (0.149)\\
			& CI  & $(434, 582)$& $(461, 719)$ & $(519, 1042)$  & $(402, 661)$ & $(487, 780)$\\
			& Coverage  & 78 & 82 & 49 & 80 & 67\\
			\multicolumn{7}{|c|}{}\\
			LLM  & $\hat{N}$(RMAE) & 496 (0.065) & 515 (0.077) & 601 (0.207) & 476 (0.102) & 569 (0.147)\\
			&  CI  & $(445, 687)$ & $(453, 713)$ & $(492, 993)$ & $(409, 740)$ & $(490, 850)$\\
			& Coverage  & 72 & 77 & 63 & 73 & 66\\
			\multicolumn{7}{|c|}{}\\
			$M_{tb}$  & $\hat{N}$(RMAE) & 464 (0.100) & 499 (0.069) & 522 (0.053) & 523 (0.087) & 498 (0.053)\\
			&  CI  & $(411, 722)$ & $(414, 785)$ & $(414, 869)$ & $(358, 899)$ & $(433, 774)$\\
			& Coverage  & 100 & 100 & 100 & 100 & 100\\
			\multicolumn{7}{|c|}{}\\
			Independent  & $\hat{N}$(RMAE) & 415 (0.171)& 416 (0.169) & 411 (0.178)  & 366 (0.267) & 431 (0.138)\\
			&  CI  & $(406, 422)$& $(407, 423)$ & $(401, 421)$  & $(354, 378)$ & $(423, 438)$\\
			& Coverage & 0 & 0 & 0 & 0 & 0\\
			NPLCM & $\hat{N}$(RMAE) & 457  (0.104) & 464  (0.092) & 482  (0.065)  & 417  (0.190) & 486  (0.051)\\
			&  CI  & $(411,523)$& $(411,539)$ & $   (408,585)$  & $(354,503)$ & $(430,567)$\\
			& Coverage & 75 & 90 & 99 & 48 & 99\\
			\multicolumn{7}{|c|}{}\\
			$M_{bh}$ & $\hat{N}$(RMAE) & 424 (0.153) & 421 (0.157) & 462 (0.104)  & 395 (0.215) & 461 (0.089)\\
			&  CI  & $(399, 492)$ & $(106, 457)$ & $(353, 728)$  & $(305, 6.9 \times 10^{7})$ & $(400, 1.6 \times 10^{8})$\\
			& Coverage & 23 & 9 & 79 & 44 & 67\\
			\multicolumn{7}{|c|}{}\\
			\hline
		\end{tabular}
	\end{center}
\end{table}

\clearpage

\section{R-programme}{\label{Code}}

\begin{minted}{R}



library(invgamma)
library(MCMCpack)
library(AR)
library(MASS)
library(gamlss.dist)
library(ramify)
library(glogis)
library(HDInterval)
library(coda)

############# Hepatitis A Data #######################
x_111=28
x_110=21
x_101=17
x_011=18
x_100=69
x_010=55
x_001=63

NN=x_111+x_110+x_101+x_011+x_100+x_010+x_001     #sum of seven ovserved cell

x=array(0,3)
x[1]=x_111+x_110+x_101+x_100
x[2]=x_111+x_110+x_001+x_010
x[3]=x_111+x_101+x_011+x_001

########## Bayesian estimate of THBM ######################################

set.seed(1)
tot=5000000    # No. of iteration in Gibbs 

########## Initial values ##########
alpha1=0.1
alpha2=0.2
alpha3=0.3  
alpha4=0.2

delta_1=0.8
delta_2=1
delta_3=1.5

lambda=array(0,3)
lambda[1]=100/delta_1
lambda[2]=100/delta_2
lambda[3]=100/delta_3

gamma=array(0,3) 
gamma[1]=delta_1/lambda[1]   # to maintain the prior mean as delta_1
gamma[2]=delta_2/lambda[2]
gamma[3]=delta_3/lambda[3]

beta=array(0.5,5) # beta parameters are 0.5 refer Jeffrey's Dirichlet prior

########## Parameter vectors ##########
alpha_1=array(NA,tot)
alpha_2=array(NA,tot)
alpha_3=array(NA,tot)
alpha_4=array(NA,tot)
alpha=array(NA,tot)

delta=array(NA,c(3,tot))
b=array(NA,c(3,tot))
p=array(NA,c(3,tot))

y111_1=array(NA,tot)
y111_2=array(NA,tot)
y111_3=array(NA,tot)
y111_4=array(NA,tot)
y111_5=array(NA,tot)

y110_1=array(NA,tot)
y110_2=array(NA,tot)

y011_1=array(NA,tot)
y011_2=array(NA,tot)

y100_1=array(NA,tot)
y100_2=array(NA,tot)

y101_1=array(NA,tot)
y101_2=array(NA,tot)

y010_1=array(NA,tot)
y010_2=array(NA,tot)

y001_1=array(NA,tot)
y001_2=array(NA,tot)

y000_1=array(NA,tot)
y000_2=array(NA,tot)
y000_3=array(NA,tot)
y000_4=array(NA,tot)
y000_5=array(NA,tot)

N_estimate=array(NA,tot)
######################################

alpha_1[1]=alpha1
alpha_2[1]=alpha2
alpha_3[1]=alpha3  
alpha_4[1]=alpha4

alpha[1]=min((alpha_1[1]+alpha_2[1]+alpha_3[1]+alpha_4[1]),1)

delta[1,1]=delta_1
delta[2,1]=delta_2
delta[3,1]=delta_3

for(t in 1:3){
b[t,1]=rglogis(1, location = 0, scale = 1, shape = delta[t,1])
p[t,1]=(1+exp(-b[t,1]))^(-1)
}
N_estimate[1]=rnbinom(1,NN,p[1,1])+NN

######### for loop starts for Gibbs sampling #########

for(h in 2:tot){
p_111=((1-alpha[h-1])*p[1,h-1]*p[2,h-1]*p[3,h-1]) +
(alpha_1[h-1]*p[1,h-1]*p[3,h-1]) +
((alpha_2[h-1]+alpha_3[h-1])*p[1,h-1]*p[2,h-1])+(alpha_4[h-1]*p[1,h-1])

Q_111_1=((1-alpha[h-1])*p[1,h-1]*p[2,h-1]*p[3,h-1])/p_111
Q_111_2=(alpha_1[h-1]*p[1,h-1]*p[3,h-1])/p_111
Q_111_3=(alpha_2[h-1]*p[1,h-1]*p[2,h-1])/p_111
Q_111_4=(alpha_3[h-1]*p[1,h-1]*p[2,h-1])/p_111
Q_111_5=(alpha_4[h-1]*p[1,h-1])/p_111
prob_y111=c(Q_111_1,Q_111_2,Q_111_3,Q_111_4,Q_111_5)
y111_vec_draw=rmultinom(1,x_111,prob_y111)

y111_1[h]=y111_vec_draw[1,1]
y111_2[h]=y111_vec_draw[2,1]
y111_3[h]=y111_vec_draw[3,1]
y111_4[h]=y111_vec_draw[4,1]
y111_5[h]=x_111-(y111_1[h]+y111_2[h]+y111_3[h]+y111_4[h])

p_110=((1-alpha[h-1])*p[1,h-1]*p[2,h-1]*(1-p[3,h-1]))+(alpha_1[h-1]*p[1,h-1]*(1-p[3,h-1]))
y110_1[h]=rbinom(1,x_110,((1-alpha[h-1])*p[1,h-1]*p[2,h-1]*(1-p[3,h-1]))/p_110)
y110_2[h]=x_110-y110_1[h]

p_011=((1-alpha[h-1])*(1-p[1,h-1])*p[2,h-1]*p[3,h-1])+(alpha_2[h-1]*(1-p[1,h-1])*p[2,h-1])
y011_1[h]=rbinom(1,x_011,((1-alpha[h-1])*(1-p[1,h-1])*p[2,h-1]*p[3,h-1])/p_011)
y011_2[h]=x_011-y011_1[h]

p_100=((1-alpha[h-1])*p[1,h-1]*(1-p[2,h-1])*(1-p[3,h-1]))+(alpha_2[h-1]*p[1,h-1]*(1-p[2,h-1]))
y100_1[h]=rbinom(1,x_100,((1-alpha[h-1])*p[1,h-1]*(1-p[2,h-1])*(1-p[3,h-1]))/p_100)
y100_2[h]=x_100-y100_1[h]

p_101=((1-alpha[h-1])*p[1,h-1]*(1-p[2,h-1])*p[3,h-1])+(alpha_3[h-1]*p[1,h-1]*(1-p[2,h-1]))
y101_1[h]=rbinom(1,x_101,((1-alpha[h-1])*p[1,h-1]*(1-p[2,h-1])*p[3,h-1])/p_101)
y101_2[h]=x_101-y101_1[h]

p_010=((1-alpha[h-1])*(1-p[1,h-1])*p[2,h-1]*(1-p[3,h-1])) +
(alpha_3[h-1]*(1-p[1,h-1])*p[2,h-1])
y010_1[h]=rbinom(1,x_010,((1-alpha[h-1])*(1-p[1,h-1])*p[2,h-1]*(1-p[3,h-1]))/p_010)
y010_2[h]=x_010-y010_1[h]

p_001=((1-alpha[h-1])*(1-p[1,h-1])*(1-p[2,h-1])*p[3,h-1]) +
(alpha_1[h-1]*(1-p[1,h-1])*p[3,h-1])
y001_1[h]=rbinom(1,x_001,((1-alpha[h-1])*(1-p[1,h-1])*(1-p[2,h-1])*p[3,h-1])/p_001)
y001_2[h]=x_001-y001_1[h]

p_000=((1-alpha[h-1])*(1-p[1,h-1])*(1-p[2,h-1])*(1-p[3,h-1])) +
(alpha_1[h-1]*(1-p[1,h-1])*(1-p[3,h-1])) +
((alpha_2[h-1]+alpha_3[h-1])*(1-p[1,h-1])*(1-p[2,h-1]))+(alpha_4[h-1]*(1-p[1,h-1]))

Q_000_1=((1-alpha[h-1])*(1-p[1,h-1])*(1-p[2,h-1])*(1-p[3,h-1]))/p_000
Q_000_2=(alpha_1[h-1]*(1-p[1,h-1])*(1-p[3,h-1]))/p_000
Q_000_3=(alpha_2[h-1]*(1-p[1,h-1])*(1-p[2,h-1]))/p_000
Q_000_4=(alpha_3[h-1]*(1-p[1,h-1])*(1-p[2,h-1]))/p_000
Q_000_5=(alpha_4[h-1]*(1-p[1,h-1]))/p_000


prob_y000=c(Q_000_1,Q_000_2,Q_000_3,Q_000_4,Q_000_5)
y000_vec_draw=rmultinom(1,(N_estimate[h-1]-NN),prob_y000)

y000_1[h]=y000_vec_draw[1,1]
y000_2[h]=y000_vec_draw[2,1]
y000_3[h]=y000_vec_draw[3,1]
y000_4[h]=y000_vec_draw[4,1]
y000_5[h]=(N_estimate[h-1]-NN)-(y000_1[h]+y000_2[h]+y000_3[h]+y000_4[h])

d1=y111_2[h]+y000_2[h]+x_110+x_001-y110_1[h]-y001_1[h]+beta[1]
d2=y111_3[h]+y000_3[h]+x_011+x_100-y011_1[h]-y100_1[h]+beta[2]
d3=y111_4[h]+y000_4[h]+x_101+x_010-y101_1[h]-y010_1[h]+beta[3]
d4=x_111+(N_estimate[h-1]-NN)-(y111_1[h]+y111_2[h]+y111_3[h]+y111_4[h])-
(y000_1[h]+y000_2[h]+y000_3[h]+y000_4[h])+beta[4]
d5=y111_1[h]+y110_1[h]+y011_1[h]+y100_1[h]+y101_1[h]+y010_1[h]+y001_1[h]+y000_1[h]+beta[5]

alpha_vec_draw=rdirichlet(1,c(d1,d2,d3,d4,d5))

alpha_1[h]=alpha_vec_draw[1,1]
alpha_2[h]=alpha_vec_draw[1,2]
alpha_3[h]=alpha_vec_draw[1,3]  
alpha_4[h]=alpha_vec_draw[1,4]
alpha[h]=min((alpha_1[h]+alpha_2[h]+alpha_3[h]+alpha_4[h]),1)

m1=x[1]
n1=(N_estimate[h-1]-x[1])

m2=y111_1[h]+y111_3[h]+y111_4[h]+y110_1[h]+x_011+x_010
n2=x_100+x_101+y001_1[h]+y000_1[h]+y000_3[h]+y000_4[h]

m3=y111_1[h]+y111_2[h]+y011_1[h]+y101_1[h]+x_001
n3=x_110+y100_1[h]+y010_1[h]+y000_1[h]+y000_2[h]

d=array(NA,3)
for(t in 1:3){
d[t]=log(1+exp(-(b[t,h-1])))
#delta[t,h]=rgamma(1,shape=(gamma[t]+1),rate=((1/lambda[t])+d[t])) # Gamma prior on delta
delta[t,h]=rexp(1,rate=d[t]) # Jeffrey's prior on delta
} # end of the t loop

b[1,h]=log(rgamma(1,(m1+delta[1,h]),1))-log(rgamma(1,(n1+1),1))
b[2,h]=log(rgamma(1,(m2+delta[2,h]),1))-log(rgamma(1,(n2+1),1))
b[3,h]=log(rgamma(1,(m3+delta[3,h]),1))-log(rgamma(1,(n3+1),1))

p[1,h]=(1+exp(-(b[1,h])))^(-1)
p[2,h]=(1+exp(-(b[2,h])))^(-1)
p[3,h]=(1+exp(-(b[3,h])))^(-1)

w=rnbinom(1,(NN+y000_1[h]+y000_2[h]+y000_3[h]+y000_4[h]),(1-alpha_4[h]*(1-p[1,h])))
N_estimate[h]=w+(NN+y000_1[h]+y000_2[h]+y000_3[h]+y000_4[h])

} # end of the 'h' loop

##################### Estimates ###############################

th=500
lw=tot*0.2
up=tot

M=N_estimate[seq(lw,up, th)]   # Posterior sample of N
Medi=median(M) # Posterior median of N
MD_Medi=sum(abs(M-Medi))/length(M)

Interval_hpd=hdi(M, credMass=0.95) #  95% highest posterior credible interval

CI_hpd_N_1=Interval_hpd[1]  # lower limit of hpd
CI_hpd_N_2=Interval_hpd[2]  # upper limit of hpd
CI_hpd_N_length=Interval_hpd[2]-Interval_hpd[1]   # length of hpd CI

delta1_estimate=median(delta[1,seq(lw,up, th)])
delta2_estimate=median(delta[2,seq(lw,up, th)])
delta3_estimate=median(delta[3,seq(lw,up, th)])

alpha_1_estimate=median(alpha_1[seq(lw,up, th)])
alpha_2_estimate=median(alpha_2[seq(lw,up, th)])
alpha_3_estimate=median(alpha_3[seq(lw,up, th)])
alpha_4_estimate=median(alpha_4[seq(lw,up, th)])
alpha_estimate=alpha_1_estimate+alpha_2_estimate+alpha_3_estimate+alpha_4_estimate

###### Results ###################

N_hat_TBMH=Medi # Estimate of the population size
MD_Median=MD_Medi  # Mean deviation about median
RMDM_N_hat_TBMH=MD_Median/N_hat_TBMH # RMDM of the estimate
hpd_lower_TBMH=CI_hpd_N_1  # Lower limit of the 95% confidence interval based on hpd
hpd_upper_TBMH=CI_hpd_N_2  # Upper limit of the 95% confidence interval based on hpd


\end{minted}


\clearpage

\bibliographystyle{apalike}
\bibliography{Bibliography_TBM}